\documentclass[12pt]{article}
\usepackage[english]{babel}
\usepackage{amsmath}
\usepackage{amsthm}
\usepackage{amssymb}
\usepackage{fullpage}
\usepackage{cite}
\usepackage{comment}
\usepackage{MnSymbol}
\usepackage{xcolor}
\usepackage[colorlinks=true]{hyperref}
\usepackage{aliascnt}
\usepackage{enumitem}

\newtheorem{thm}{Theorem}[section]

\newtheorem*{thm*}{Theorem}
\newaliascnt{cor}{thm}\newtheorem{cor}[cor]{Corollary}\aliascntresetthe{cor}

\newaliascnt{lem}{thm}\newtheorem{lem}[lem]{Lemma}\aliascntresetthe{lem}

\newaliascnt{prop}{thm}\newtheorem{prop}[prop]{Proposition}\aliascntresetthe{prop}

\newtheorem*{prop*}{Proposition}
\theoremstyle{definition}
\newaliascnt{defn}{thm}\newtheorem{defn}[defn]{Definition}\aliascntresetthe{defn}

\newaliascnt{ex}{thm}\aliascntresetthe{ex}

\theoremstyle{remark}
\newaliascnt{rem}{thm}\newtheorem{rem}[rem]{Remark}\aliascntresetthe{rem}

\addto\extrasenglish{}
\numberwithin{equation}{section}

\newcommand{\be}{\begin{equation}}
\newcommand{\ee}{\end{equation}}

\newcommand{\betac}{\beta_c}

\newcommand{\E}{\mathbb{E}}

\newcommand{\Pp}{\mathbb{P}}
\newcommand{\1}{\mathbf{1}}

\newcommand{\dd}{\,\mathrm{d}}
\newcommand{\stepref}[1]{\hyperref[sec:step#1]{Step~#1}}

\title{Exponential mixing time for the Davies dynamics of the quantum random energy model}
\author{Giacomo De Palma\thanks{Department of Mathematics, University of Bologna, Piazza di Porta San Donato 5, 40126 Bologna, Italy. Email: \href{mailto:giacomo.depalma@unibo.it}{\texttt{giacomo.depalma@unibo.it}}}
\and Emanuele Mingione\thanks{Department of Mathematics, University of Bologna, Piazza di Porta San Donato 5, 40126 Bologna, Italy. Email: \href{mailto:emanuele.mingione2@unibo.it}{\texttt{emanuele.mingione2@unibo.it}}}}

\begin{document}
\maketitle

\begin{abstract}
We prove exponentially slow thermalization for the quantum random energy model (QREM) under Davies dynamics. The QREM is the $n$-qubit Hamiltonian $H=\Gamma T+U$, where $U$ assigns independent Gaussian energies with variance equal to the system size to computational-basis configurations and $T$ is the transverse field generated by single-qubit Pauli $X$ operators. Let $\beta_c=\sqrt{2\log 2}$ be the freezing inverse temperature of the classical random energy model. For every inverse temperature $\beta>\beta_c$ and every transverse-field strength $0\leq\Gamma<\Gamma_c(\beta)$, where $e^{\beta\beta_c}=2\cosh(\beta\Gamma_c(\beta))$, we show that the mixing time of the dissipative Davies dynamics associated with single-qubit Pauli system--bath couplings grows exponentially with the system size, almost surely; the bound is unchanged if the standard zero-Bohr-frequency dephasing terms are included. The result is uniform over all Davies rates satisfying detailed balance and a polynomial upper bound and does not assume primitivity.

In the spin-glass phase, the Gibbs weight is carried by a sparse family of deep classical configurations that are separated by an extensive Hamming distance. A localization theorem for the QREM [Manai, Warzel, Comm. Math. Phys. 402, 1259 (2023)] transfers this geometry to the corresponding low-energy quantum eigenstates. Consequently, the Gibbs weight of the intermediate spectral sector is exponentially small, while single-qubit bath operators have exponentially small matrix elements between the two macroscopic sides of the bottleneck. A Davies-specific flux estimate then converts these two facts into an exponential lower bound on the mixing time. This gives a rigorous open-system dynamical manifestation of the frozen, localized low-energy structure of the QREM throughout its equilibrium spin-glass phase below the quantum transition line.
\end{abstract}

\section{Introduction}
Thermalization in a disordered quantum many-body system is controlled by two logically distinct ingredients: the equilibrium distribution of spectral weight and the dynamical mechanisms by which a system coupled to an environment can move between the relevant states. The quantum random energy model (QREM) is a particularly sharp setting in which to separate these ingredients. It combines the extreme-energy statistics and freezing transition of Derrida's random energy model (REM) \cite{Derrida80,Derrida81} with a transverse field that creates genuine quantum tunnelling. Despite the absence of geometric locality in its random diagonal potential, the QREM has a tractable equilibrium phase diagram and a rigorously established low-energy localization regime \cite{Manai_2020,Manai_2023}. The purpose of this work is to show that this localized glassy structure has a direct consequence for weak-coupling open-system dynamics: in the equilibrium spin-glass phase, a Davies thermalization dynamics generated by single-qubit bath couplings mixes only on an exponentially long time scale.

The classical REM may be viewed as an energy landscape on the Boolean hypercube $Q_n=\{0,1\}^n$ in which every configuration $x$ receives an independent Gaussian energy $U(x)$ of variance $n$. At inverse temperature $\beta$ the classical Gibbs measure assigns weight proportional to $e^{-\beta U(x)}$ to the configuration $x$. The critical inverse temperature
\begin{equation}
\beta_c:=\sqrt{2\log 2}
\end{equation}
marks the freezing transition: for $\beta>\beta_c$, an order-one fraction of the Gibbs weight is concentrated on a subexponential collection of exceptionally low-energy configurations. Since those rare configurations are typically separated by Hamming distance proportional to the system size, any dynamics built from one-spin moves encounters a natural geometric bottleneck. This picture underlies the very slow low-temperature dynamics and aging phenomena known for classical REM dynamics \cite{FontesIsopiKohayakawaPicco1998,BenArousBovierGayrard2002,CernyWassmer2017,BenArousBovierGayrard2003a,BenArousBovierGayrard2003b,Gayrard2019}.

The quantum ingredient is equally concrete. On $n$ qubits, let $\{|x\rangle:x\in Q_n\}$ be the computational basis and let $U$ be diagonal in this basis with the REM energies $U(x)$. The QREM Hamiltonian is
\begin{equation}
H=\Gamma T+U,\qquad T=-\sum_{l=1}^n\sigma_1^{(l)},
\end{equation}
where $\sigma_1^{(l)}$ is the Pauli $X$ matrix acting on qubit $l$ and $\Gamma\geq0$ is the transverse-field strength. The transverse field connects configurations at Hamming distance one, so the energy eigenvectors are no longer basis configurations. Nevertheless, at sufficiently low energies the relevant eigenvectors remain exponentially concentrated in Hamming balls around deep REM configurations \cite{Manai_2023}. This localization is the bridge that allows the classical REM bottleneck to survive in the quantum problem.

We couple the QREM weakly to a thermal environment and consider its Davies semigroup \cite{davies1976quantum,Davies1974}. A Davies semigroup is a quantum Markov semigroup obtained in the weak-coupling limit; its generator, called a Lindbladian, satisfies quantum detailed balance and leaves the Gibbs state
\begin{equation}
\omega:=\frac{e^{-\beta H}}{\mathrm{Tr}\,e^{-\beta H}}
\end{equation}
invariant. We take the system--bath coupling operators to be the single-qubit Pauli matrices. Thus the bath is allowed to induce energy-changing transitions, but only through observables acting on one qubit at a time. The relevant dynamical quantity is the mixing time $T_{\mathrm{mix}}$ \cite{TemmeEtAl2010,KastoryanoTemme2013}, namely the time after which the semigroup has lost, in trace norm, an order-one amount of dependence on the initial state. The precise convention is given in \autoref{def:mixing}. This is the quantum analogue of asking how long a reversible continuous-time Markov dynamics takes to approach its Gibbs distribution; the important difference is that transitions are between energy eigenstates and are weighted by matrix elements of the local system--bath operators.

\subsection{Main result and proof strategy}
The equilibrium QREM has a spin-glass region, predicted by Goldschmidt \cite{Goldschmidt1990} and rigorously established by Manai and Warzel \cite{Manai_2020}, characterized by
\begin{equation}
\beta>\beta_c,\qquad 0\leq\Gamma<\Gamma_c(\beta),
\end{equation}
where the transition field $\Gamma_c(\beta)$ is the unique positive solution of
\begin{equation}
e^{\beta\beta_c}=2\cosh\!\bigl(\beta\Gamma_c(\beta)\bigr).
\end{equation}
Our main theorem states that throughout this region the Davies mixing time is exponentially large: there is a deterministic constant $c_{\mathrm{mix}}>0$ such that, for almost every realization of the disorder,
\begin{equation}
\liminf_{n\to\infty}\frac1n\log T_{\mathrm{mix}}\geq c_{\mathrm{mix}}.
\end{equation}
The theorem allows arbitrary nonnegative Davies transition rates satisfying detailed balance, provided their maximum grows at most polynomially with the system size. No lower bound on the rates or primitivity assumption is imposed; cases in which the mixing criterion is never met are covered by the convention $T_{\mathrm{mix}}=+\infty$.

The proof has three ingredients, each with a simple interpretation.

\subsubsection{A classical frozen bottleneck}\label{sec:intro-classical}
We first isolate two subsets of deep REM configurations. One subset carries a non-exponentially-small Gibbs weight, while the other carries strictly more than half of the Gibbs weight; the two subsets are separated by a Hamming distance of order the system size. The construction combines elementary extreme-value estimates for the REM with the deep-hole geometry of low-energy configurations \cite[Lemma~4.2]{Manai_2023}. This is the classical input to the argument; it is carried out in \autoref{sec:classical}.

\subsubsection{Localization turns Hamming separation into dynamical suppression}\label{sec:intro-loc}
A localization theorem of Manai and Warzel \cite[Theorem~1.5]{Manai_2023}, which we use in the precise form of \autoref{thm:Manai} below (see \autoref{rem:MW-threshold}), associates with every deep REM configuration its own QREM eigenstate, exponentially localized around that configuration, and shows that every eigenstate whose energy lies below the corresponding threshold by more than an order-one constant arises in this way. We use this result to convert the two classical regions into spectral subspaces $A$ and $C$ of the QREM. The remaining spectral subspace is denoted by $B$. Because a one-qubit Pauli operator changes computational-basis Hamming distance by at most one, it cannot directly connect the separated localization cores. The exponentially small tails of the localized eigenvectors then imply that the total Davies rate from $A$ to $C$ is exponentially small. Separately, the fact that every eigenstate not associated with a deep configuration has energy at least the deep-configuration threshold minus an order-one constant, together with the QREM pressure formula \cite{Manai_2020}, shows that the Gibbs weight of the intermediate sector $B$ is exponentially small compared with that of $A$. This is the content of Sections~\ref{sec:loc}--\ref{sec:Pi}.

\subsubsection{A direct continuous-time flux bound}\label{sec:intro-flux}
The final step is a bottleneck estimate tailored to Davies generators. Start the semigroup from the Gibbs state conditioned on $A$. Positivity and stationarity prevent the evolving state from accumulating more than the Gibbs-proportional amount of weight in $B$. Hence probability can enter $C$ only through the exponentially small direct $A$--$C$ rate or through the exponentially small Gibbs mass of $B$. Since equilibrium puts more than one half of its mass in $C$, reaching trace-distance mixing requires an exponentially long time. This direct argument is deliberately kept in continuous time. General quantum bottleneck principles are available \cite{gamarnik2024slowmixingquantumgibbs,RakovszkyPlackeBreuckmannKhemani2026}, and our estimate is closely related to the continuous-time bottleneck lemma of \cite[Corollary~3.5]{gamarnik2024slowmixingquantumgibbs}, which starts from the same conditioned Gibbs state (see \autoref{rem:classical-bottleneck}).
For the present Davies generator, the elementary flux estimate gives exactly the needed quantitative bound: direct transitions from $A$ to $C$ are quantified by an explicit rate, and the resulting bound is explicit for every $n$. For energy-diagonal initial states, the Davies dynamics reduces to a reversible classical Markov chain on the energy eigenbasis (\autoref{rem:populations}), and in this language our estimate is a variant of the classical bottleneck-ratio bound. This estimate is proved in \autoref{sec:main}.

Two features of the argument are worth emphasizing. First, the obstruction is not simply a small Hamiltonian spectral gap: it is a bottleneck in the Gibbs-weighted transition structure of the open-system dynamics. Second, the result does not assume that the QREM itself is a geometrically local Hamiltonian. What is local is the bath coupling, through single-qubit Pauli operators. The low-energy eigenvector localization theorem supplies the geometric information needed to exploit that locality even though the random potential is defined independently on all computational-basis configurations.

\subsection{Related works}
The QREM was introduced as a solvable mean-field quantum spin-glass model by Goldschmidt \cite{Goldschmidt1990}, as the $p\to\infty$ limit of $p$-spin glasses in a transverse field; for the quantum $p$-spin glasses themselves see, e.g., \cite{DobrosavljevicThirumalai1990,NieuwenhuizenRitort1998,ObuchiNishimoriSherrington2007}. Its quantum transition and its implications for quantum annealing were studied by J\"org, Krzakala, Kurchan, and Maggs \cite{JorgKrzakalaKurchanMaggs2008,JorgEtAl2010,BapstEtAl2013}; Adame and Warzel proved that, with probability tending to one, the spectral gap above the ground state becomes exponentially small in $n$ along the path that interpolates between the transverse field and the REM \cite{AdameWarzel2015}. The QREM also became a paradigmatic model for many-body localization in Fock space, beginning with the mobility-edge analysis of Laumann, Pal, and Scardicchio \cite{LaumannPalScardicchio2014} and the phase-diagram study of Baldwin, Laumann, Pal, and Scardicchio \cite{BaldwinLaumannPalScardicchio2016}. Subsequent work investigated nonergodic extended behavior, nonequilibrium regimes, and the mobility edge in greater detail \cite{FaoroFeigelmanIoffe2019,BiroliFacoettiSchiroTarziaVivo2021,BalducciEtAl2025}; see also \cite{BaldwinLaumann2018,ParoliniMossi2020} for population transfer and energy matching, and \cite{ScoquartGornyiMirlin2024,ScoquartGornyiMirlin2025} for comparisons with other Fock-space models of many-body localization. All these works concern the closed-system (unitary) dynamics of the QREM. On the rigorous side, the free energy of the QREM is the $p\to\infty$ limit of that of the quantum $p$-spin glass: this was first proved in a joint limit in which $p$ grows with $n$ \cite{ManaiWarzel2021RMP}, and recently, for all $\beta,\Gamma\geq0$, in the iterated limit $n\to\infty$ followed by $p\to\infty$ \cite{KouraichManaiWarzel2025}. Free energies and phase diagrams of hierarchical generalizations of the QREM were obtained in \cite{ManaiWarzelGREM2022,ManaiWarzelAT2022}; for rigorous results on other mean-field quantum spin glasses, see, e.g., \cite{Crawford2007,AdhikariBrennecke2020,LeschkeManaiRuderWarzel2021,ManaiWarzelParisi2025}.

The equilibrium results used here are due to Manai and Warzel. Their rigorous pressure formula establishes the QREM thermodynamic phase diagram \cite{Manai_2020}. Their later spectral analysis \cite{Manai_2023} establishes a localization--delocalization transition for the low-energy eigenvectors: in the paramagnetic regime these eigenvectors are delocalized, while for $\Gamma<\beta_c$ the analysis proves Poisson low-energy statistics, a correspondence between sufficiently low QREM eigenstates and deep REM configurations, and exponential localization of those eigenvectors; the energy thresholds in this correspondence require some care, see \autoref{rem:MW-threshold}. It also identifies the order-one correction to the free energy in the spin-glass phase, which we use in \autoref{sec:Pi}. The latter paper also proves the deep-hole geometry of low-energy REM configurations that we invoke in the classical bottleneck construction. These results are crucial because they allow a statement about local bath operators in the computational basis to be converted into a statement about transitions between energy eigenstates.

On the classical side, metastable and glassy REM dynamics has a long history. Fontes, Isopi, Kohayakawa, and Picco determined the exponential scale of the spectral gap for Metropolis dynamics \cite{FontesIsopiKohayakawaPicco1998}. Ben Arous, Bovier, and Gayrard analyzed the metastable motion of a Glauber dynamics on the extreme states and established aging below the critical temperature \cite{BenArousBovierGayrard2002,BenArousBovierGayrard2003a,BenArousBovierGayrard2003b}; the universality of REM-like aging for mean-field spin glasses was shown in \cite{BenArousBovierCerny2008}, and aging of the Metropolis dynamics was established in \cite{CernyWassmer2017,Gayrard2019}; see \cite{BenArousCerny2006} for a review in terms of trap models. For mean-field spin glasses with finite $p$, exponentially small spectral gaps at low temperature have been derived from bottleneck estimates \cite{BenArousJagannath2018,GheissariJagannath2019,KouraichWarzel2026}. The large deviations of the energy along trajectories of classical REM dynamics have been related to the spectral theory of QREM-type Hamiltonians \cite{GarrahanManaiWarzel2023,ManaiWarzel2025dyn}. Our classical estimates serve a different purpose: rather than analyze a particular classical Markov chain, we construct a Gibbs-geometric partition that can be transferred, through quantum localization, to the spectral decomposition relevant for Davies dynamics.

For quantum Gibbs samplers, Davies generators are the canonical weak-coupling thermal dynamics \cite{davies1976quantum,Davies1974}. Mixing properties of Gibbs samplers have been studied extensively for local and especially commuting Hamiltonians; see, for example, the correlation--mixing results of Kastoryano and Brand\~ao \cite{KastoryanoBrandao2016}, as well as \cite{TemmeEtAl2010,KastoryanoTemme2013,Temme2013,Temme2017,BardetEtAl2023,KochanowskiEtAl2025}. More recent work develops exactly detailed-balanced or KMS-symmetric Gibbs samplers and thermalization generators for general noncommuting Hamiltonians, and emphasizes that their algorithmic cost is controlled by the mixing time \cite{ChenKastoryanoGilyen2023,DingLiLin2025,GilyenChenDoriguelloKastoryano2024,ScandiAlhambra2026}; see also \cite{TemmeEtAl2011,RallWangWocjan2023,ChenKastoryanoBrandaoGilyen2025} for earlier and related quantum algorithms for thermal state preparation. Rapid mixing of such samplers has been proved at sufficiently high temperature \cite{RouzeFrancaAlhambra2026,BakshiEtAl2024,BertoniDePalma2026} and for one-dimensional systems at all temperatures \cite{BergamaschiChen2026}. These developments make lower bounds on thermalization times directly relevant to both open-system physics and quantum algorithms for Gibbs-state preparation, although the algorithmic samplers of \cite{ChenKastoryanoGilyen2023,DingLiLin2025,GilyenChenDoriguelloKastoryano2024} are not Davies generators, so that our theorem does not apply to them directly (see \autoref{sec:discussion}).

Bottleneck obstructions to fast quantum Gibbs sampling have recently been formulated in general terms. Exponentially long relaxation times of Davies generators have been established, e.g., for commuting stabilizer Hamiltonians with macroscopic energy barriers such as Kitaev's four-dimensional toric code \cite{AlickiHorodecki2010}; see also \cite{HongGuoLucas2025,PlackeEtAl2024} for more recent code-based constructions of thermally stable or glassy quantum order. Gamarnik, Kiani, and Zlokapa derive slow-mixing criteria for quantum Gibbs samplers in the form of a quantum bottleneck lemma for general Lindbladians that leave the Gibbs state invariant, with the separation between the two sides of the bottleneck measured either through jumps in the Bohr spectrum or through operator locality. They apply it to classical spin glasses and random constraint satisfaction problems, to good stabilizer codes, to Hamiltonians with a classical part of bounded degree perturbed by a weak transverse field, and to the two-dimensional transverse-field Ising model in its ferromagnetic phase \cite{gamarnik2024slowmixingquantumgibbs}. Rakovszky, Placke, Breuckmann, and Khemani give a complementary channel-based bottleneck theory for finite-temperature phases and quasi-local thermalization dynamics, which they apply to classical spin models and spin glasses, to quantum codes, and to weak perturbations of commuting-projector Hamiltonians with extensive energy barriers \cite{RakovszkyPlackeBreuckmannKhemani2026}. The present setting differs in two respects. The QREM is a genuinely noncommuting random Hamiltonian whose diagonal disorder is not generated by a bounded-degree spatial interaction, and its transverse field is not a small perturbation: our result covers every $\Gamma$ below the first-order transition line $\Gamma_c(\beta)$. Moreover, any bottleneck for a dynamics with energy-resolved jump operators must be expressed in the energy eigenbasis, which for a noncommuting random Hamiltonian is in general not accessible. Indeed, mixing-time lower bounds for noncommuting random quantum Hamiltonians such as quantum spin glasses are listed as an open problem in \cite[Section~1.3]{gamarnik2024slowmixingquantumgibbs}, precisely because the eigenbasis, and hence the jump operators, are not readily available for such Hamiltonians. The low-energy localization theorem provides precisely the missing bridge for the QREM. To our knowledge, the resulting exponential lower bound is the first rigorous mixing-time lower bound for a thermal open-system dynamics of the QREM, and it covers the full equilibrium spin-glass region; we are not aware of earlier results of this kind for any noncommuting random quantum spin-glass Hamiltonian.

\subsection{Organization}
\autoref{sec:setting} introduces the model and the notation used throughout the paper. \autoref{sec:davies} defines the Davies generator, its detailed-balance property and the mixing time. \autoref{sec:main} states the main result, \autoref{thm:main}, proves the continuous-time bottleneck estimate on which it rests, and outlines how the three subspaces $A$, $B$ and $C$ are constructed and controlled. \autoref{sec:classical} constructs the classical bottleneck for the REM, and \autoref{sec:loc} states the low-energy localization theorem for the QREM. \autoref{sec:subsp} combines the two to define $A$, $B$ and $C$. \autoref{sec:Ltilde} bounds the direct Davies transitions from $A$ to $C$, \autoref{sec:Pi} estimates the Gibbs weights of the three subspaces, and \autoref{sec:proof} completes the proof of \autoref{thm:main}. \autoref{sec:discussion} discusses extensions of the result and open problems. \autoref{app:deg} shows that the QREM almost surely has nondegenerate spectrum and nondegenerate gaps, \autoref{app:tails} collects Gaussian and binomial estimates, and \autoref{app:loc} contains the proof of the localization theorem.

\section{Setting and notation}\label{sec:setting}
In this section we fix the model and the notation used throughout the paper.

\subsection{Configurations and Hamming geometry}
Let $Q_n:=\{0,1\}^n$. The Hilbert space of $n$ qubits is $(\mathbb C^2)^{\otimes n}\cong\ell^2(Q_n)$, with computational basis $\{|x\rangle:x\in Q_n\}$. We equip $Q_n$ with the Hamming distance
\begin{equation}\label{eq:hamming-distance}
d(x,x'):=\sum_{l=1}^n|x_l-x_l'|\,,\qquad x,x'\in Q_n,
\end{equation}
and, for $y\in Q_n$ and nonempty $S,S'\subseteq Q_n$, we write
\begin{equation}\label{eq:set-distance}
d(y,S):=\min_{x\in S}d(y,x),\qquad d(S,S'):=\min_{x\in S,\,x'\in S'}d(x,x').
\end{equation}
For $x\in Q_n$ and $r\geq0$,
\begin{equation}\label{eq:ball}
B_r(x):=\{y\in Q_n:d(x,y)\leq r\}
\end{equation}
denotes the Hamming ball of radius $r$ around $x$. We denote by $\sigma_1$, $\sigma_2$ and $\sigma_3$ the Pauli matrices, and by $\sigma^{(l)}_k$ the Pauli matrix $\sigma_k$ acting on the $l$-th qubit. In the computational basis, $\sigma_1^{(l)}$ and $\sigma_2^{(l)}$ map $|x\rangle$ to a multiple of $|x'\rangle$, where $x'$ is obtained from $x$ by flipping its $l$-th coordinate, whereas $\sigma_3^{(l)}$ is diagonal. Hence a single-qubit Pauli operator changes the Hamming distance by at most one:
\begin{equation}\label{eq:pauli-local}
\langle y|\sigma_k^{(l)}|x\rangle=0\qquad\text{whenever }d(x,y)\geq2.
\end{equation}

\subsection{The random energy model}
\begin{defn}[Random energy model]\label{def:REM}
For each $n$, the random potential $U=U_n$ assigns to every configuration $x\in Q_n$ the energy
\begin{equation}
U_n(x)=\sqrt n\,G(x),\qquad G(x)\overset{\mathrm{iid}}{\sim}\mathcal N(0,1),
\end{equation}
so that the energies $U_n(x)$, $x\in Q_n$, are independent Gaussian random variables with mean zero and variance $n$. For different values of $n$, the Gaussian families are assumed independent, and all almost-sure asymptotic statements refer to the resulting product probability space. We write $U_n$ instead of $U$ when we want to stress the dependence on $n$.
\end{defn}
For $\beta>0$, the Gibbs measure and the partition function of the REM at inverse temperature $\beta$ are
\begin{equation}\label{eq:mu-Z}
\mu_{n,\beta}(x):=\frac{e^{-\beta U(x)}}{Z_{n,\beta}},\qquad x\in Q_n,\qquad Z_{n,\beta}:=\sum_{y\in Q_n}e^{-\beta U(y)},
\end{equation}
and for $S\subseteq Q_n$ we write
\begin{equation}
\mu_{n,\beta}(S):=\sum_{x\in S}\mu_{n,\beta}(x).
\end{equation}
The freezing inverse temperature of the REM is
\begin{equation}\label{eq:betac}
\beta_c:=\sqrt{2\log2},
\end{equation}
and, by \autoref{lem:Umin} below, the ground-state energy $\min_{x}U(x)$ equals $-\beta_cn(1+o(1))$ almost surely. For $\delta\in(0,\beta_c)$, we consider the level sets
\begin{equation}\label{eq:R}
\mathcal R_{n,\delta}:=\{x\in Q_n:U(x)\leq-(\beta_c-\delta)n\},
\end{equation}
whose elements we call \emph{deep configurations}: to leading order, their energy lies within $\delta n$ of the ground-state energy.

\subsection{The quantum random energy model}
We use the same symbol $U$ for the diagonal operator $U|x\rangle=U(x)|x\rangle$ on $(\mathbb C^2)^{\otimes n}$. The transverse field and the QREM Hamiltonian are
\begin{equation}\label{eq:qREM}
T:=-\sum_{l=1}^n\sigma_1^{(l)},\qquad H:=\Gamma\,T+U,
\end{equation}
where $\Gamma\geq0$ is the transverse-field strength. For an inverse temperature $\beta>0$, the Gibbs state of $H$ is
\begin{equation}\label{eq:gibbs-state}
\omega:=\frac{e^{-\beta H}}{\mathrm{Tr}\,e^{-\beta H}}.
\end{equation}
For $\beta\geq\beta_c$, we denote by $\Gamma_c(\beta)$ the unique positive solution of
\begin{equation}\label{eq:Gammac}
e^{\beta\beta_c}=2\cosh\bigl(\beta\,\Gamma_c(\beta)\bigr).
\end{equation}
The solution exists and is unique because $\cosh$ is increasing on $[0,\infty)$ and
\begin{equation}
e^{\beta\beta_c}\geq e^{\beta_c^2}=4>2\cosh0;
\end{equation}
moreover, $\Gamma_c(\beta)<\beta_c$, because $2\cosh(\beta\Gamma)>e^{\beta\Gamma}$. The region $\beta>\beta_c$, $0\leq\Gamma<\Gamma_c(\beta)$ is the spin-glass phase of the QREM \cite{Goldschmidt1990,Manai_2020}.
\begin{rem}
    The function $\beta\mapsto\Gamma_c(\beta)$ is increasing on $[\beta_c,\infty)$ with
\begin{equation}
\Gamma_c(\beta_c) = \frac{\mathrm{arccosh}2}{\beta_c}\qquad\text{and}\qquad\lim_{\beta\to\infty}\Gamma_c(\beta) = \beta_c;
\end{equation}
the last two facts are recorded in \cite[Section~1]{Manai_2023}. For $\beta\geq\beta_c$, our definition of $\Gamma_c(\beta)$ coincides with the definition
\begin{equation}
\Gamma_c(\beta):=\beta^{-1}\mathrm{arccosh}\,e^{p_{\mathrm{REM}}(\beta)}
\end{equation}
of \cite{Manai_2020,Manai_2023}, where
\begin{equation}
p_{\mathrm{REM}}(\beta)=\beta\beta_c-\log2
\end{equation}
is the normalized REM pressure; for $\beta<\beta_c$ the definition in \cite{Manai_2020,Manai_2023} is different, but it is not needed here. To see the monotonicity, write $\Gamma_c(\beta)=g(\beta)/\beta$ with $g(\beta):=\mathrm{arccosh}\,y$ and $y:=e^{\beta\beta_c}/2\geq2$. Then
\begin{equation}
g'(\beta)=\frac{\beta_cy}{\sqrt{y^2-1}}>\beta_c,\qquad g(\beta)<\log(2y)=\beta\beta_c;
\end{equation}
hence $\beta g'(\beta)-g(\beta)>0$, i.e., $\Gamma_c'(\beta)>0$.
\end{rem}

\subsection{Asymptotic notation}
Unless stated otherwise, asymptotic statements refer to $n\to\infty$. A statement holds for all sufficiently large $n$ if it holds for all $n\geq n_0$; when the statement is asserted almost surely, $n_0$ may depend on the realization of the disorder. We write $O_{\Gamma,\delta}(\cdot)$ for a bound whose constant depends only on $\Gamma$ and $\delta$, and $\gamma=O(\mathrm{poly}\,n)$ if $\gamma\leq Kn^p$ for constants $K,p>0$ that do not depend on $n$.

\section{The Davies dynamics}\label{sec:davies}
Quantum Markov semigroups describe the evolution of an open quantum system in contact with an external environment.
Their generators, called Lindbladians, have the form \cite{Lindblad1976,GoriniKossakowskiSudarshan1976}
\begin{equation}
\mathcal{L} = \sum_L \mathcal{L}_L\,,
\end{equation}
where the sum runs over a finite family of linear operators $L$ on $\left(\mathbb{C}^2\right)^{\otimes n}$, called Lindblad operators, and
\begin{equation}
    \mathcal{L}_L(\cdot) = L\,\cdot\,L^\dag - \frac{1}{2}\left\{L^\dag\,L\,,\;\cdot\right\}
\end{equation}
are the associated Lindbladians. Here $L^\dag$ denotes the adjoint of $L$.
We omit the Hamiltonian part $-i[K,\,\cdot\,]$ of the general form, which will play no role; see the discussion of the Lamb shift below.

The main physics-motivated quantum Markov semigroup that describes the dynamics of an open quantum system in contact with a thermal bath is the Davies semigroup \cite{davies1976quantum}, which arises from the system--bath Hamiltonian dynamics in the weak-coupling (van Hove) limit \cite{Davies1974,Spohn1980,AlickiLendi2007}.
Let $\left\{|\psi_1\rangle,\,\ldots,\,|\psi_{2^n}\rangle\right\}$ be an orthonormal eigenbasis of the $n$-qubit Hamiltonian $H$ of \eqref{eq:qREM}, and let $E_1<\ldots<E_{2^n}$ be the associated eigenvalues, which we assume to be nondegenerate and with nondegenerate gaps, i.e., such that
\begin{equation}\label{eq:gap-nondeg}
E_i+E_j=E_{i'}+E_{j'}\quad\Longrightarrow\quad\{i,j\}=\{i',j'\}.
\end{equation}
By \autoref{lem:deg} in \autoref{app:deg}, both assumptions hold almost surely for the QREM. A \emph{spectral subspace} of $H$ is the span of a subset of the eigenbasis $\{|\psi_i\rangle\}$.
We use the dissipative Davies dynamics generated by the population-changing components of the single-qubit Pauli couplings. Its Lindbladian is
\begin{equation}\label{eq:defL}
    \mathcal{L} = \sum_{i\neq j=1}^{2^n}\sum_{k=1}^3\sum_{l=1}^n\mathcal{L}_{L_{ijkl}}
\end{equation}
with Lindblad operators
\begin{equation}\label{eq:jump-operators}
\begin{split}
&L_{ijkl} = \sqrt{\gamma_{ijkl}}\,|\psi_i\rangle\langle\psi_i|\sigma^{(l)}_k|\psi_j\rangle\langle\psi_j|\,,\\
&i\neq j=1,\,\ldots,\,2^n\,,\quad k=1,\,2,\,3\,,\quad l=1,\,\ldots,\,n\,,
\end{split}
\end{equation}
where $\sigma^{(l)}_k$ is as in \autoref{sec:setting}, and where the coefficients $\gamma_{ijkl}$ satisfy
\begin{equation}\label{eq:rates}
\begin{split}
&0\le\gamma_{ijkl}\le\gamma_{\max} = O(\mathrm{poly}\,n)\,,\quad \gamma_{jikl} = e^{\beta\left(E_i-E_j\right)}\gamma_{ijkl}\\
&\forall\;i\neq j=1,\,\ldots,\,2^n,\; k=1,\,2,\,3,\; l=1,\,\ldots,\,n\,.
\end{split}
\end{equation}
A standard Davies generator may additionally contain zero-Bohr-frequency dephasing terms. We omit them from \eqref{eq:defL}: because the spectrum is nondegenerate, their jump operators are diagonal in the energy basis, so the corresponding dissipators annihilate every energy-diagonal state and their adjoints annihilate every spectral projector. Hence the population estimates below, and in particular the mixing-time lower bound obtained from the energy-diagonal initial state used in \autoref{thm:direct-bottleneck}, are unchanged if such terms are included. We omit the Lamb-shift Hamiltonian by convention; since it commutes with $H$ and the spectrum is nondegenerate, it is diagonal in the energy eigenbasis, so the corresponding term $-i[H_{\mathrm{LS}},\,\cdot\,]$ also annihilates every energy-diagonal state, its adjoint annihilates every spectral projector, and it leaves the Gibbs state invariant. Hence including it does not affect any of the estimates below either.
The generator \eqref{eq:defL} satisfies the following condition, called Kubo--Martin--Schwinger quantum detailed balance, which ensures that the Gibbs state of $H$ at inverse temperature $\beta$ is a fixed point:
\begin{defn}[KMS quantum detailed balance]
The Lindbladian $\mathcal{L}$ satisfies the Kubo--Martin--Schwinger quantum detailed balance with respect to the state $\omega$ if
\begin{equation}
\sqrt{\omega}\,\mathcal{L}^\dag(\cdot)\,\sqrt{\omega} = \mathcal{L}\left(\sqrt{\omega}\,\cdot\,\sqrt{\omega}\right)\,,
\end{equation}
where $\mathcal{L}^\dag$ is the adjoint of $\mathcal{L}$ with respect to the Hilbert--Schmidt inner product.
In this case, $\mathcal{L}(\omega)=0$, i.e., $\omega$ is a fixed point of $\mathcal{L}$.
\end{defn}
Equivalently, $\mathcal L^\dag$ is self-adjoint with respect to the KMS inner product \cite{KossakowskiFrigerioGoriniVerri1977,FagnolaUmanita2007,CarlenMaas2017}
\begin{equation}
\langle X,Y\rangle_\omega:=\mathrm{Tr}(X^\dag\sqrt\omega\,Y\sqrt\omega)\,.
\end{equation}
Davies generators satisfy the stronger GNS detailed-balance condition \cite{Alicki1976,KossakowskiFrigerioGoriniVerri1977}. The only consequence of detailed balance used in the proof of our main result is the stationarity $\mathcal L(\omega)=0$.

\begin{prop}\label{prop:fixed}
The generator of the Davies semigroup \eqref{eq:defL} satisfies KMS quantum detailed balance for the Gibbs state $\omega$ of $H$ at inverse temperature $\beta$, defined in \eqref{eq:gibbs-state}.
\end{prop}

\begin{proof}
This is the standard detailed-balance property of Davies generators \cite{davies1976quantum}; we record the short verification in the present notation. The rate relation gives, for every $i\neq j,k,l$,
\begin{equation}
\gamma_{ijkl}e^{-\beta E_j}=\gamma_{jikl}e^{-\beta E_i},
\end{equation}
and therefore
\begin{equation}
\sqrt{\omega}\,L_{ijkl}^\dag(\,\cdot\,)L_{ijkl}\sqrt{\omega}
=L_{jikl}\sqrt{\omega}(\,\cdot\,)\sqrt{\omega}\,L_{jikl}^\dag.
\end{equation}
Moreover,
\begin{equation}
L_{ijkl}^\dag L_{ijkl}
=\gamma_{ijkl}|\langle\psi_i|\sigma_k^{(l)}|\psi_j\rangle|^2|\psi_j\rangle\langle\psi_j|,
\end{equation}
which commutes with $\omega$. Pairing the jump terms with indices $(i,j)$ and $(j,i)$ and using this commutation for the anticommutator terms yields
\begin{equation}
\sqrt\omega\,\mathcal L^\dag(\,\cdot\,)\sqrt\omega=\mathcal L(\sqrt\omega(\,\cdot\,)\sqrt\omega).
\end{equation}
\end{proof}

\begin{rem}[Population dynamics]\label{rem:populations}
The generator \eqref{eq:defL} maps energy-diagonal operators to energy-diagonal operators: a direct computation gives, for every $j$,
\begin{equation}\label{eq:L-on-diagonal}
\mathcal L\bigl(|\psi_j\rangle\langle\psi_j|\bigr)=\sum_{i\neq j}q_{j\to i}\bigl(|\psi_i\rangle\langle\psi_i|-|\psi_j\rangle\langle\psi_j|\bigr),\qquad
q_{j\to i}:=\sum_{k=1}^3\sum_{l=1}^n\gamma_{ijkl}\bigl|\langle\psi_i|\sigma_k^{(l)}|\psi_j\rangle\bigr|^2.
\end{equation}
Hence, for an initial state that is diagonal in the energy eigenbasis, the populations
\begin{equation}
p_j(t):=\langle\psi_j|e^{t\mathcal L}(\rho)|\psi_j\rangle
\end{equation}
evolve as a continuous-time Markov chain on the set of eigenvectors with jump rates $q_{j\to i}$. The rate relation in \eqref{eq:rates} gives
\begin{equation}
e^{-\beta E_j}q_{j\to i}=e^{-\beta E_i}q_{i\to j},
\end{equation}
i.e., this chain is reversible with respect to the Gibbs weights $\langle\psi_j|\omega|\psi_j\rangle$. The quantum nature of the problem enters only through the rates $q_{j\to i}$, that is, through the matrix elements of the single-qubit Pauli operators between eigenvectors of $H$; these are controlled in \autoref{sec:Ltilde} by means of the localization of the low-energy eigenvectors. For $\Gamma=0$ the eigenbasis is the computational basis, $\sigma_3^{(l)}$ induces no transitions, and the chain is a reversible single-spin-flip dynamics of the classical REM with rates
\begin{equation}
q_{x\to x'}=\begin{cases}\gamma_{x'x1l}+\gamma_{x'x2l}&\text{if $x'$ is obtained from $x$ by flipping the spin $l$},\\
0&\text{if }d(x,x')\neq1.\end{cases}
\end{equation}
\end{rem}

Finally, we define the mixing time. We denote by $\mathcal D\bigl((\mathbb C^2)^{\otimes n}\bigr)$ the set of density operators on $(\mathbb C^2)^{\otimes n}$ and by $\|\cdot\|_1$ the trace norm.
\begin{defn}[Mixing time]\label{def:mixing}
The mixing time of a Lindbladian $\mathcal L$ on $(\mathbb C^2)^{\otimes n}$ is
\begin{equation}\label{eq:Tmix}
T_{\mathrm{mix}}(\mathcal L):=\inf\left\{t\geq0:\left\|e^{t\mathcal L}(\rho-\rho')\right\|_1\leq1\quad\forall\;\rho,\rho'\in\mathcal D\bigl((\mathbb C^2)^{\otimes n}\bigr)\right\},
\end{equation}
with the convention $\inf\varnothing=+\infty$.
\end{defn}
Since $e^{t\mathcal L}$ is contractive in trace norm, $T_{\mathrm{mix}}(\mathcal L)$ is the time after which the trace distance between any two evolved states is at most one half:
\begin{equation}
\frac12\left\|e^{t\mathcal L}(\rho)-e^{t\mathcal L}(\rho')\right\|_1\leq\frac12\qquad\forall\;t\geq T_{\mathrm{mix}}(\mathcal L),\;\rho,\rho'\in\mathcal D\bigl((\mathbb C^2)^{\otimes n}\bigr);
\end{equation}
see \cite{TemmeEtAl2010,KastoryanoTemme2013} for this notion of mixing time. No primitivity assumption is needed: if the criterion is never met, then $T_{\mathrm{mix}}(\mathcal L)=+\infty$.

\section{\texorpdfstring{Main result and the bottleneck estimate}{Main result and the bottleneck estimate}}\label{sec:main}
We can now state our main result.
\begin{thm}[Exponential mixing time for the Davies semigroup for the quantum REM]\label{thm:main}
Let $H=\Gamma T+U$ be the QREM \eqref{eq:qREM}, with the random potential $U$ of \autoref{def:REM}, and let $\mathcal L$ be the generator \eqref{eq:defL} of the dissipative Davies dynamics, with rates satisfying \eqref{eq:rates} (by \autoref{lem:deg}, $H$ almost surely has nondegenerate spectrum and nondegenerate energy gaps, so that $\mathcal L$ is almost surely well defined). Then, for any $\beta>\beta_c$ and any $0\le\Gamma<\Gamma_c(\beta)$, there is a deterministic constant $c_{\mathrm{mix}}>0$ such that the mixing time \eqref{eq:Tmix} of $\mathcal{L}$ almost surely satisfies
\begin{equation}
\liminf_{n\to\infty}\frac{\mathop{\log}\nolimits T_{\mathrm{mix}}(\mathcal{L})}{n} \ge c_{\mathrm{mix}}\,.
\end{equation}
\end{thm}
The assumption that $\gamma_{\max}$ grows at most polynomially can be weakened to subexponential growth, and the threshold $1$ in \autoref{def:mixing} can be replaced by any threshold in $(0,2)$; see \autoref{sec:discussion}.

The proof of \autoref{thm:main} uses the following direct continuous-time bottleneck estimate for the Davies generator. It is a Davies-specific version of the quantum bottleneck principle developed in \cite{gamarnik2024slowmixingquantumgibbs,RakovszkyPlackeBreuckmannKhemani2026}; in particular, it is closely related to \cite[Proposition~3.4 and Corollary~3.5]{gamarnik2024slowmixingquantumgibbs}, see \autoref{rem:classical-bottleneck}. For the present generator, the direct flux argument below is shorter than translating the general frameworks and gives exactly the quantitative estimate needed later.
\begin{thm}[Direct continuous-time bottleneck estimate]\label{thm:direct-bottleneck}
Let $A,B,C$ be mutually orthogonal spectral subspaces of $H$ whose direct sum is the full Hilbert space, with projectors $\Pi_A,\Pi_B,\Pi_C$. Let $\omega$ be the Gibbs state and set
\begin{equation}
w_A:=\mathrm{Tr}(\Pi_A\omega)>0,\qquad {w_B}:=\mathrm{Tr}(\Pi_B\omega),\qquad {w_C}:=\mathrm{Tr}(\Pi_C\omega)>\frac12.
\end{equation}
For the Davies generator \eqref{eq:defL}, define
\begin{align}
r_A&:=\max_{j:\,|\psi_j\rangle\in A}\sum_{i:\,|\psi_i\rangle\in C}\sum_{k=1}^3\sum_{l=1}^n\gamma_{ijkl}\left|\langle\psi_i|\sigma_k^{(l)}|\psi_j\rangle\right|^2,\\
r_B&:=\max_{j:\,|\psi_j\rangle\in B}\sum_{i:\,|\psi_i\rangle\in C}\sum_{k=1}^3\sum_{l=1}^n\gamma_{ijkl}\left|\langle\psi_i|\sigma_k^{(l)}|\psi_j\rangle\right|^2.
\end{align}
Then,
\begin{equation}\label{eq:direct-bottleneck}
T_{\mathrm{mix}}(\mathcal L)\geq \frac{w_C-\frac12}{r_B\,w_B/w_A+r_A}.
\end{equation}
Moreover,
\begin{equation}\label{eq:rB-bound}
r_B\leq 3n\gamma_{\max}.
\end{equation}
\end{thm}
\begin{proof}
The idea is to start the dynamics from the Gibbs state conditioned on $A$ and to show that the weight of $C$ can grow at most linearly in time, with a small slope, whereas the mixing criterion requires this weight to reach at least $w_C-\frac12$.
Define
\begin{equation}
\rho_0:=\frac{\Pi_A\omega}{w_A}.
\end{equation}
Since $\Pi_A$ commutes with $\omega$,
\begin{equation}
0\leq\rho_0\leq\frac{\omega}{w_A}.
\end{equation}
The completely positive semigroup $e^{t\mathcal L}$ preserves operator order and fixes $\omega$ (by \autoref{prop:fixed}, $\mathcal L(\omega)=0$), hence
\begin{equation}
\rho_t:=e^{t\mathcal L}(\rho_0)\leq \frac{\omega}{w_A}\quad(t\geq0),
\end{equation}
so
\begin{equation}
\mathrm{Tr}(\Pi_B\rho_t)\leq\frac{w_B}{w_A}.
\end{equation}
Let
\begin{equation}
p_C(t):=\mathrm{Tr}(\Pi_C\rho_t),\qquad\text{so that}\qquad p_C'(t)=\mathrm{Tr}\bigl(\mathcal L^\dag(\Pi_C)\rho_t\bigr).
\end{equation}
A direct computation from \eqref{eq:defL} gives the explicit diagonal formula
\begin{equation}\label{eq:LdagPiC}
\begin{split}
\mathcal L^\dag(\Pi_C)
&=\sum_{j:\,|\psi_j\rangle\in A\cup B}\left(\sum_{i:\,|\psi_i\rangle\in C}\sum_{k=1}^3\sum_{l=1}^n
\gamma_{ijkl}|\langle\psi_i|\sigma_k^{(l)}|\psi_j\rangle|^2\right)|\psi_j\rangle\langle\psi_j|\\
&\quad-\sum_{j:\,|\psi_j\rangle\in C}\left(\sum_{i:\,|\psi_i\rangle\in A\cup B}\sum_{k=1}^3\sum_{l=1}^n
\gamma_{ijkl}|\langle\psi_i|\sigma_k^{(l)}|\psi_j\rangle|^2\right)|\psi_j\rangle\langle\psi_j|.
\end{split}
\end{equation}
In this formula, the terms with source index in $C$ are nonpositive, while the terms with source index in $A$ and $B$ give
\begin{equation}
p_C'(t)\leq r_A\,\mathrm{Tr}(\Pi_A\rho_t)+r_B\,\mathrm{Tr}(\Pi_B\rho_t)
\leq r_A+r_B\frac{w_B}{w_A}.
\end{equation}
Since $p_C(0)=0$,
\begin{equation}\label{eq:pC-upper}
p_C(t)\leq t\left(r_A+r_B\frac{w_B}{w_A}\right).
\end{equation}
For any $t$ satisfying the defining mixing criterion, stationarity of $\omega$ gives
\begin{equation}
\|\rho_t-\omega\|_1\leq1.
\end{equation}
Write $\rho_t-\omega=X_+-X_-$ with $X_\pm\geq0$ and $X_+X_-=0$. Since $\mathrm{Tr}(\rho_t-\omega)=0$, we have
\begin{equation}
\mathrm{Tr}\,X_-=\frac12\|\rho_t-\omega\|_1\leq\frac12,
\end{equation}
hence
\begin{equation}
\mathrm{Tr}\bigl(\Pi_C(\rho_t-\omega)\bigr)\geq-\mathrm{Tr}(\Pi_CX_-)\geq-\mathrm{Tr}\,X_-\geq-\frac12.
\end{equation}
Therefore
\begin{equation}
p_C(t)\geq \mathrm{Tr}(\Pi_C\omega)-\frac12=w_C-\frac12.
\end{equation}
Combining this with \eqref{eq:pC-upper} and taking the infimum over all such $t$ proves \eqref{eq:direct-bottleneck}. Finally, by completeness of the energy eigenbasis and unitarity of each Pauli matrix,
\begin{equation}
r_B\leq \gamma_{\max}\max_{j:\,|\psi_j\rangle\in B}\sum_{i=1}^{2^n}\sum_{k=1}^3\sum_{l=1}^n
|\langle\psi_i|\sigma_k^{(l)}|\psi_j\rangle|^2\leq 3n\gamma_{\max}.
\end{equation}
\end{proof}
\begin{rem}[Relation to classical bottleneck bounds]\label{rem:classical-bottleneck}
The initial state $\rho_0$ in the proof is diagonal in the energy eigenbasis, so by \autoref{rem:populations} the proof of \autoref{thm:direct-bottleneck} only concerns the reversible Markov chain of the populations, and $r_A$ and $r_B$ are the maximal total jump rates from $A$ and from $B$ into $C$. In this language, \eqref{eq:direct-bottleneck} is a continuous-time variant of the classical bottleneck-ratio (conductance) lower bound on mixing times of reversible Markov chains \cite[Theorem~7.4]{LevinPeresWilmer2017}, see also \cite{LawlerSokal1988,SinclairJerrum1989}: with $S:=A\oplus B$, the Gibbs-weighted escape rate from $S$ into $C$ satisfies
\begin{equation}
\frac{1}{w_A+w_B}\sum_{j\in A\cup B}\langle\psi_j|\omega|\psi_j\rangle\sum_{i\in C}q_{j\to i}\leq\frac{w_Ar_A+w_Br_B}{w_A+w_B}\leq r_A+r_B\frac{w_B}{w_A}.
\end{equation}
The three-set formulation is convenient because the eigenvectors in $B$ are not localized, so that $r_B$ is only bounded by the polynomial quantity \eqref{eq:rB-bound}; this is compensated by the exponential smallness of $w_B/w_A$. The same initial state $\rho_0$ and the same domination $\rho_t\leq\omega/w_A$ are used in the quantum bottleneck lemma of Gamarnik, Kiani, and Zlokapa \cite[Proposition~3.4 and Corollary~3.5]{gamarnik2024slowmixingquantumgibbs}, which applies to general Lindbladians with Gibbs fixed point. There, direct transitions from $A$ to $C$ are excluded for an auxiliary generator that approximates the given one in diamond norm, and the conclusion is asymptotic; for the Davies generator, the reduction to populations allows us to quantify these transitions directly by $r_A$ and to obtain the explicit bound \eqref{eq:direct-bottleneck}.
\end{rem}
\subsection{\texorpdfstring{Strategy of the proof of \autoref{thm:main}}{Strategy of the proof of the main theorem}}\label{sec:strategy}
We will construct $A,B,C$ so that $w_C>1/2$, $w_B/w_A$ is exponentially small, and $r_A$ is exponentially small.
We summarize how the subspaces are chosen and how the three requirements of \autoref{thm:direct-bottleneck} are verified in Sections~\ref{sec:classical}--\ref{sec:Pi}. Fix a small $\delta>0$ and consider the deep configurations, i.e., the elements of the level set $\mathcal R_{n,\delta}$ of \eqref{eq:R}. With high probability, distinct deep configurations are at Hamming distance of order $n$ (\autoref{prop:deep-hole}), and, by the localization theorem (\autoref{thm:Manai} below), each deep configuration $x$ carries its own eigenvector $\psi_x$ of $H$, exponentially localized in a Hamming ball around $x$. We split the deep configurations into the very deepest ones, $\mathcal C_n'$, whose energy lies within $O(\log n)$ of the typical ground-state energy, and the remaining ones, $\mathcal A_n'$ (\autoref{thm:classical}). The subspaces $A$ and $C$ are spanned by the eigenvectors associated with $\mathcal A_n'$ and $\mathcal C_n'$, respectively, and $B$ is spanned by all the other eigenvectors. Then:
\begin{itemize}
\item $w_C>\frac12$, because the classical Gibbs weight of $\mathcal C_n'$ is eventually at least $\frac34$ (\autoref{thm:classical}) and the quantum Gibbs weights of $A$ and $C$ are asymptotically at least the classical weights of $\mathcal A_n'$ and $\mathcal C_n'$ (\autoref{lem:Pi});
\item $w_B/w_A$ is exponentially small, because all eigenvalues in $B$ lie above $-(\beta_c-\delta)n-O(1)$, whereas for $\beta>\beta_c$ the Gibbs state is carried by energies close to $-\beta_cn$ (\autoref{lem:PiB}), and because $w_A$ is not exponentially small (\autoref{cor:PiA});
\item $r_A$ is exponentially small, because a single-qubit Pauli operator changes the Hamming distance by at most one, the localization centres in $\mathcal A_n'$ and in $\mathcal C_n'$ are at distance of order $n$, the localized eigenvectors have exponentially small tails, and $\mathcal C_n'$ has subexponential cardinality (\autoref{lem:epsilon}).
\end{itemize}
The intermediate set $\mathcal A_n'$ is needed because the eigenvectors in $B$ are not known to be localized: for them we only have the polynomial bound \eqref{eq:rB-bound} on the rate into $C$, which must be compensated by the smallness of $w_B$ relative to $w_A$.

\section{The classical REM}\label{sec:classical}
In this section, $\Gamma$ plays no role: we work with the classical REM of \autoref{def:REM}, which is the QREM at $\Gamma=0$, where $H=U$ is diagonal. The main result of this section is the following theorem, which establishes a bottleneck for the Gibbs measure $\mu_{n,\beta}$ of \eqref{eq:mu-Z} in the low-temperature regime $\beta>\beta_c$. Its almost-sure consequence, which is the form used in the rest of the paper, is \autoref{thm:classical} at the end of the section.

 \begin{thm}[Classical Bottleneck]\label{prop:two-separated-regions}
Let $\beta>\beta_c$, $c\in(0,1)$,
$\theta>0$ and $\delta$ such that   

\be
0<\delta<\beta_c\left(1-\frac{1}{\sqrt{2}}\right).
\label{eq:delta-condition-separated-regions}
\ee
Then there exist $a\in(0,\frac12)$   and $\kappa>0$ with the following
property:  if
\be
\mathcal C_n':=\mathcal{R}_{n,\delta_{n}},
\qquad
\mathcal A_n':=\mathcal{R}_{n,\delta}\setminus \mathcal{R}_{n,\delta_{n}}
\label{eq:AC-separated-regions}
\ee
where 
\be
\delta_{n}:=\frac{\log n}{2\beta_cn}+\frac{\kappa \log n}{n},
\label{eq:xn-separated-regions}
\ee

then 

\begin{enumerate}

\item\label{it:sep-prob} there exist events $\Omega_n'\equiv\Omega'_{n,\beta,\delta,c,\theta}$ and a constant $C>0$ such
that, 

\be
\Pp(\Omega_n')
\geq 1-Cn^{-\theta}.
\label{eq:probability-separated-regions}
\ee
\item\label{it:sep-dist} on $\Omega_n'$  and for all sufficiently large $n$ one has  
\be
d(\mathcal A_n',\mathcal C_n')\geq an.
\label{eq:AC-distance}
 \ee

\item\label{it:sep-card} almost surely,
\be
\limsup_{n\to\infty}
\frac{1}{n}\log|\mathcal C_n'|=0,
\label{eq:C-subexponential}
\ee
and
\be
\lim_{n\to\infty}
\frac{1}{n}\log\mu_{n,\beta}(\mathcal A_n')=0.
\label{eq:A-nonexponential-mass}
\ee
\item\label{it:sep-mass} If $\theta>1$, then almost surely
\be
\liminf_{n\to\infty}
\mu_{n,\beta}(\mathcal C_n')\geq c.
\label{eq:C-positive-mass-as}
\ee
\end{enumerate}
\end{thm}

 The proof, given in \autoref{subsec:proof-classical}, requires some intermediate results. It combines three ingredients, each providing one of the three events whose intersection is $\Omega_n'$: (i) the level set $\mathcal R_{n,\delta_n}$, which consists of the configurations whose energy lies within $\kappa\log n$ above the typical ground-state energy $-\beta_cn+\frac{\log n}{2\beta_c}$ (see \eqref{eq:cn} below), has subexponential cardinality and carries Gibbs mass at least $c$ with probability $1-O(n^{-\theta})$ (\autoref{prop:splitting-support}); (ii) distinct configurations of $\mathcal R_{n,\delta}$ are at Hamming distance of order $n$ with probability exponentially close to one (\autoref{prop:deep-hole}); (iii) with overwhelming probability, $\mathcal A_n'$ contains a configuration whose energy exceeds the typical ground-state energy by at most $\kappa\log n+1$, which forces $\mu_{n,\beta}(\mathcal A_n')$ to be at most polynomially small (the shell argument in the proof of \autoref{it:sep-prob}).

\subsection{\texorpdfstring{Free energy and ground-state energy}{Free energy and ground-state energy}}
\begin{prop}[REM free energy]\label{prop:free-energy-limit}
Define
\begin{equation}
f_{n,\beta}:=-(\beta n)^{-1}\log Z_{n,\beta}.
\end{equation}
The thermodynamic limit of the free energy density at any $\beta>0$ is given by
\be
\lim_{n\to\infty} -\frac{1}{\beta n}\log Z_{n,\beta}
=
\begin{cases}
-\dfrac{\beta}{2}-\dfrac{\log 2}{\beta},&\beta\leq\beta_c,\\[1ex]
-\beta_c,&\beta>\beta_c.
\end{cases}
\label{eq:thermodynamic-free-energy}
\ee
 In particular, the limiting ground-state energy density is
\be
\lim_{\beta\to\infty}\lim_{n\to\infty}f_{n,\beta}=-\beta_c.
\label{eq:ground-state-energy}
\ee
All the above convergences are a.s. and in $L^p$ for $1\leq p<\infty$.
\end{prop}
The above result  goes back to \cite{Derrida80,Derrida81}, in this form a proof can be found in \cite{Picco, BovierBook}. Equation~\eqref{eq:thermodynamic-free-energy} suggests that for large $n$ and low temperatures the Gibbs measure concentrates on a sub-exponential number of spin configurations whose energy density $U_n(x)/n$ is close to the limiting ground-state energy density $-\beta_c$ of \eqref{eq:ground-state-energy}. Much more is known: for $\beta>\beta_c$ the Gibbs measure is asymptotically carried by the configurations whose energies lie within $O(1)$ of the minimum, and the ranked Gibbs weights converge in distribution to the Poisson--Dirichlet law with parameter $\beta_c/\beta$ \cite{Ruelle87,BovierKurkovaLowe2002,BovierBook,PY97}. \autoref{prop:splitting-support} below is a quantitative, almost-sure counterpart of the (weak) part of this picture that we need: a level set of subexponential cardinality that carries a prescribed fraction of the Gibbs mass, with summable failure probabilities so that the Borel--Cantelli lemma applies.

The limit \eqref{eq:ground-state-energy} identifies the leading order of the ground-state energy of the REM; we record it in the following form.
\begin{lem}[Ground-state energy of the REM]\label{lem:Umin}
Let
\begin{equation}
U_{\min}:=\min_{x\in Q_n}U(x).
\end{equation}
Then
\begin{equation}\label{eq:Umin-limit}
\lim_{n\to\infty}\frac{U_{\min}}{n}=-\beta_c\qquad\text{almost surely}.
\end{equation}
\end{lem}
\begin{proof}
For every fixed $t>\beta_c$,
\begin{equation}
 e^{-tU_{\min}}\le Z_{n,t}\le 2^n e^{-tU_{\min}},
\end{equation}
so
\begin{equation}
 \frac{U_{\min}}n-\frac{\log2}{t}\le f_{n,t}\le\frac{U_{\min}}n.
\end{equation}
For each fixed $t>\beta_c$, \autoref{prop:free-energy-limit} gives $f_{n,t}\to-\beta_c$ almost surely. Taking a countable sequence $t_m\to\infty$ and intersecting the corresponding probability-one events, the two inequalities imply
\begin{equation}
\liminf_{n\to\infty}\frac{U_{\min}}n\ge-\beta_c,\qquad\limsup_{n\to\infty}\frac{U_{\min}}n\le-\beta_c+\frac{\log2}{t_m}\qquad\text{for every }m;
\end{equation}
hence \eqref{eq:Umin-limit} holds.
\end{proof}

\subsection{\texorpdfstring{Level sets carrying the Gibbs mass}{Level sets carrying the Gibbs mass}}
For $\beta>0$ and $\delta\in(0,\beta_c)$, consider the truncated partition function:
\be
{Z_\delta:=\sum_{x\in Q_n}e^{-\beta U_n(x)}\1(x\in\mathcal R_{n,\delta}).}
\label{eq:truncated-partition}
\ee
{Its deviation from $Z_{n,\beta}$ is controlled by the following lemma.}

\begin{lem}\label{lem:truncated-partition-tail}
Let $\beta>\beta_c$ and $\delta\in(0,\beta_c)$. For any $\varepsilon>0$ one has
\be
\Pp(Z_{n,\beta}-Z_\delta\geq\varepsilon)
\leq
\frac{e^{ng(\delta,\beta)}}{\varepsilon},
\label{eq:truncation-markov}
\ee
where
\be
g(\delta,\beta):=
\beta\beta_c-\frac{\delta^2}{2}
-\delta(\beta-\beta_c).
\label{eq:g-function}
\ee
\end{lem}

\begin{proof}
By Markov's inequality it is enough to show that
\be
\E(Z_{n,\beta}-Z_\delta)
\leq e^{ng(\delta,\beta)}.
\label{eq:EZ-bound}
\ee
Since each $U_n(x)$ has law $\mathcal N(0,n)$ one has 
\begin{align}
\E(Z_{n,\beta}-Z_\delta)
&=
\E\sum_x e^{-\beta U_n(x)}
\1\bigl(-U_n(x)\leq(\beta_c-\delta)n\bigr)\\
&=
2^n\E\left[
e^{\beta(-U_n(x))}
\1\bigl(-U_n(x)\leq(\beta_c-\delta)n\bigr)
\right]\\
&=
\frac{2^n}{\sqrt{2\pi}}
\int_{-(\beta_c-\delta)\sqrt n}^{\infty}
e^{-\beta\sqrt n s}e^{-s^2/2}\dd s.
\end{align}
Setting
\begin{equation}
u=s+(\beta_c-\delta)\sqrt n,
\end{equation}
one has
\begin{align}
\E(Z_{n,\beta}-Z_\delta)
&=
2^ne^{n[\beta(\beta_c-\delta)-\frac12(\beta_c-\delta)^2]}
\frac1{\sqrt{2\pi}}
\int_0^\infty
e^{-u^2/2+\sqrt n u(\beta_c-\delta-\beta)}\dd u.
\end{align}
Notice that $\delta\in(0,\beta_c)$ and $\beta>\beta_c$ imply
$\beta_c-\delta-\beta<0$, hence the remaining integral is bounded by
\begin{equation}
\int_0^\infty e^{-u^2/2}\dd u<\sqrt{2\pi},
\end{equation}
and
\be
\E(Z_{n,\beta}-Z_\delta)
\leq
e^{n[\frac{\beta_c^2}{2}
+\beta(\beta_c-\delta)
-\frac12(\beta_c-\delta)^2]}
=
e^{ng(\delta,\beta)}.
\ee
\end{proof}

\subsubsection{Centred energies}
We measure energies from the reference value $-c_n$, where
\begin{equation}\label{eq:cn}
\ell_n:=\frac{\log n}{2\beta_c},\qquad c_n:=\beta_cn-\ell_n,
\end{equation}
and set
\begin{equation}\label{eq:tx}
t^{(n)}_x:=c_n+U_n(x),\qquad t^{(n)}_{(1)}:=\min_{x\in Q_n}t^{(n)}_x.
\end{equation}
Up to an additive constant, $-c_n$ is the standard centring of the ground-state energy of the REM (see, e.g., \cite{BovierBook}): the expected number of configurations with $t^{(n)}_x\leq D$ is of order $e^{\beta_cD}$, see \eqref{eq:asymptotoclambdan} below. For a sequence $(D_n)_{n\geq1}$ we put
\begin{equation}\label{eq:crticalscaling}
\delta_n:=\frac{\ell_n+D_n}{n}=\frac{\log n}{2\beta_cn}+\frac{D_n}{n},\qquad\text{so that}\qquad\mathcal R_{n,\delta_n}=\{x:t^{(n)}_x\leq D_n\}.
\end{equation}
The next two results show that, when $D_n$ grows logarithmically, the level set $\mathcal R_{n,\delta_n}$ has subexponential cardinality and, for a suitable choice of the logarithmic growth rate, retains any prescribed fraction of the Gibbs mass.

\begin{lem}\label{lem:subexponential-cardinality}
Let $D_n\to\infty$ with $D_n=O(\log n)$, and let $\delta_n$ be as in \eqref{eq:crticalscaling}. Then

\be
\limsup_{n\to\infty}\frac1n\log |\mathcal R_{n,\delta_n}|=0
\qquad\text{a.s.}
\ee

and

\be\label{eq:logscaletail}
\Pp(t^{(n)}_{(1)}>D_n)
\leq\exp\left(-\frac{(1+o(1))e^{\beta_cD_n}}{\beta_c\sqrt{2\pi}}\right).
\ee
\end{lem}

\begin{proof}
Set
\begin{equation}
\lambda_n:=2^np_n(\delta_n)=\E|\mathcal R_{n,\delta_n}|,\qquad p_n(\delta):=\Pp(U_n(x)\leq-(\beta_c-\delta)n),
\end{equation}
where $p_n(\delta)$ does not depend on $x\in Q_n$ (see \eqref{eq:pnx}). 
By the two-sided Mills-ratio estimate \eqref{eq:mills-application} one gets
\begin{equation}
\frac{2^n}{\sqrt{2\pi}t_n}
\left(1-\frac{1}{t_n^2}\right)
e^{-t_n^2/2}
\leq
\lambda_n
\leq
\frac{2^n}{\sqrt{2\pi}t_n}
e^{-t_n^2/2},
\label{eq:lambda-mills}
\end{equation}
where
\begin{equation}
t_n:=\sqrt{n}(\beta_c-\delta_n).
\end{equation} 
Since $\beta_c^2=2\log2$ one has 

\be
n\log2-\frac{t_n^2}{2}
=
n\beta_c \delta_n-\frac{n\delta_n^2}{2}.
\ee

Therefore
\begin{equation}
2^n e^{-t_n^2/2}
=
\sqrt n\,
e^{\beta_c D_n}
e^{-n\delta_n^2/2}.
\label{eq:lambda-exponential}
\end{equation}

Moreover,
\begin{equation}
n\delta_n^2
=
\frac{\left(\frac{\log n}{2\beta_c}+D_n\right)^2}{n}
=
O\!\left(\frac{(\log n)^2}{n}\right)
=o(1),
\end{equation}
because \(D_n=O(\log n)\). Hence
\begin{equation}
e^{-n\delta_n^2/2}=1+o(1).
\end{equation}

We also have
\begin{equation}
\frac{\sqrt n}{t_n}
=
\frac{1}{\beta_c-\delta_n}
=
\frac{1}{\beta_c}\bigl(1+o(1)\bigr),
\end{equation}
because $\delta_n\to0$. Furthermore,
\begin{equation}
\frac1{t_n^2}=O(n^{-1})=o(1).
\end{equation}

Substituting these estimates into \eqref{eq:lambda-mills} and using
\eqref{eq:lambda-exponential}, we conclude that
\be\label{eq:asymptotoclambdan}
\lambda_n
=
\frac{e^{\beta_cD_n}}
{\beta_c\sqrt{2\pi}}
\bigl(1+o(1)\bigr),
\ee
{Notice that} $\lambda_n\to\infty$ and
\be
\frac{\log\lambda_n}{n}=\frac{\beta_cD_n+O(1)}{n}\to0.
\ee

Thus conditions \hyperlink{cond:bin-i}{i)} and \hyperlink{cond:bin-ii}{ii)} of \autoref{lem:binomialcontrol} hold, and hence
\be
\limsup_{n\to\infty}\frac1n\log|\mathcal R_{n,\delta_n}|=0
\qquad\text{a.s.}
\ee

Since $\delta_n$ satisfies $(\beta_c-\delta_n)n=c_n-D_n$, i.e., $n\delta_n=\ell_n+D_n$, we have
\begin{equation}
\{t^{(n)}_{(1)}>D_n\}=\{|\mathcal{R}_{n,\delta_n}|=0\}.
\end{equation}
Since
$|\mathcal R_{n,\delta_n}|\sim\mathrm{Bin}(2^n,p_n(\delta_n))$,
\be
\Pp(|\mathcal R_{n,\delta_n}|=0)
=(1-p_n(\delta_n))^{2^n}
\leq\exp(-2^np_n(\delta_n)).
\ee
Using  the asymptotic \eqref{eq:asymptotoclambdan} for  $\lambda_n\mathrel{=}2^{n}p_n(\delta_n)$ one obtains

\begin{align}
\lambda_n
\geq \frac{e^{\beta_cD_n}}{\beta_c\sqrt{2\pi}}(1+o(1)),
\end{align}
 proving the claim.
\end{proof}

The next result shows that, by properly choosing the threshold energy, it is possible to split the support of the Gibbs measure in such a way that the level set is subexponential while retaining a prescribed fraction of the Gibbs mass.
\begin{prop}[Splitting the support]\label{prop:splitting-support}
For any  $\beta>\beta_c$ let    $c\in(0,1)$ and $\theta\in (0,\infty)$. Fix  $c'>0$ and define
\be\label{eq:defsplitting}
\kappa:=\frac{\theta+\beta c'+\frac12}{\beta-\beta_c},
\qquad
D_{n}:=\kappa\log n,
\qquad
\delta_{n}  :=\frac{\log n}{2\beta_cn}+\frac{D_{n}}{n}.
\ee
Then:
\be
\text{\hypertarget{cond:split-i}{(i)} }\limsup_{n\to\infty}\frac1n\log| \mathcal{R}_{n,\delta_{n}}|=0\ \text{a.s.},
\qquad
\text{\hypertarget{cond:split-ii}{(ii)} }\Pp(\mu_{n,\beta}(\mathcal R_{n,\delta_{n}})\leq c)=O(n^{-\theta}).
\ee
{Thus $\delta_n$ is of the form \eqref{eq:crticalscaling} with $D_n=\kappa\log n$.}
\end{prop}

\begin{proof}
Fix  $\theta,c'>0$  and the quantities $\kappa,D_{n}$ and $\delta_{n}$ as in \eqref{eq:defsplitting}. Set also 
\be
D_n':=c'\log n \,,\quad \delta'_n=\frac{\log n}{2\beta_cn}+\frac{D'_n}{n}.
\ee
{Notice that}
\be
\kappa-c'
=\frac{\theta+\frac12+\beta_c c'}{\beta-\beta_c}>0,
\ee
implies that  $D_n'<D_n$ and then $\mathcal R_{n,\delta'_n}\mathrel{\subseteq} \mathcal R_{n,\delta_n}$ for every $n\geq2$.

Let $x^*$ be a minimizer of
$t_x^{(n)}$, namely a ground state configuration.
On the event $\{t_{(1)}^{(n)}\leq D_n'\}$, since $D_n'<D_n$, one has
$x^*\in \mathcal{R}_{n,\delta_n}$. Therefore,
\be
Z_{\delta_n}
\geq e^{-\beta U_n(x^*)}
=e^{\beta c_n-\beta t_{(1)}^{(n)}}
\geq e^{\beta c_n-\beta D_n'}\mathrel{=:}w_n
\ee

Now if in addition

\be
Z_{n,\beta}-Z_{\delta_n}<\frac{1-c}{c}w_n,
\ee 

and using the fact that the map $z\mapsto \frac{z}{z+r}$ is increasing for any fixed $r$, one gets 
\be
\mu_{n,\beta}(\mathcal R_{n,\delta_n})
=\frac{Z_{\delta_n}}{Z_{\delta_n}+(Z_{n,\beta}-Z_{\delta_n})}
\geq
\frac{w_n}{w_n+(Z_{n,\beta}-Z_{\delta_n})}>c.
\ee

Therefore

\be
\Pp\bigl(\mu_{n,\beta}(\mathcal R_{n,\delta_n})\leq c\bigr)
\leq
\Pp\bigl(t_{(1)}^{(n)}>D_n'\bigr)
+
\Pp\left(
Z_{n,\beta}-Z_{\delta_n}\geq\frac{1-c}{c}w_n
\right).
\label{eq:splitting-union-bound}
\ee

By \eqref{eq:logscaletail} 
\be
\Pp\bigl(t_{(1)}^{(n)}>D_n'\bigr)
\leq
\exp\left(
-\frac{1+o(1)}{\beta_c\sqrt{2\pi}}
n^{\beta_c c'}
\right)
=o(n^{-\theta}).
\ee

Moreover since $w_n$ is deterministic, \autoref{lem:truncated-partition-tail}, applied with
$\delta=\delta_n$ and $\varepsilon=(1-c)w_n/c$, for all $n$ so large that $\delta_n\in(0,\beta_c)$, gives
\be
\Pp\left(
Z_{n,\beta}-Z_{\delta_n}\geq\frac{1-c}{c}w_n
\right)
\leq
\frac{c}{1-c}\frac{e^{ng(\delta_n,\beta)}}{w_n}.
\ee
Using $n\delta_n=\ell_n+D_n$ and $n\delta_n^2=o(1)$, we obtain
\be
ng(\delta_n,\beta)
=n\beta\beta_c-(\beta-\beta_c)(\ell_n+D_n)+o(1).
\ee
{Notice that}
\be
n\beta\beta_c-(\beta-\beta_c)\ell_n
=\beta c_n+\frac12\log n.
\ee
Hence
\be
ng(\delta_n,\beta)
=\beta c_n+\frac12\log n-(\beta-\beta_c)D_n+o(1).
\ee
Since $w_n=e^{\beta c_n-\beta D_n'}$, it follows that
\begin{align}
\frac{e^{ng(\delta_n,\beta)}}{w_n}
&=\sqrt n\,
 e^{-(\beta-\beta_c)D_n+\beta D_n'}(1+o(1))\\
&=n^{\frac12-(\beta-\beta_c)\kappa+\beta c'}(1+o(1))\\
&=n^{-\theta}(1+o(1)).
\end{align}
Therefore,
\be
\Pp\left(
Z_{n,\beta}-Z_{\delta_n}\geq\frac{1-c}{c}w_n
\right)
=O(n^{-\theta}).
\ee
Combining this estimate with \eqref{eq:splitting-union-bound} proves
\be
\Pp\bigl(\mu_{n,\beta}(\mathcal R_{n,\delta_n})\leq c\bigr)
=O(n^{-\theta}).
\ee
Finally, since $D_n=\kappa \log n$ satisfies $D_n\to\infty$ and
$D_n=O(\log n)$, \autoref{lem:subexponential-cardinality} yields
\be
\limsup_{n\to\infty}\frac1n\log |\mathcal R_{n,\delta_n}|=0
\qquad\text{a.s.}
\ee

\end{proof}

\subsection{\texorpdfstring{Separation of deep configurations}{Separation of deep configurations}}

The last geometric input concerns the separation of low-energy REM configurations  in the Hamming distance \eqref{eq:hamming-distance}. It is the source of the bottleneck, because a single-qubit Pauli operator changes the Hamming distance by at most one (see \autoref{sec:setting}).

\begin{prop}[Separation of low-energy REM configurations]\label{prop:deep-hole}
Assume
\begin{equation}
0<\delta<\beta_c\left(1-\frac{1}{\sqrt2}\right).
\end{equation}
Then there exist deterministic constants $a_{\mathrm{dh}}\in(0,1/2)$ and $C_{\mathrm{dh}}>0$ such that, for all sufficiently large $n$,
\begin{equation}
\mathcal D_n:=\{d(x,x')\geq a_{\mathrm{dh}}n\ \text{for all distinct }x,x'\in\mathcal R_{n,\delta}\}
\label{eq:deep-hole-event}
\end{equation}
satisfies
\begin{equation}
\mathbb P(\mathcal D_n)\geq1-e^{-C_{\mathrm{dh}}n}.
\label{eq:deep-hole-probability}
\end{equation}
\end{prop}
\begin{proof}
This follows directly from the global deep-hole scenario in \cite[Lemma~4.2]{Manai_2023}, whose statement is repeated in \autoref{lem:MW-deep-hole}, after the isometric relabeling of $\{-1,1\}^n$ by $Q_n=\{0,1\}^n$. The displayed restriction on $\delta$ is equivalent to the existence of $\varepsilon\in(0,\beta_c-\delta)$ such that
\begin{equation}
\delta(2\beta_c-\delta)<\varepsilon^2.
\end{equation}
Choose such an $\varepsilon$ and then choose the Hamming-ball parameter $\alpha\in(0,1/6]$ (so that $3\alpha\leq1/2$, as needed in the proof of \cite[Lemma~4.2]{Manai_2023}) small enough that
\begin{equation}
2h_2(3\alpha)+\delta(2\beta_c-\delta)<\varepsilon^2,
\end{equation}
where
\begin{equation}\label{eq:binary-entropy}
h_2(t):=-t\log t-(1-t)\log(1-t)
\end{equation}
is the binary entropy. The cited lemma gives, with probability at least $1-e^{-C_{\mathrm{dh}}n}$, pairwise disjoint Hamming balls $B_{\alpha n}(x)$, $x\in\mathcal R_{n,\delta}$. Hence distinct centers have distance at least
\begin{equation}
2\lfloor\alpha n\rfloor+1>2\alpha n-1;
\end{equation}
choosing any $a_{\mathrm{dh}}\in(0,\min\{2\alpha,1/2\})$ proves the claim, since $2\alpha n-1\geq a_{\mathrm{dh}}n$ for all sufficiently large $n$.
\end{proof}

\subsection{\texorpdfstring{Proof of \autoref{prop:two-separated-regions}}{Proof of Theorem 5.1}}\label{subsec:proof-classical}

\begin{proof}[Proof of \autoref{prop:two-separated-regions}, \autoref{it:sep-prob}]

The event $\Omega_n'$ will be the intersection of three events. We first use \autoref{prop:deep-hole}. Let $a_{\mathrm{dh}}\in(0,1/2)$ and $C_{\mathrm{dh}}>0$ be the deterministic constants supplied there, and fix any deterministic $a\in(0,a_{\mathrm{dh}})$. Then the event $\mathcal D_n$ in \eqref{eq:deep-hole-event} satisfies
\be
\Pp(\mathcal D_n^c)\leq e^{-C_{\mathrm{dh}}n}
\label{eq:deep-hole-failure}
\ee
for all sufficiently large $n$. On $\mathcal D_n$, any two disjoint subsets of $\mathcal R_{n,\delta}$ are separated by at least $an$.

Now we define the second event. For  fixed  $\theta,c'>0$,  recall $\kappa$ from \eqref{eq:defsplitting} and $\delta_{n}$ from \eqref{eq:xn-separated-regions}. Hence if 

\be\mathcal G_{n}
\equiv \mathcal G_{n,  \beta,c,\theta}
:=\left\{\mu_{n,\beta}(\mathcal{R}_{n,\delta_{n}})\geq c
\right\}\ee
then \autoref{prop:splitting-support}
implies 
\be
\Pp(\mathcal G_n^c)\leq C_1n^{-\theta}.
\label{eq:good-mass-event-bound}
\ee
for some $C_1>0$ and all sufficiently large $n$,  and also 
\be
\limsup_{n\to\infty}
\frac1n\log|\mathcal{R}_{n,\delta_{n}}|=0
\qquad\text{a.s.}
\label{eq:level-set-subexponential}
\ee
Since $\mathcal C_n'=\mathcal{R}_{n,\delta_{n}}$, this proves
\eqref{eq:C-subexponential}.\\

For the third event we recall the centred energies $t_x^{(n)}$ from \eqref{eq:tx}, with $c_n$ as in \eqref{eq:cn}.

By the definition of $\delta_n$ and $ c_{n}$ one verifies 
\be
\mathcal{R}_{n,\delta_{n}}=
\left\{x:t_x^{(n)}\leq \kappa\log n
\right\}.
\label{eq:Rxn-centered}
\ee
Consider the shell
\be
S_{n}
:=
\left\{
x:
\kappa\log n<t_x^{(n)}\leq \kappa\log n+1
\right\}.
\label{eq:thin-shell}
\ee

We claim that, for all sufficiently large $n$,
\be
S_{n}\subseteq
\mathcal{R}_{n,\delta}\setminus \mathcal{R}_{n,\delta_{n}}= \mathcal A_n'.
\label{eq:shell-in-A}
\ee
Indeed, by \eqref{eq:Rxn-centered}, every point of $S_{n}$ lies outside
$\mathcal{R}_{n,\delta_{n}}$. Moreover,
\be
x\in \mathcal{R}_{n,\delta}
\quad\Longleftrightarrow\quad
t_x^{(n)}
\leq
\delta n-\frac{\log n}{2\beta_c}.
\label{eq:Rdelta-centered}
\ee
Since
\begin{equation}
\kappa \log n +1 < \delta n-\frac{\log n}{2\beta_c}
\end{equation}  for all sufficiently large
$n$, 
the inclusion \eqref{eq:shell-in-A} follows.  Since  
the energies are independent, $|S_n|\sim\mathrm{Bin}(2^n,q_n)$ where
\be
q_n
:
=\Pp \Big(U_n(x)\in I_{n}\Big) \,,\,\,\quad I_n:=( \kappa\log n-c_n, \kappa \log n-c_n+1) 
\label{eq:shell-probability}
\ee
 We claim that there exists $c_0>0$ such that 
\be
 2^nq_n
= c_0 n^{\betac \kappa}
(1+o(1)).
\ee

Indeed since
\begin{equation}
\kappa\log n-c_n=-\beta_c n+(\kappa+\frac{1}{2\beta_c})\log n,
\end{equation}
we may write 
\be
\mathbb P(U_n(x)\in I_n)
=
\frac1{\sqrt{2\pi n}}
\int_{-\beta_c n+(\kappa+\frac{1}{2\beta_c})\log n}^{-\beta_c n+(\kappa+\frac{1}{2\beta_c})\log n+1}
e^{-u^2/(2n)}\dd u.
\ee

By the change of variables
\begin{equation}
u=-\beta_c n+(\kappa+\frac{1}{2\beta_c})\log n+s
\end{equation}
one gets 
\be
\mathbb P(U_n(x)\in I_n)=
\frac{e^{-\beta_c^2n/2}\,n^{\beta_c(\kappa+\frac{1}{2\beta_c})}}{\sqrt{2\pi n}}
\int_0^1 e^{\beta_c s}
\exp\!\left(-\frac{\bigl((\kappa+\frac{1}{2\beta_c})\log n+s\bigr)^2}{2n}\right)\dd s
\ee

Since
\begin{equation}
\frac{1}{2n}\bigl((\kappa+\frac{1}{2\beta_c})\log n+s\bigr)^2\longrightarrow0
\end{equation}
uniformly for \(s\in[0,1]\) one obtains
\be
2^n \mathbb P(U_n(x)\in I_n)=
\frac{e^{\beta_c}-1}{\beta_c\sqrt{2\pi}}\,n^{\beta_c\kappa}(1+o(1))
\ee
{and using $\beta_c^2=2\log 2$ the claim is proved.} Now using $1-u\leq e^{-u}$, we obtain
\be
\Pp(S_n=\varnothing)
=
(1-q_n)^{2^n}
\leq
\exp(-2^nq_n)
\leq
\exp\left(-\frac{c_0}{2}n^{\beta_c\kappa}\right)
\label{eq:shell-empty-bound}
\ee
for $n$ large enough. The third  event is defined as 
\be
\mathcal S_n:=\{S_n\neq\varnothing\}.
\label{eq:shell-nonempty-event}
\ee
and then 
\be
\Pp(\mathcal S_n^c)
\leq
\exp\left(-\frac{c_0}{2}n^{\beta_c\kappa}\right).
\label{eq:shell-event-bound}
\ee

We are ready to construct the main event

\be
\Omega_n'\equiv \Omega'_{n,\beta,\delta,c,\theta}
:=
\mathcal G_n\cap\mathcal D_n\cap\mathcal S_n.
\label{eq:Omega-prime-definition}
\ee
By the union bound,
\be
\Pp\bigl((\Omega_n')^c\bigr)
\leq
\Pp(\mathcal G_n^c)
+\Pp(\mathcal D_n^c)
+\Pp(\mathcal S_n^c)
\leq
C_1n^{-\theta}
+e^{-C_{\mathrm{dh}}n}
+\exp\left(-\frac{c_0}{2}n^{\beta_c\kappa}\right).
\label{eq:Omega-prime-failure}
\ee
The last two terms are $o(n^{-\theta})$ so 
\be
\Pp(\Omega_n')
\geq1-Cn^{-\theta}
\label{eq:Omega-prime-bound}
\ee
for all sufficiently large $n$.
\end{proof}
\begin{proof}[Proof of \autoref{prop:two-separated-regions}, \autoref{it:sep-dist}]
Because $\delta_n\to0$, one has $\delta_n<\delta$ for all sufficiently large
$n$, hence
\be
\mathcal{R}_{n,\delta_n}\subseteq \mathcal{R}_{n,\delta}.
\label{eq:level-set-inclusion}
\ee
The definitions in \eqref{eq:AC-separated-regions} therefore imply that 
\be
\mathcal A_n'\mathbin{
\cup}\mathcal C_n'=
\mathcal{R}_{n,\delta}.
\ee

Now observe that on $\mathcal{G}_n$
\be
\mu_{n,\beta}(\mathcal C_n')
\equiv \mu_{n,\beta}(\mathcal{R}_{n,\delta_n})
\geq c 
\ee
Furthermore, on $\mathcal S_n$, the set $\mathcal A_n'$ is nonempty,
whereas on $\mathcal G_n$, the set $\mathcal C_n'$ is nonempty. Finally, both $\mathcal A_n'$ and $\mathcal C_n'$ are subsets of
$\mathcal{R}_{n,\delta}$. Hence, on $\mathcal D_n$
\be
d(\mathcal A_n',\mathcal C_n')
\mathrel{=}
\min_{\substack{x\in\mathcal A_n'\\
x'\in\mathcal C_n'}}
d(x,x')
\geq an.
\ee
This proves all the assertions that hold on
$\Omega_n'$.
\end{proof}

\begin{proof}[Proof of \autoref{prop:two-separated-regions}, \autoref{it:sep-card}]

Since $\mathcal{C}'_n=\mathcal{R}_{n,\delta_n}$ then \eqref{eq:C-subexponential} follows  directly from \autoref{prop:splitting-support}.
It remains  to prove  that  $\mathcal A_n'$ cannot have exponentially small Gibbs mass.
By \eqref{eq:shell-empty-bound} one has 
\be
\sum_{n=1}^{\infty}\Pp(S_n=\varnothing)<\infty.
\label{eq:shell-empty-summable}
\ee
The first Borel--Cantelli lemma therefore implies that, almost surely there exists $m$ such that 
$S_n\neq\varnothing$ for all large $n\geq m$. For every such $n$, choose $x_n\in S_n$,  hence  by
\eqref{eq:shell-in-A} one has 
$x_n\in\mathcal A_n'$. Moreover by  \eqref{eq:thin-shell}
\be
\kappa\log n<t_{x_n}^{(n)}\leq \kappa \log n+1,
\label{eq:sigman-shell-energy}
\ee
so that
\be
t_{x_n}^{(n)}=O(\log n)
\quad\rightarrow\quad
\frac{t_{x_n}^{(n)}}{n}\longrightarrow0.
\label{eq:sigman-centered-small}
\ee
Recall that 
\be
U_n(x_n)
=-c_n+t_{x_n}^{(n)},
\ee
with $c_n/n\to\beta_c$, and \autoref{prop:free-energy-limit} gives
\be
\lim_{n\to\infty}
\frac1n\log Z_{n,\beta}=\beta\beta_c \,,\quad\text{a.s.}
\label{eq:partition-free-energy-limit}
\ee 

Therefore
\begin{align}
\frac1n\log\mu_{n,\beta}(x_n)
&=
\frac{\beta c_n}{n}
-\frac{\beta t_{x_n}^{(n)}}{n}
-\frac1n\log Z_{n,\beta}\to 0 \quad\text{a.s.}
\label{eq:single-weight-log}
\end{align}

Since $x_n\in\mathcal A_n'$,
\be
\mu_{n,\beta}(x_n)\leq \mu_{n,\beta}(\mathcal A_n')
\leq 1 
\ee

and then almost surely
\be
0\leq \liminf_{n\to\infty}
\frac1n\log\mu_{n,\beta}(\mathcal A_n')
\leq
\limsup_{n\to\infty}
\frac1n\log\mu_{n,\beta}(\mathcal A_n')
\leq 0
\ee
which proves \eqref{eq:A-nonexponential-mass}.

\end{proof}

\begin{proof}[Proof of \autoref{prop:two-separated-regions}, \autoref{it:sep-mass}]
Assume now that $\theta>1$, from
\eqref{eq:good-mass-event-bound}
\be
\sum_{n=1}^{\infty}
\Pp\left(
\mu_{n,\beta}(\mathcal C_n')<c
\right)
<\infty.
\ee
By the first Borel--Cantelli lemma, almost surely,
\be
\mu_{n,\beta}(\mathcal C_n')\geq c
\ee
for all sufficiently large $n$. Hence
\be
\liminf_{n\to\infty}
\mu_{n,\beta}(\mathcal C_n')\geq c
\qquad\text{almost surely}.
\ee
This completes the proof.
\end{proof}

\subsection{\texorpdfstring{The almost-sure classical bottleneck}{The almost-sure classical bottleneck}}
\begin{cor}[Almost-sure classical bottleneck]\label{thm:classical}
Let $\beta>\beta_c$, let $\delta$ satisfy \eqref{eq:delta-condition-separated-regions}, and let $c_\star:=3/4$. There exist regions $\mathcal A_n',\mathcal C_n'\mathrel{\subseteq}Q_n$ and a deterministic constant $a>0$ such that, for all sufficiently large $n$,
\begin{equation}\label{eq:Bott-partition}
\mathcal A_n'\cup\mathcal C_n'=\mathcal R_{n,\delta},
\end{equation}
and almost surely
\begin{align}
&\lim_{n\to\infty}\frac{\mathop{\log}\nolimits\mu_{n,\beta}(\mathcal A_n')}{n}=0,\label{eq:Bott1}\\
&\liminf_{n\to\infty}\mu_{n,\beta}(\mathcal C_n')\geq c_\star>\frac12,\label{eq:Bott2}\\
&d(\mathcal A_n',\mathcal C_n')\geq an\quad\text{for all sufficiently large $n$},\label{eq:Bott3}\\
&\limsup_{n\to\infty}\frac{\mathop{\log}\nolimits|\mathcal C_n'|}{n}=0.\label{eq:Bott4}
\end{align}
\end{cor}
\begin{proof}
Apply \autoref{prop:two-separated-regions} with $c=c_\star$ and any fixed $\theta>1$. Let $a\in(0,1/2)$ and $\kappa>0$ be the deterministic constants furnished there, and define $\mathcal A_n',\mathcal C_n'$ by \eqref{eq:AC-separated-regions}. Since $\delta_n\to0$ (see \eqref{eq:xn-separated-regions}), one has $\mathcal R_{n,\delta_n}\subseteq\mathcal R_{n,\delta}$ for all sufficiently large $n$, and hence \eqref{eq:Bott-partition} holds.
Equation \eqref{eq:A-nonexponential-mass} gives \eqref{eq:Bott1}; \eqref{eq:C-subexponential} gives \eqref{eq:Bott4}; and, because $\theta>1$, \eqref{eq:C-positive-mass-as} gives \eqref{eq:Bott2}. Finally, by \eqref{eq:probability-separated-regions}, the complements of the events $\Omega_n'$ of \autoref{prop:two-separated-regions} have summable probabilities, so the first Borel--Cantelli lemma implies that $\Omega_n'$ occurs eventually almost surely. Equation \eqref{eq:AC-distance} then gives the deterministic eventual bound \eqref{eq:Bott3}.
\end{proof}

\section{\texorpdfstring{Low-energy localization}{Low-energy localization}}\label{sec:loc}
Our choice of the subspaces $A,\,B,\,C\subseteq\left(\mathbb{C}^2\right)^{\otimes n}$ is based on the following result, which connects the low-energy eigenvectors of $H$ to the low-energy configurations of $U$.

\begin{thm}[Low-energy localization; cf.\ {\cite[Theorem~1.5]{Manai_2023}}]\label{thm:Manai}
Fix $0\leq\Gamma<\beta_c=\sqrt{2\mathop{\log}\nolimits2}$ and set
\begin{equation}\label{eq:KGamma}
K_\Gamma:=\frac{2\Gamma^2}{\beta_c}+1.
\end{equation}
There is $\bar\delta=\bar\delta(\Gamma)>0$ such that, for every $\delta\in(0,\bar\delta)$, there are events $\Omega^{\mathrm{loc}}_{n,\Gamma,\delta}$ and a constant $c(\Gamma,\delta)>0$ with
\begin{equation}
\mathbb P\!\left(\Omega^{\mathrm{loc}}_{n,\Gamma,\delta}\right)\geq1-e^{-c(\Gamma,\delta)n}\qquad\text{for all sufficiently large $n$},
\end{equation}
such that, on $\Omega^{\mathrm{loc}}_{n,\Gamma,\delta}$ and for all $n$ larger than a deterministic threshold depending only on $\Gamma$ and $\delta$ (and, in \autoref{it:loc-decay}, on $\alpha$), the following hold, where $\mathcal R_{n,\delta}$ is the level set \eqref{eq:R}.
\begin{enumerate}
\item\label{it:loc-label} (Labelling.) To each $x\in\mathcal R_{n,\delta}$ is associated a simple eigenvalue $E(x)$ of $H$, with normalized eigenvector $|\psi_x\rangle$, in such a way that distinct configurations are associated with distinct eigenvalues. Moreover,
\begin{equation}\label{eq:E-below-U}
E(x)\leq U(x)\leq-(\beta_c-\delta)n.
\end{equation}
\item\label{it:loc-energy} (Energies.) Uniformly for $x\in\mathcal R_{n,\delta}$,
\begin{equation}\label{eq:E(x)}
E(x)=U(x)+\frac{\Gamma^2n}{U(x)}+O_{\Gamma,\delta}\!\left(n^{-1/4}\right).
\end{equation}
\item\label{it:loc-decay} (Localization.) For every fixed $0<\alpha<1$ there exists a deterministic $c_\alpha=c_\alpha(\Gamma,\delta)>0$ such that, uniformly for $x\in\mathcal R_{n,\delta}$,
\begin{equation}\label{eq:concentration}
\sum_{y\in {Q_n}\setminus B_{\alpha n}(x)}|\langle y|\psi_x\rangle|^2\leq e^{-c_\alpha n}\,,
\end{equation}
with $B_{\alpha n}(x)$ the Hamming ball \eqref{eq:ball}.
\item\label{it:loc-threshold} (Threshold.) Every eigenvalue of $H$ that is not of the form $E(x)$ with $x\in\mathcal R_{n,\delta}$ is at least $-(\beta_c-\delta)n-K_\Gamma$.
\end{enumerate}
\end{thm}
\begin{rem}[Relation to {\cite[Theorem~1.5]{Manai_2023}}]\label{rem:MW-threshold}
\autoref{thm:Manai} contains the parts of \cite[Theorem~1.5]{Manai_2023} that are needed below, with the energy thresholds stated in a slightly different form, and its proof in \autoref{app:loc} follows closely the proof given by Manai and Warzel in \cite[Section~4.4]{Manai_2023}. All the main ingredients of the proof, and in particular the localization mechanism, are due to them. To make it clear which parts of their argument we rewrite, and why, we list the differences here. We use our notation, in which the set $\mathcal L_{\beta_c-\delta}$ of \cite{Manai_2023} is $\mathcal R_{n,\delta}$, and $N=n$.
\begin{enumerate}[label=(\alph*)]
\item\label{it:mw-statement} \emph{Thresholds in the statement.} In \cite[Theorem~1.5, item~1]{Manai_2023}, the eigenvalues of $H$ below $-(\beta_c-\delta)n$ are stated to be in one-to-one correspondence with the low-energy configurations, which, by the last sentence of that item, are those in $\mathcal R_{n,\delta}$; the eigenvalue associated with $x$ is
\begin{equation}
U(x)+\frac{\Gamma^2n}{U(x)}+O\bigl(n^{-1/4}\bigr).
\end{equation}
When $\Gamma>0$, this formulation needs a small adjustment near the threshold. Indeed, the shift $\Gamma^2n/U(x)$ is negative and of order one, so that a configuration $x\notin\mathcal R_{n,\delta}$ with $U(x)$ slightly above $-(\beta_c-\delta)n$ also produces an eigenvalue below $-(\beta_c-\delta)n$ (by items~\ref{it:loc-label} and~\ref{it:loc-energy} of \autoref{thm:Manai} applied with some $\delta'\in(\delta,\bar\delta)$ in place of $\delta$, which assign distinct eigenvalues to all the configurations in $\mathcal R_{n,\delta'}\supseteq\mathcal R_{n,\delta}$), and with probability tending to one such configurations exist. Hence, with probability tending to one, $H$ has more eigenvalues below $-(\beta_c-\delta)n$ than there are configurations in $\mathcal R_{n,\delta}$. For this reason, \autoref{thm:Manai} uses two thresholds: the eigenvalues labelled by $\mathcal R_{n,\delta}$ are at most $-(\beta_c-\delta)n$ (\autoref{it:loc-label}), all other eigenvalues are at least $-(\beta_c-\delta)n-K_\Gamma$ (\autoref{it:loc-threshold}), and the window between these two thresholds may contain eigenvalues of both kinds.
\item\label{it:mw-proof} \emph{Thresholds in the proof.} In \cite[Section~4.4]{Manai_2023}, the same parameter $\delta$ is used for the deep-hole scenario and as the threshold parameter of the statement. With this choice, the Schur complement analysis there determines the eigenvalues of $H$ below $-(\beta_c-\delta/8)n$ (paragraph ``Spectrum'' on \cite[p.~1289]{Manai_2023}), and the eigenvector estimate concerns the configurations in $\mathcal R_{n,s}$ with $s\in(0,\delta/8]$ so small that $2\beta_cs<c$, where $c$ is the decay rate in \cite[(4.25)]{Manai_2023} (paragraph ``Eigenvectors'' on \cite[pp.~1289--1290]{Manai_2023}). Thus, the conclusions are obtained below thresholds that are lower than $-(\beta_c-\delta)n$. Since $\delta$ may be taken small in \cite[Theorem~1.5]{Manai_2023}, this amounts to a relabelling of the parameters. We make it explicit: in \autoref{app:loc}, the deep-hole parameter is denoted by $\delta_p$, and the threshold parameter $\delta$ of \autoref{thm:Manai} satisfies $\delta<\delta_p/8$.
\item\label{it:mw-nonres} \emph{Non-resonance.} For $x\in\mathcal R_{n,s}$, the eigenvector estimate \cite[(4.27)]{Manai_2023} divides by the non-resonance scale $r(s)$ of \cite[p.~1289]{Manai_2023}. It therefore requires the ground-state energy of the ball around $x$ to be at distance at least of order $r(s)$ from the ground-state energies of all the other balls whose ground states belong to the range of the projection \cite[(4.22)]{Manai_2023}. The non-resonance event \cite[(4.26)]{Manai_2023} provides this separation for the balls around the configurations in $\mathcal R_{n,s}$. However, the ball around a configuration $x'\notin\mathcal R_{n,s}$ with $U(x')$ slightly above $-(\beta_c-s)n$ may have its ground-state energy below $-(\beta_c-s)n$, and closer than $r(s)$ to the ground-state energy of the ball around a configuration $x\in\mathcal R_{n,s}$ with $U(x)$ close to $-(\beta_c-s)n$; such pairs are not covered by \cite[(4.26)]{Manai_2023}. This can be remedied by imposing the non-resonance condition on a slightly larger set. We impose it on $\mathcal R_{n,2\delta}$, where $\delta$ is the threshold parameter of \autoref{thm:Manai} (which plays the role of $s$); the probability of the resulting event is controlled by the same spectral-averaging estimate \cite[Lemma~4.4(2)]{Manai_2023}, and we treat the remaining configurations by means of the energy asymptotics of \cite[Lemma~4.3]{Manai_2023}; see the event \ref{it:E3} and \stepref{4} in \autoref{app:loc}.
\end{enumerate}
Two further points, of a more technical nature, are explained in \autoref{app:loc}: the hypotheses of \cite[Lemma~4.2]{Manai_2023}, which are satisfied for the parameters used in \cite[Section~4.4]{Manai_2023} (see \autoref{lem:MW-deep-hole} and the comment after it), and the disjointness of slightly enlarged balls (item~\ref{it:rw-balls} at the beginning of \autoref{app:loc}). With these modifications, the argument of \cite[Section~4.4]{Manai_2023} proves \autoref{thm:Manai}.
\end{rem}
Because
\begin{equation}
\sum_n\mathbb P((\Omega^{\mathrm{loc}}_{n,\Gamma,\delta})^c)<\infty,
\end{equation}
the first Borel--Cantelli lemma yields, for fixed $\Gamma$ and $\delta$, a probability-one event $\Omega^{\mathrm{loc}}_\infty$ on which the conclusions of \autoref{thm:Manai} hold for every sufficiently large $n$.

\section{Choice of the subspaces}\label{sec:subsp}
From now on, and up to the end of \autoref{sec:proof}, the parameters $\beta>\beta_c$ and $0\leq\Gamma<\Gamma_c(\beta)$ are fixed as in \autoref{thm:main}.

We next combine the localization theorem, \autoref{thm:Manai}, with a classical REM bottleneck construction.

Choose $\delta\in(0,\bar\delta(\Gamma))$, with $\bar\delta(\Gamma)$ from \autoref{thm:Manai}, so small that also $\delta<\beta_c(1-1/\sqrt2)$, i.e., \eqref{eq:delta-condition-separated-regions} holds. Let $\mathcal A_n'$, $\mathcal C_n'$ and $a>0$ be given by \autoref{thm:classical} for this $\delta$.
Intersecting the probability-one event on which the conclusions of \autoref{thm:classical} hold with $\Omega^{\mathrm{loc}}_\infty$, we henceforth work on a single probability-one event $\Omega_\infty$ on which both the localization conclusions and \eqref{eq:Bott1}--\eqref{eq:Bott4} hold for every sufficiently large $n$. The finitely many exceptional values of $n$ may be defined arbitrarily without affecting any asymptotic statement.

For such $n$, define
\begin{equation}\label{eq:A}
\mathcal A_n:=\{|\psi_x\rangle:x\in\mathcal A_n'\},\qquad
\mathcal C_n:=\{|\psi_x\rangle:x\in\mathcal C_n'\},\qquad
\mathcal B_n:=\{|\psi_1\rangle,\ldots,|\psi_{2^n}\rangle\}\setminus(\mathcal A_n\cup\mathcal C_n).
\end{equation}
Here, since the eigenvalues $E(x)$ are simple and pairwise distinct (\autoref{it:loc-label} of \autoref{thm:Manai}), each $|\psi_x\rangle$ coincides up to a phase with an element of the eigenbasis $\{|\psi_i\rangle\}$, and we choose the phases of that basis so that $|\psi_x\rangle$ belongs to it. By \eqref{eq:Bott-partition}, the set $\mathcal B_n$ consists precisely of the eigenvectors whose eigenvalue is not of the form $E(x)$ with $x\in\mathcal R_{n,\delta}$. We use the same symbols $\mathcal A_n,\mathcal B_n,\mathcal C_n$ for the corresponding index sets when no confusion can arise, and set
\begin{equation}
A:=\mathrm{span}\,\mathcal A_n,\qquad B:=\mathrm{span}\,\mathcal B_n,\qquad C:=\mathrm{span}\,\mathcal C_n.
\end{equation}
By construction, $A,B,C$ are mutually orthogonal spectral subspaces of $H$, their direct sum is the full Hilbert space, and their projectors commute with $H$.

\section{\texorpdfstring{Direct transitions from $A$ to $C$}{Direct transitions from A to C}}\label{sec:Ltilde}
We define the maximal total $A$-to-$C$ rate
\begin{equation}\label{eq:eta-def}
\eta_n:=\max_{j\in\mathcal A_n}\sum_{i\in\mathcal C_n}\sum_{k=1}^3\sum_{l=1}^n
\gamma_{ijkl}|\langle\psi_i|\sigma_k^{(l)}|\psi_j\rangle|^2.
\end{equation}
Thus $\eta_n$ is the rate $r_A$ of \autoref{thm:direct-bottleneck} for the present subspaces $A$ and $C$.

\begin{lem}\label{lem:epsilon}
There is a deterministic $\tilde c>0$ such that, almost surely,
\begin{equation}
\limsup_{n\to\infty}\frac1n\log\eta_n\leq-\tilde c.
\end{equation}
\end{lem}
\begin{proof}
We enlarge $\mathcal A_n'$ and $\mathcal C_n'$ to Hamming neighbourhoods $\widetilde{\mathcal A}_n$ and $\widetilde{\mathcal C}_n$ that are still at distance at least two from each other, so that no single-qubit Pauli operator connects them. Each matrix element $\langle\psi_i|\sigma_k^{(l)}|\psi_j\rangle$ is then bounded by the norms of the parts of $\psi_i$ and $\psi_j$ lying outside these neighbourhoods, which are exponentially small by the localization estimate \eqref{eq:concentration}.
Work on the probability-one event $\Omega_\infty$. By \eqref{eq:Bott3}, the deterministic constant $a>0$ satisfies $d(\mathcal A_n',\mathcal C_n')\geq an$ for all sufficiently large $n$. Fix once and for all a deterministic $\lambda\in(0,a)$, for instance $\lambda=a/2$, and set
\begin{equation}
\begin{aligned}
\widetilde{\mathcal A}_n&:=\{y:d(y,\mathcal A_n')\leq\lambda n/2-1\},\\
\widetilde{\mathcal C}_n&:=\{y:d(y,\mathcal C_n')\leq\lambda n/2-1\}.
\end{aligned}
\end{equation}
For all sufficiently large $n$,
\begin{equation}
d(\widetilde{\mathcal A}_n,\widetilde{\mathcal C}_n)>1.
\end{equation}
\par
Let $\Pi_{\widetilde{\mathcal A}_n}$ and $\Pi_{\widetilde{\mathcal C}_n}$ denote the two corresponding computational-basis projectors.

For $j\in\mathcal A_n$ and $i\in\mathcal C_n$, let $x_j\in\mathcal A_n'$ and $x_i\in\mathcal C_n'$ denote the associated configurations, and define
\begin{equation}
|\phi_j\rangle:=(I-\Pi_{\widetilde{\mathcal A}_n})|\psi_j\rangle,\qquad
|\phi_i\rangle:=(I-\Pi_{\widetilde{\mathcal C}_n})|\psi_i\rangle.
\end{equation}
Apply \eqref{eq:concentration} with the strictly smaller radius $\alpha=\lambda/3$. For all sufficiently large $n$,
\begin{equation}
B_{\lambda n/3}(x_j)\subseteq\widetilde{\mathcal A}_n,\qquad B_{\lambda n/3}(x_i)\subseteq\widetilde{\mathcal C}_n,
\end{equation}
hence
\begin{equation}
\|\phi_i\|^2,\|\phi_j\|^2\leq e^{-c_{\lambda/3}n}.
\end{equation}
Because an $X$ or $Y$ Pauli changes the ordinary Hamming distance by one and a $Z$ Pauli does not change it,
\begin{equation}
\Pi_{\widetilde{\mathcal C}_n}\sigma_k^{(l)}\Pi_{\widetilde{\mathcal A}_n}=0.
\end{equation}
Therefore
\begin{equation}
|\langle\psi_i|\sigma_k^{(l)}|\psi_j\rangle|
\leq\|\phi_i\|+\|\phi_j\|+\|\phi_i\|\|\phi_j\|
\leq3e^{-c_{\lambda/3}n/2}.
\end{equation}
Consequently
\begin{equation}\label{eq:eta-bound}
\eta_n\leq27n\gamma_{\max}|\mathcal C_n'|e^{-c_{\lambda/3}n}.
\end{equation}
Since $\gamma_{\max}$ is polynomial and \eqref{eq:Bott4} gives $|\mathcal C_n'|=e^{o(n)}$, any fixed $\tilde c<c_{\lambda/3}$ satisfies the claimed bound for all sufficiently large $n$.
\end{proof}

\section{\texorpdfstring{Gibbs weights of the three subspaces}{Gibbs weights of the three subspaces}}\label{sec:Pi}
We now estimate the Gibbs weights $w_A$, $w_B$ and $w_C$ of the three subspaces. We use two results of Manai and Warzel on the partition function of the QREM.
\begin{thm}[{\cite[Theorem~1.1]{Manai_2020}}]\label{thm:Z}
For $t\geq\beta_c$ and $0\leq\Gamma\leq\Gamma_c(t)$, almost surely
\begin{equation}\label{eq:normalized-pressure}
\lim_{n\to\infty}\frac1n\log\left(2^{-n}\mathrm{Tr}\,e^{-tH}\right)=t\beta_c-\frac{\beta_c^2}{2}.
\end{equation}
Since $\beta_c^2/2=\log2$, equivalently
\begin{equation}\label{eq:unnormalized-pressure}
\lim_{n\to\infty}\frac1n\log\mathrm{Tr}\,e^{-tH}=t\beta_c.
\end{equation}
\end{thm}

\begin{thm}[{\cite[Theorem 1.10]{Manai_2023}}]\label{thm:ZQC}
Let $Z_{n,\beta}$ be the partition function of the classical Hamiltonian $U$ at inverse temperature $\beta$, see \eqref{eq:mu-Z}.
If $\beta>\beta_c$ and $\Gamma<\Gamma_c(\beta)$, we have almost surely
    \begin{equation}
        \lim_{n\to\infty}\mathop{\log}\nolimits\frac{\mathrm{Tr}\,e^{-\beta H}}{Z_{n,\beta}} = \frac{\Gamma^2\,\beta}{\beta_c}\,.
    \end{equation}
\end{thm}

\begin{lem}\label{lem:Pi}
    Almost surely, $\mu_{n,\beta}(\mathcal A_n')>0$ and $\mu_{n,\beta}(\mathcal C_n')>0$ for all sufficiently large $n$, and on that eventual range
    \begin{equation}
\liminf_{n\to\infty}\frac{\mathrm{Tr}(\Pi_A\omega)}{\mu_{n,\beta}(\mathcal{A}_n')}\ge1\,,\qquad \liminf_{n\to\infty}\frac{\mathrm{Tr}(\Pi_C\omega)}{\mu_{n,\beta}(\mathcal{C}_n')}\ge1\,.
    \end{equation}
\end{lem}

\begin{proof}
The eventual positivity of both denominators follows from \eqref{eq:A-nonexponential-mass} and \eqref{eq:C-positive-mass-as} (with the choice $c=c_\star>0$). Let $U_{\min}$ be the ground-state energy of $U$, as in \autoref{lem:Umin}. We have for large enough $n$
\begin{equation}\label{eq:PiA-lower}
\begin{split}
    \mathrm{Tr}(\Pi_A\omega) &= \frac{1}{\mathrm{Tr}\,e^{-\beta H}}\sum_{i\in\mathcal{A}_n} e^{-\beta\,E_i} = \frac{1}{\mathrm{Tr}\,e^{-\beta H}}\sum_{x\in\mathcal{A}_n'} e^{-\beta\,E(x)}\\
    &\overset{\hypertarget{step:PiA-a}{\mathrm{(a)}}}{\ge}  \frac{1}{\mathrm{Tr}\,e^{-\beta H}}\sum_{x\in\mathcal{A}_n'} \exp\left(-\beta\left(U(x) + \frac{\Gamma^2\,n}{U_{\min}} + O_{\Gamma,\delta}\left(n^{-\frac{1}{4}}\right)\right)\right)\\
    &= \frac{Z_{n,\beta}\,\mu_{n,\beta}(\mathcal{A}_n')}{\mathrm{Tr}\,e^{-\beta H}}\exp\left(-\beta\left(\frac{\Gamma^2\,n}{U_{\min}} + O_{\Gamma,\delta}\left(n^{-\frac{1}{4}}\right)\right)\right)\,,
\end{split}
\end{equation}
where \hyperlink{step:PiA-a}{(a)} holds since from \autoref{it:loc-energy} of \autoref{thm:Manai} we have
\begin{equation}
    E(x)\le U(x) + \frac{\Gamma^2\,n}{U_{\min}} + O_{\Gamma,\delta}\left(n^{-\frac{1}{4}}\right)\,.
\end{equation}
Then,
\begin{align}
\liminf_{n\to\infty}\mathop{\log}\nolimits\frac{\mathrm{Tr}(\Pi_A\omega)}{\mu_{n,\beta}(\mathcal{A}_n')} &\ge \liminf_{n\to\infty}\left(\mathop{\log}\nolimits\frac{Z_{n,\beta}}{\mathrm{Tr}\,e^{-\beta H}} -\beta\left(\frac{\Gamma^2\,n}{U_{\min}} + O_{\Gamma,\delta}\left(n^{-\frac{1}{4}}\right)\right)\right) = 0\,,
\end{align}
where in the last step we have used \autoref{thm:ZQC} and the ground-state limit \eqref{eq:Umin-limit}.
The claim for $A$ follows.
The proof of the claim for $C$ is analogous.
\end{proof}

\begin{cor}\label{cor:PiA}
       We have almost surely
    \begin{equation}
\lim_{n\to\infty}\frac{1}{n}\mathop{\log}\nolimits\mathrm{Tr}(\Pi_A\omega) = 0\,.
    \end{equation}
\end{cor}
\begin{proof}
    Since from \autoref{thm:classical} we have almost surely
    \begin{equation}
        \lim_{n\to\infty}\frac{\mathop{\log}\nolimits\mu_{n,\beta}(\mathcal{A}_n')}{n} = 0\,,
    \end{equation}
    and, on the eventual range where $\mu_{n,\beta}(\mathcal A_n')>0$,
    \begin{equation}
    \frac1n\mathop{\log}\nolimits\mathrm{Tr}(\Pi_A\omega)
    =\frac1n\mathop{\log}\nolimits\frac{\mathrm{Tr}(\Pi_A\omega)}{\mu_{n,\beta}(\mathcal A_n')}
    +\frac1n\mathop{\log}\nolimits\mu_{n,\beta}(\mathcal A_n').
    \end{equation}
    By \autoref{lem:Pi}, the liminf of the first term on the right is nonnegative, while the second term converges to zero. Hence
    \begin{equation}
    \liminf_{n\to\infty}\frac1n\mathop{\log}\nolimits\mathrm{Tr}(\Pi_A\omega)\ge0.
    \end{equation}
    Since $\mathrm{Tr}(\Pi_A\omega)\le1$, the corresponding limsup is at most zero, proving the claim.
\end{proof}

\begin{lem}\label{lem:PiB}
For any $\beta>\beta_c$ and $0\leq\Gamma<\Gamma_c(\beta)$, there is a deterministic constant $c_B>0$ such that, almost surely,
\begin{equation}
\limsup_{n\to\infty}\frac1n\log\mathrm{Tr}(\Pi_B\omega)\leq-c_B.
\end{equation}
\end{lem}
\begin{proof}
The idea is the following. All eigenvalues in $B$ lie at least $\delta n-O(1)$ above the energy scale $-\beta_cn$ that carries the Gibbs state. Factorizing $e^{-\beta E}=e^{-sE}e^{-(\beta-s)E}$ with a slightly smaller inverse temperature $s$ that is still in the spin-glass phase, the second factor is bounded on $B$ by $e^{(\beta-s)(\beta_c-\delta)n+O(1)}$, while the sum of the first factors is at most the partition function at $s$. The pressure formula of \autoref{thm:Z} at $s$ and at $\beta$ then converts the energy gap $\delta n$ into the exponentially small factor $e^{-(\beta-s)\delta n}$.

Set
\begin{equation}
\chi(t):=t\beta_c-\log2-\log\cosh(t\Gamma).
\end{equation}
Since $\chi(\beta)>0$ and $\chi$ is continuous, choose $s\in(\beta_c,\beta)$ sufficiently close to $\beta$ that $\chi(s)>0$, equivalently $\Gamma<\Gamma_c(s)$. By \autoref{it:loc-threshold} of \autoref{thm:Manai} (recall that $\mathcal B_n$ consists of the eigenvectors whose eigenvalue is not of the form $E(x)$ with $x\in\mathcal R_{n,\delta}$), every $i\in\mathcal B_n$ satisfies
\begin{equation}
E_i\geq-(\beta_c-\delta)n-K_\Gamma
\end{equation}
for all sufficiently large $n$, with the constant $K_\Gamma$ defined there. Thus
\begin{equation}
e^{-\beta E_i}=e^{-sE_i}e^{-(\beta-s)E_i}
\leq {e^{(\beta-s)((\beta_c-\delta)n+K_\Gamma)}}e^{-sE_i}.
\end{equation}
Summing over $i\in\mathcal B_n$ and dividing by $\mathrm{Tr}\,e^{-\beta H}$ gives
\begin{equation}
\mathrm{Tr}(\Pi_B\omega)\leq {e^{(\beta-s)((\beta_c-\delta)n+K_\Gamma)}}
\frac{\mathrm{Tr}\,e^{-sH}}{\mathrm{Tr}\,e^{-\beta H}}.
\end{equation}
Using \eqref{eq:unnormalized-pressure} at $s$ and $\beta$, and $K_\Gamma/n\to0$, yields
\begin{equation}
\limsup_{n\to\infty}\frac1n\log\mathrm{Tr}(\Pi_B\omega)
\leq(\beta-s)(\beta_c-\delta)+s\beta_c-\beta\beta_c=-(\beta-s)\delta.
\end{equation}
Set
\begin{equation}
c_B:=(\beta-s)\delta>0.
\end{equation}
\end{proof}

\section{Proof of \autoref{thm:main}}\label{sec:proof}
Set, as in \autoref{thm:direct-bottleneck},
\begin{equation}
w_A:=\mathrm{Tr}(\Pi_A\omega),\qquad {w_B}:=\mathrm{Tr}(\Pi_B\omega),\qquad {w_C}:=\mathrm{Tr}(\Pi_C\omega);
\end{equation}
these quantities depend on $n$. By \eqref{eq:Bott2} and \autoref{lem:Pi},
\begin{equation}
\liminf_{n\to\infty}w_C\geq c_\star=\frac34.
\end{equation}
Hence $w_C-1/2\geq1/8$ for all sufficiently large $n$. By \autoref{cor:PiA},
\begin{equation}
\frac1n\log {w_A}\longrightarrow0,
\end{equation}
and by \autoref{lem:PiB},
\begin{equation}
\limsup_{n\to\infty}\frac1n\log {w_B}\leq-c_B.
\end{equation}
In \autoref{thm:direct-bottleneck}, the $A$-to-$C$ rate $r_A$ is exactly $\eta_n$ from \eqref{eq:eta-def}, while $r_B\leq3n\gamma_{\max}$ by \eqref{eq:rB-bound}. Therefore, for all sufficiently large $n$,
\begin{equation}\label{eq:final-Tmix-bound}
T_{\mathrm{mix}}(\mathcal L)\geq
\frac{1}{8\,(3n\gamma_{\max}\,w_B/w_A+\eta_n)}.
\end{equation}
The polynomial factor $3n\gamma_{\max}$ and the subexponential factor $1/w_A$ do not affect exponential rates. Together with \autoref{lem:epsilon}, \eqref{eq:final-Tmix-bound} yields
\begin{equation}
\liminf_{n\to\infty}\frac1n\log T_{\mathrm{mix}}(\mathcal L)
\geq\min\{c_B,\tilde c\}>0.
\end{equation}
Thus \autoref{thm:main} holds with
\begin{equation}
c_{\mathrm{mix}}:=\min\{c_B,\tilde c\}.
\end{equation}

\section{\texorpdfstring{Extensions and open problems}{Extensions and open problems}}\label{sec:discussion}
We conclude with some comments on the assumptions of \autoref{thm:main} and on questions that this work leaves open.

\subsection{Growth of the rates}
The proof only uses that $\gamma_{\max}$ grows subexponentially, i.e.,
\begin{equation}
\limsup_{n\to\infty}\frac1n\log\gamma_{\max}\leq0,
\end{equation}
since the rates enter only through the factors $3n\gamma_{\max}$ in \eqref{eq:final-Tmix-bound} and $27n\gamma_{\max}$ in \eqref{eq:eta-bound}.
\subsection{Other mixing thresholds}
For $\varepsilon\in(0,2)$, let $T^{(\varepsilon)}_{\mathrm{mix}}$ be defined as $T_{\mathrm{mix}}$ in \autoref{def:mixing}, with the threshold $1$ replaced by $\varepsilon$. The proof of \autoref{thm:direct-bottleneck} gives
\begin{equation}
T^{(\varepsilon)}_{\mathrm{mix}}(\mathcal L)\geq\frac{w_C-\frac\varepsilon2}{r_B\,w_B/w_A+r_A},
\end{equation}
and \autoref{thm:classical} and its proof remain valid if $c_\star=\frac34$ is replaced by any $c_\star\in(\frac\varepsilon2,1)$, since \autoref{prop:two-separated-regions} holds for every $c\in(0,1)$. Hence the conclusion of \autoref{thm:main} holds for $T^{(\varepsilon)}_{\mathrm{mix}}$ as well.
\subsection{The exponent}
We have made no attempt to optimize $c_{\mathrm{mix}}$, which depends on the localization rates of \autoref{thm:Manai} and on the choice of $\delta$. For $\Gamma=0$, \autoref{rem:populations} shows that the Davies dynamics reduces to a reversible single-spin-flip dynamics of the classical REM, for which much sharper information is available: for the Metropolis dynamics, the exponential scale of the spectral gap was determined in \cite{FontesIsopiKohayakawaPicco1998}. Identifying the optimal exponent for $\Gamma>0$ is an open problem.
\subsection{Other thermal Lindbladians}
Our result concerns the Davies generator, whose jump operators are exactly resolved in the energy eigenbasis. The KMS-detailed-balance Lindbladians constructed for quantum Gibbs sampling algorithms \cite{ChenKastoryanoGilyen2023,DingLiLin2025,GilyenChenDoriguelloKastoryano2024} have jump operators that are smoothed in energy and create coherences between eigenvectors, so that the reduction to population dynamics of \autoref{rem:populations} is not available, and \autoref{thm:direct-bottleneck} does not apply to them directly. We expect the same localization mechanism to produce slow mixing for these dynamics as well, but we do not pursue this here.

\appendix

\section{\texorpdfstring{Nondegeneracy of the spectrum}{Nondegeneracy of the spectrum}}\label{app:deg}

\begin{lem}\label{lem:deg}
The spectrum and the energy gaps of $H=\Gamma T+U$ are almost surely nondegenerate, i.e., \eqref{eq:gap-nondeg} holds almost surely.
\end{lem}
\begin{proof}
For fixed $n$, let $\mathcal P$ be the finite set of unordered pairs of eigenvalue indices, with repetitions allowed, and define the squared pair-sum discriminant
\begin{equation}
F(H):=\prod_{\substack{P,Q\in\mathcal P\\ P<Q}}\bigl(s_P-s_Q\bigr)^2,\qquad s_{\{i,j\}}:=E_i+E_j,
\end{equation}
where $<$ is any fixed ordering of $\mathcal P$. Then $F(H)=0$ exactly when two distinct unordered pairs have the same energy sum. Because the square makes this expression invariant under permutations of the eigenvalues, $F$ is a symmetric polynomial in the eigenvalues. By the fundamental theorem of symmetric polynomials, it is a polynomial in the coefficients of the characteristic polynomial of $H$, hence a polynomial in the matrix entries of $H$ and therefore, for $H=\Gamma T+U$, a polynomial in the diagonal entries $\{U(x)\}_{x\in {Q_n}}$.
It remains to show that this polynomial is not identically zero, i.e., to exhibit one diagonal potential $\widehat U$ such that $\Gamma T+\widehat U$ has no pair-sum degeneracy. We first find such a potential for $\Gamma=0$, where the eigenvalues of $\widehat U$ are simply its diagonal entries, and then treat $\Gamma T$ as a perturbation.

We identify a diagonal potential $\widehat U$ with the vector $(\widehat U(x))_{x\in Q_n}\in\mathbb R^{Q_n}\cong\mathbb R^{2^n}$. Let $\mathcal M_n$ be the set of unordered pairs $\{x,y\}$ of configurations $x,y\in Q_n$, where $x=y$ is allowed (i.e., $\mathcal M_n$ is the set of multisets of size two in $Q_n$). For $p=\{x,y\}$ and $q=\{z,w\}$ in $\mathcal M_n$ with $p\neq q$, the potential $\widehat U$, viewed as a Hamiltonian with $\Gamma=0$, has the pair-sum degeneracy labelled by $(p,q)$ if and only if $\widehat U$ belongs to
\begin{equation}
\mathcal H_{p,q}:=\bigl\{\widehat U\in\mathbb R^{Q_n}:\ \ell_{p,q}(\widehat U)=0\bigr\},\qquad \ell_{p,q}(\widehat U):=\widehat U(x)+\widehat U(y)-\widehat U(z)-\widehat U(w).
\end{equation}
The map $\ell_{p,q}:\mathbb R^{Q_n}\to\mathbb R$ is linear, and it is not identically zero: the coefficient of $\widehat U(u)$ in $\ell_{p,q}(\widehat U)$ is the number of occurrences of $u$ in $p$ minus the number of occurrences of $u$ in $q$, and since $p\neq q$ as multisets, this difference is nonzero for at least one $u\in Q_n$. (For instance, $p=\{x,x\}$ and $q=\{x,z\}$ with $z\neq x$ give $\ell_{p,q}(\widehat U)=\widehat U(x)-\widehat U(z)$.) Hence $\mathcal H_{p,q}$, the kernel of a nonzero linear functional, is a hyperplane through the origin, i.e., a linear subspace of $\mathbb R^{Q_n}$ of dimension $2^n-1$. In particular, $\mathcal H_{p,q}$ has $2^n$-dimensional Lebesgue measure zero. Since $\mathcal M_n$ is finite (it has $2^{n-1}(2^n+1)$ elements), there are only finitely many such hyperplanes, and their union
\begin{equation}
\mathcal H:=\bigcup_{\substack{p,q\in\mathcal M_n\\ p\neq q}}\mathcal H_{p,q}
\end{equation}
is a finite union of Lebesgue-null sets, hence itself Lebesgue-null. In particular, $\mathcal H\neq\mathbb R^{Q_n}$: there exist diagonal potentials (in fact, Lebesgue-almost all of them) that have no pair-sum degeneracy when $\Gamma=0$. An explicit example is $\widehat U(x):=4^{k(x)}$, where $k:Q_n\to\{0,\dots,2^n-1\}$ is any bijection: the base-$4$ expansion of $4^a+4^b$ has the digit $1$ at the positions $a$ and $b$ if $a\neq b$, and the digit $2$ at the position $a$ if $a=b$, so the sum $4^a+4^b$ determines the multiset $\{a,b\}$.

Choose a diagonal realization $\widehat U$ outside $\mathcal H$ and let
\begin{equation}
\Delta:=\min_{\{x,y\}\neq\{z,w\}}|\widehat U(x)+\widehat U(y)-\widehat U(z)-\widehat U(w)|>0
\end{equation}
(a minimum of finitely many nonzero numbers, hence positive). Since every $\mathcal H_{p,q}$ is a linear subspace, $\lambda\widehat U\notin\mathcal H$ for every $\lambda>0$, and replacing $\widehat U$ by $\lambda\widehat U$ multiplies $\Delta$ by $\lambda$. After multiplying $\widehat U$ by a sufficiently large positive constant, we may therefore assume $\Delta>4\Gamma\|T\|$ (note that $\|T\|=n$). Weyl's eigenvalue perturbation bound \cite[Corollary~III.2.6]{Bhatia1997} gives a labeling $\{\widehat E(x)\}_{x\in Q_n}$ of the eigenvalues of $\Gamma T+\widehat U$ such that
\begin{equation}
|\widehat E(x)-\widehat U(x)|\leq\Gamma\|T\|,\qquad x\in Q_n.
\end{equation}
Hence, for distinct unordered pairs,
\begin{align}
|\widehat E(x)+\widehat E(y)-\widehat E(z)-\widehat E(w)|
&\geq|\widehat U(x)+\widehat U(y)-\widehat U(z)-\widehat U(w)|-4\Gamma\|T\|\\
&\geq\Delta-4\Gamma\|T\|>0.
\end{align}
Thus $F(\Gamma T+\widehat U)\neq0$, so the polynomial is not identically zero. Its zero set has Lebesgue measure zero; this is the standard fact that the zero set of a real polynomial in $N$ variables that is not identically zero is a Lebesgue-null subset of $\mathbb R^N$, which follows by induction on $N$ and Fubini's theorem. Since the Gaussian vector of diagonal entries of $U$ has an absolutely continuous law, $\mathbb P(F(H)=0)=0$. Taking $i=j$ also excludes eigenvalue degeneracy. This argument holds for each fixed $n$; since $n\in\mathbb N$ is countable, intersecting the corresponding probability-one events yields simultaneous nondegeneracy of the spectrum and all energy gaps for every $n$ almost surely.
\end{proof}

\section{\texorpdfstring{Gaussian and binomial estimates}{Gaussian and binomial estimates}}\label{app:tails}

The following are standard estimates for the  tail of Gaussian and Binomial random variables. Let $G\sim\mathcal N(0,1)$; we denote by $\varphi$ and $\Phi$ its probability density and cumulative distribution function respectively. In particular,
\be
\Pp(G>t)=1-\Phi(t)=\Phi(-t)=:\overline\Phi(t).
\ee

\begin{lem}[Mills ratio bounds]\label{lem:mills-ratio}
For every $t>0$,
\be
\varphi(t)\left(\frac1t-\frac1{t^3}\right)<\overline\Phi(t)<\frac{\varphi(t)}{t}.
\label{eq:mills-ratio}
\ee
\end{lem}

\begin{lem}[Chernoff upper tail for Binomial]\label{lem:chernoff-binomial}
If $X\sim\mathrm{Bin}(N,p)$ and $\lambda:=Np$, then for every $t>\lambda$,
\be
\Pp(X\geq t)\leq(e\lambda/t)^t.
\label{eq:chernoff-binomial}
\ee
\end{lem}

\subsection{The cardinality of the level sets}

Recall from \eqref{eq:R} that the level sets are

\be
\mathcal{R}_{n,\delta}=\{x\in Q_n:-U_n(x)\geq(\beta_c-\delta)n\}.
\ee
 Since the energies $U_n(x)$, $x\in Q_n$, are i.i.d., the cardinality  $|\mathcal{R}_{n,\delta}|$ has a binomial distribution $\mathrm{Bin}(2^n,p_n(\delta))$ where
\be
p_n(\delta):=\Pp(x\in \mathcal{R}_{n,\delta})
=\Pp(-U_n(x)\geq(\beta_c-\delta)n)
\label{eq:pnx}
\ee
is nothing but the tail of a Gaussian random variable. Hence by \eqref{eq:mills-ratio}, for any $\delta<\beta_c$ and all sufficiently large $n$,
\be
\frac1{\sqrt{2\pi}}\frac1t(1-t^{-2})e^{-\frac n2(\beta_c-\delta)^2}
\leq p_n(\delta)\leq
\frac1{\sqrt{2\pi}}\frac1t e^{-\frac n2(\beta_c-\delta)^2},
\qquad t=(\beta_c-\delta)\sqrt n.
\label{eq:mills-application}
\ee

The next lemma is a result for a sequence of independent Binomial random variables.

\begin{lem}\label{lem:binomialcontrol}
Let $(p_n)_{n\geq1}\subset(0,1)$ and let $(X_n)_{n\geq1}$ be a sequence of independent random variables with
$X_n\sim\mathrm{Bin}(2^n,p_n)$. Set
\begin{equation}
\lambda_n:=2^np_n=\E[X_n],
\end{equation}
and define
\be
Y_n:=\frac1n\log X_n,
\ee
with the convention $\log0=-\infty$. If
\be
\text{\hypertarget{cond:bin-i}{i)} }\limsup_{n\to\infty}\frac{\log\lambda_n}{n}\leq0,
\qquad
\text{\hypertarget{cond:bin-ii}{ii)} }\sum_{n=1}^\infty\lambda_n=\infty,
\label{eq:binomial-conditions}
\ee
then
\be
\limsup_{n\to\infty}Y_n=0
\qquad\text{almost surely}.
\ee
\end{lem}

\begin{proof}
We will show that \hyperlink{cond:bin-i}{i)} forces $\limsup_nY_n\leq0$ a.s., while \hyperlink{cond:bin-ii}{ii)} forces $\limsup_nY_n\geq0$ a.s.

Fix $\varepsilon>0$ and set $t_n=e^{n\varepsilon}$. By the Chernoff bound of \autoref{lem:chernoff-binomial}, for every $t>\lambda_n$,
\be
\Pp(X_n\geq t)\leq\exp\bigl(t(1+\log\lambda_n-\log t)\bigr).
\ee
Condition \hyperlink{cond:bin-i}{i)} implies that, for all sufficiently large $n$,
\be
\log\lambda_n\leq\frac{\varepsilon n}{2}.
\ee
In particular $t_n>\lambda_n$ and, for all sufficiently large $n$,
\be
1+\log\lambda_n-n\varepsilon
\leq1-\frac{\varepsilon n}{2}
\leq-\frac{\varepsilon n}{4}.
\ee
Therefore
\be
\Pp(X_n\geq e^{n\varepsilon})
\leq\exp\left(-\frac{\varepsilon n}{4}e^{n\varepsilon}\right),
\ee
and hence
\be
\sum_{n=1}^\infty\Pp(X_n\geq e^{n\varepsilon})<\infty.
\ee
By the first Borel--Cantelli lemma,
\begin{equation}
\Pp\bigl(X_n\geq e^{n\varepsilon}\ \mathrm{i.o.}\bigr)=0,
\end{equation}
which yields
$\limsup_nY_n\leq\varepsilon$ a.s. Since $\varepsilon>0$ is arbitrary, $\limsup_nY_n\leq0$ a.s.

Let $\mathcal E_n:=\{X_n\geq1\}$. Since $X_n\sim\mathrm{Bin}(2^n,p_n)$,
\be
\Pp(\mathcal E_n)=1-(1-p_n)^{2^n}\geq1-e^{-2^np_n}=1-e^{-\lambda_n},
\ee
using $1-x\leq e^{-x}$. Notice that
\be
1-e^{-x}\geq(1-e^{-1})\min(x,1)\geq\frac12\min(x,1),
\qquad x\geq0,
\ee
and, since $\sum_n\lambda_n=\infty$,
\begin{equation}
\sum_{n=1}^\infty\min(\lambda_n,1)=\infty.
\end{equation}
Therefore
\be
\sum_{n=1}^\infty\Pp(\mathcal E_n)
\geq\frac12\sum_{n=1}^\infty\min(\lambda_n,1)=\infty.
\ee
Since the $X_n$ are independent, the events $\mathcal E_n$ are independent, and by the second Borel--Cantelli lemma,
\begin{equation}
\Pp(\mathcal E_n\ \mathrm{i.o.})=1.
\end{equation}
Thus almost surely $X_n\geq1$ for infinitely many $n$, so $\limsup_nY_n\geq0$ a.s.
\end{proof}

\section{Proof of \texorpdfstring{\autoref{thm:Manai}}{the localization theorem}}\label{app:loc}
The proof below is a rewriting of the proof of \cite[Theorem~1.5]{Manai_2023} given by Manai and Warzel in \cite[Section~4.4]{Manai_2023}. We reproduce their argument in order to keep track of the energy thresholds and to impose the non-resonance condition on a set that is large enough for the eigenvector estimate; cf.\ \autoref{rem:MW-threshold}. All the main ingredients of the proof are due to Manai and Warzel. Specifically, the following parts are taken over from \cite{Manai_2023}: \cite[Lemmas~4.2--4.4 and Proposition~4.5]{Manai_2023}, which we use as stated (see \ref{it:M1}--\ref{it:M4} below, and the comment on the hypotheses of \cite[Lemma~4.2]{Manai_2023} after its statement in \autoref{lem:MW-deep-hole}); the choice of the parameters $\varepsilon$ and $\alpha$ on \cite[p.~1288]{Manai_2023}, with the modifications described in~\ref{it:rw-deephole} and~\ref{it:rw-balls} below; the block decomposition based on the auxiliary Hamiltonian \cite[(4.20)]{Manai_2023} (\stepref{2}); the Schur complement analysis \cite[(4.22)--(4.25)]{Manai_2023} (Steps~\hyperref[sec:step2]{2} and~\hyperref[sec:step3]{3}); and the eigenvector estimate \cite[(4.24) and (4.27)]{Manai_2023} (\stepref{6}). The parts that we rewrite are the following:
\begin{enumerate}[label=(\roman*)]
\item\label{it:rw-deephole} the deep-hole parameter, denoted by $\delta_p$ below, is chosen larger than the threshold parameter $\delta$ of \autoref{thm:Manai}; see the choice of the parameters below and \autoref{it:mw-proof} of \autoref{rem:MW-threshold};
\item\label{it:rw-balls} the deep-hole scenario is required on balls of twice the radius (event \ref{it:E1} below). This guarantees that the balls around distinct deep configurations (i.e., configurations in the set $\mathcal R=\mathcal R_{n,\delta_p}$ defined below) remain pairwise disjoint when their radius is increased by one (\stepref{1}), so that no edge of the hypercube joins two distinct balls and the hopping terms leaving different balls act on disjoint sets (\stepref{2}). In \cite[Section~4.4]{Manai_2023}, these properties are used, in the notation there, in the representation $H-H'=-\Gamma\bigoplus_\sigma A_\sigma$ and in the equality $\|A\|=\max_\sigma\|A_\sigma\|$ on \cite[p.~1287]{Manai_2023}, and in the identity $P_\delta AP_\delta=0$ on \cite[p.~1289]{Manai_2023}. They follow from the proof of \cite[Lemma~4.2]{Manai_2023}, which, under the additional condition \hyperlink{cond:C1}{(C1)} below and outside the exceptional event considered there, shows that the distance between two distinct deep configurations exceeds three times the radius of the balls, and this implies these properties for $N$ large; they do not follow from the disjointness of the balls required in \cite[Definition~4.1]{Manai_2023} alone, which allows two balls to be joined by an edge;
\item\label{it:rw-nonres} the non-resonance event is imposed on $\mathcal R_{n,2\delta}$ (event \ref{it:E3} below), and the ground-state energies of the balls around the remaining deep configurations are separated by means of the energy asymptotics of \cite[Lemma~4.3]{Manai_2023} (\stepref{4}); see \autoref{it:mw-nonres} of \autoref{rem:MW-threshold};
\item\label{it:rw-counting} the comparison of the eigenvalues of $H$ with the ground-state energies of the balls, which on \cite[p.~1289]{Manai_2023} is obtained from a standard perturbation result, is formulated as the counting inequality \eqref{eq:counting} (\stepref{3}), from which the two thresholds in items~\ref{it:loc-label} and~\ref{it:loc-threshold} of \autoref{thm:Manai} are read off (\stepref{5}); see \autoref{it:mw-statement} of \autoref{rem:MW-threshold}.
\end{enumerate}
Other changes, which are made for convenience only, are indicated where they occur: for instance, the projection $P$ of \stepref{2} includes the ground states of the balls around all deep configurations, whereas the projection \cite[(4.22)]{Manai_2023} includes only those with ground-state energy below $E_{\delta/2}$ (in the notation there), and the hopping terms between the balls are bounded by the Schur test instead of \cite[(4.21)]{Manai_2023}.
As in \autoref{prop:deep-hole}, we identify $\{-1,1\}^n$ with $Q_n$; the results of \cite{Manai_2023} are stated with $N$ in place of $n$.

\subsection{Structure of the proof}
Steps~\hyperref[sec:step1]{1} and~\hyperref[sec:step2]{2} construct an auxiliary block-diagonal Hamiltonian $H'$, in which the Hamming balls of radius $m\approx\alpha_0n$ around the deep configurations are decoupled from each other and from the rest of the hypercube. The ground state $\varphi_x$ of the ball around $x$ is localized near $x$, its energy $e_x$ is close to $U(x)+\Gamma^2n/U(x)$, and the rest of the spectrum of $H'$ lies above the relevant energies. \stepref{3} treats the coupling $V=H-H'$ by a Schur complement (Feshbach) argument and shows that, below the energy $\Lambda_{\delta_p/2}$, the eigenvalue counting function of $H$ agrees with that of the ball energies $\{e_x\}$ up to an exponentially small energy shift $\nu_n$ (the counting inequality \eqref{eq:counting}). \stepref{4} shows that each ball energy $e_x$ with $x\in\mathcal R_{n,\delta}$ is separated from all the other ball energies by $r_n\gg\nu_n$ (non-resonance). \stepref{5} deduces items~\ref{it:loc-label}, \ref{it:loc-energy} and~\ref{it:loc-threshold} from Steps~\hyperref[sec:step3]{3} and~\hyperref[sec:step4]{4}, and \stepref{6} deduces \autoref{it:loc-decay} by showing that each eigenvector $\psi_x$ is exponentially close to $\varphi_x$.

\subsection{Notation}
Configurations are denoted by $x,x',y\in Q_n=\{0,1\}^n$, and $|y\rangle$ is the corresponding computational-basis vector, so that $(\mathbb C^2)^{\otimes n}\cong\ell^2(Q_n)$. For $x\in Q_n$ and $m\in\mathbb N_0$ we write
\begin{equation}
S_m(x):=\{y:d(x,y)=m\},
\end{equation}
and for real $r\geq0$ we have $B_r(x)=B_{\lfloor r\rfloor}(x)$. For $S\subseteq Q_n$ let
\begin{equation}
\Pi_S:=\sum_{y\in S}|y\rangle\langle y|,
\end{equation}
and let $H_S$ denote the restriction of $\Pi_SH\Pi_S$ to $\mathrm{Ran}\,\Pi_S$; vectors in $\mathrm{Ran}\,\Pi_S$ are regarded as vectors in $\ell^2(Q_n)$. For $t\in(0,\beta_c)$ we set
\begin{equation}
\Lambda_t:=-(\beta_c-t)n,\qquad\text{so that}\qquad\mathcal R_{n,t}=\{x:U(x)\leq\Lambda_t\}.
\end{equation} Finally, $h_2$ denotes the binary entropy \eqref{eq:binary-entropy}, which is increasing on $[0,1/2]$.

\subsection{\texorpdfstring{Input from \cite{Manai_2023}}{Input from Manai and Warzel}}
Let $\varepsilon_0,t>0$ and $\alpha\in(0,1/2)$. Following \cite[Definition~4.1]{Manai_2023}, $U$ satisfies the \emph{local $(\varepsilon_0,t,\alpha)$-deep-hole scenario at $x\in\mathcal R_{n,t}$} if
\begin{equation}
|U(y)|\leq\varepsilon_0n\quad\text{for all }y\in B_{\alpha n}(x)\setminus\{x\}\qquad\text{and}\qquad n^{-2}\sum_{y\in S_1(x)}|U(y)|\leq n^{-1/4};
\end{equation}
it satisfies the \emph{global $(\varepsilon_0,t,\alpha)$-deep-hole scenario} if the local one holds at every $x\in\mathcal R_{n,t}$ and the balls $B_{\alpha n}(x)$, $x\in\mathcal R_{n,t}$, are pairwise disjoint. Consider the conditions
\begin{equation}\label{eq:MW-conditions}
\text{\hypertarget{cond:C1}{(C1)}}\ \ \varepsilon_0<\beta_c-t,\qquad
\text{\hypertarget{cond:C2}{(C2)}}\ \ 2h_2(3\alpha)+t(2\beta_c-t)<\varepsilon_0^2,\qquad
\text{\hypertarget{cond:C3}{(C3)}}\ \ 2\Gamma\sqrt{\alpha(1-\alpha)}+\varepsilon_0<\beta_c-2t.
\end{equation}
Conditions \hyperlink{cond:C2}{(C2)} and \hyperlink{cond:C3}{(C3)} are \cite[(4.1) and (4.7)]{Manai_2023}, and \hyperlink{cond:C3}{(C3)} implies \hyperlink{cond:C1}{(C1)}. For the reader's convenience, we repeat the statement of \cite[Lemma~4.2]{Manai_2023} in the present notation; in \cite{Manai_2023}, $\varepsilon_0$, $t$, $n$, $h_2$ and $\mathcal R_{n,t}$ are denoted by $\varepsilon$, $\delta$, $N$, $\gamma$ and $\mathcal L_{\beta_c-\delta}$, respectively.
\begin{lem}[{\cite[Lemma~4.2]{Manai_2023}}]\label{lem:MW-deep-hole}
Let $\varepsilon_0,t>0$ and $\alpha\in(0,1/2)$ be such that \hyperlink{cond:C2}{(C2)} holds. Then $U$ satisfies the global $(\varepsilon_0,t,\alpha)$-deep-hole scenario with probability exponentially close to one, i.e., there is some $c(\varepsilon_0,t,\alpha)>0$ such that, for all sufficiently large $n$,
\begin{equation}\label{eq:MW-deep-hole-prob}
\mathbb P\bigl(U\text{ satisfies the global }(\varepsilon_0,t,\alpha)\text{-deep-hole scenario}\bigr)\geq1-e^{-c(\varepsilon_0,t,\alpha)n}.
\end{equation}
\end{lem}
The statement of \autoref{lem:MW-deep-hole} assumes only \hyperlink{cond:C2}{(C2)} and $\alpha\in(0,1/2)$, but its proof in \cite{Manai_2023} also uses \hyperlink{cond:C1}{(C1)}, which guarantees that outside the exceptional event $\widehat\Omega_N(\varepsilon,\delta,\alpha)$ considered there, for every $x\in\mathcal R_{n,t}$, no other configuration of $\mathcal R_{n,t}$ lies in $B_{3\alpha n}(x)$ (whence the disjointness of the balls), and $3\alpha\leq1/2$, which is needed for the volume asymptotics
\begin{equation}
\log|B_{3\alpha n}(x)|=n(h_2(3\alpha)+o(1)).
\end{equation}
Both conditions hold for the parameters used in \cite[Section~4.4]{Manai_2023}, where, in the notation there, $\varepsilon=\beta_c/2$ and $\delta<\beta_c/8$, so that $\varepsilon<\beta_c-\delta$, and where $\alpha$ can be chosen so small that $3\alpha\leq1/2$; they also hold for all parameters used below. From now on, $t\in(0,\beta_c)$ and $\alpha\in(0,1/6]$, so that $3\alpha\leq1/2$. We use the following four results.
\begin{enumerate}[label=(M\arabic*)]
\item\label{it:M1} \cite[Lemma~4.2]{Manai_2023}, i.e., \autoref{lem:MW-deep-hole}, with the additional hypothesis \hyperlink{cond:C1}{(C1)} used in its proof. If \hyperlink{cond:C1}{(C1)} and \hyperlink{cond:C2}{(C2)} hold, then the global $(\varepsilon_0,t,\alpha)$-deep-hole scenario has probability at least $1-e^{-cn}$ for some $c=c(\varepsilon_0,t,\alpha)>0$ and all sufficiently large $n$.
\item\label{it:M2} \cite[Lemma~4.3]{Manai_2023}. Assume \hyperlink{cond:C3}{(C3)}, and let $U$ satisfy the local $(\varepsilon_0,t,\alpha)$-deep-hole scenario at $x\in\mathcal R_{n,t}$. Then, for all $n$ larger than a deterministic threshold depending only on $(\Gamma,\varepsilon_0,t,\alpha)$, the ground-state energy $e_x$ of $H_{B_{\alpha n}(x)}$ is a simple eigenvalue, and all other eigenvalues of $H_{B_{\alpha n}(x)}$ are at least $\Lambda_t$. (In \cite{Manai_2023} this is phrased as: the spectrum of $H_{B_{\alpha n}(x)}$ below $\Lambda_t$ consists of one simple eigenvalue; recall that $e_x\leq\langle x|H|x\rangle=U(x)\leq\Lambda_t$.) Uniformly in such $x$,
\begin{equation}\label{eq:MW-ball-energy}
e_x=U(x)+\frac{\Gamma^2n}{U(x)}+O_{\Gamma,t}\!\left(n^{-1/4}\right),
\end{equation}
and, for every $\alpha'\in(0,\alpha]$, the normalized ground state $|\varphi_x\rangle$ of $H_{B_{\alpha n}(x)}$ satisfies
\begin{equation}\label{eq:MW-ball-decay}
\sum_{y\notin B_{\alpha'n}(x)}|\langle y|\varphi_x\rangle|^2\leq C\,n\,e^{-c(\alpha,\alpha')n},
\end{equation}
with $C=C(\Gamma,t)<\infty$ and $c(\alpha,\alpha')>0$.
\item\label{it:M3} \cite[Lemma~4.4(2)]{Manai_2023}. If \hyperlink{cond:C1}{(C1)}--\hyperlink{cond:C3}{(C3)} hold, then there are $c>0$ and $n_0$, depending only on $(\Gamma,\varepsilon_0,t,\alpha)$, such that for all $n\geq n_0$ and all $r>0$ the probability that there exist distinct $x,x'\in\mathcal R_{n,t}$ with
\begin{equation}
\mathrm{dist}\Bigl(\mathrm{spec}\,H_{B_{\alpha n}(x)}\cap(-\infty,\Lambda_t),\ \mathrm{spec}\,H_{B_{\alpha n}(x')}\cap(-\infty,\Lambda_t)\Bigr)\leq r
\end{equation}
is at most
\begin{equation}
4r\,e^{(2\beta_ct-t^2)n}+e^{-cn}.
\end{equation}
\item\label{it:M4} \cite[Proposition~4.5]{Manai_2023}. If $\Gamma<\beta_c-t$, then, outside an event of exponentially small probability,
\begin{equation}
\inf\mathrm{spec}\bigl(\Gamma T+U\1_{\{U\geq-(\beta_c-t)n\}}\bigr)\geq-(\beta_c-t)n-o_{\Gamma,t}(n),
\end{equation}
where $U\1_{\{U\geq-(\beta_c-t)n\}}$ is the diagonal operator with entries $U(y)\1\bigl(U(y)\geq-(\beta_c-t)n\bigr)$. (For $\Gamma=0$ this is trivial.)
\end{enumerate}

\subsection{Choice of the parameters}
Fix $\Gamma\in[0,\beta_c)$ and set
\begin{equation}
\varepsilon_0:=\frac{\beta_c}2,\qquad \delta_p:=\frac12\min\Bigl\{\beta_c-\Gamma,\frac{\beta_c}8\Bigr\}.
\end{equation}
Thus $0<\delta_p\leq\beta_c/16$ and $\Gamma<\beta_c-\delta_p$. Since
\begin{equation}
\delta_p(2\beta_c-\delta_p)<2\beta_c\cdot\beta_c/8=\varepsilon_0^2
\end{equation}
and $h_2(u)\to0$ as $u\downarrow0$, we can fix $\alpha_0\in(0,1/12)$ such that
\begin{equation}\label{eq:alpha0-choice}
2h_2(6\alpha_0)+\delta_p(2\beta_c-\delta_p)<\varepsilon_0^2,\qquad
\Gamma\sqrt{\alpha_0(1-\alpha_0)}<\frac{\delta_p}{16}.
\end{equation}
Then \hyperlink{cond:C1}{(C1)}--\hyperlink{cond:C3}{(C3)} hold for $(\varepsilon_0,t,\alpha_0)$ for every $t\in(0,\delta_p]$, and \hyperlink{cond:C1}{(C1)}--\hyperlink{cond:C2}{(C2)} hold for $(\varepsilon_0,\delta_p,2\alpha_0)$. Indeed, \hyperlink{cond:C1}{(C1)} follows from $t\leq\delta_p<\beta_c/2$; \hyperlink{cond:C2}{(C2)} follows from \eqref{eq:alpha0-choice}, because $h_2$ is increasing on $[0,1/2]$, $6\alpha_0<1/2$, and $t\mapsto t(2\beta_c-t)$ is increasing on $[0,\beta_c]$; and \hyperlink{cond:C3}{(C3)} follows from
\begin{equation}
2\Gamma\sqrt{\alpha_0(1-\alpha_0)}<\delta_p/8<\beta_c/2-2\delta_p\leq\beta_c/2-2t,
\end{equation}
where we used $\delta_p\leq\beta_c/16$. These are the choices made by Manai and Warzel in the proof of \cite[Theorem~1.5]{Manai_2023} (see \cite[p.~1288]{Manai_2023}), where $\varepsilon=\beta_c/2$, $\delta\in(0,\min\{\beta_c-\Gamma,\beta_c/8\})$, and $\alpha$ is such that \cite[(4.1)]{Manai_2023} holds and $2\Gamma\sqrt{\alpha(1-\alpha)}<\delta/8$; the only differences are that we fix the particular value $\delta_p$ of their parameter $\delta$, and that we also require \cite[(4.1)]{Manai_2023} with $2\alpha_0$ in place of $\alpha_0$.

Let
\begin{equation}
C_{\mathrm b}:=C(\Gamma,\delta_p),\qquad c_{\mathrm b}:=c(\alpha_0,\alpha_0/2)>0
\end{equation}
be the constants in \eqref{eq:MW-ball-decay} for $t=\delta_p$, $\alpha=\alpha_0$ and $\alpha'=\alpha_0/2$, and define
\begin{equation}
\bar\delta(\Gamma):=\min\Bigl\{\frac{\delta_p}{8},\frac{c_{\mathrm b}}{8\beta_c}\Bigr\}.
\end{equation}
(The conditions $\delta<\delta_p/8$ and $\delta<c_{\mathrm b}/(8\beta_c)$ play the role of the conditions $s\leq\delta/8$ and $2\beta_cs<c$ imposed on the parameter $s$ in \cite[p.~1289]{Manai_2023}.) Fix $\delta\in(0,\bar\delta(\Gamma))$ and put
\begin{equation}
m:=\lfloor\alpha_0n\rfloor,\qquad \hat\delta:=2\delta,\qquad r_n:=e^{-c_{\mathrm b}n/2},\qquad \mathcal R:=\mathcal R_{n,\delta_p},\qquad M:=|\mathcal R|.
\end{equation}
Note that $\delta<\hat\delta<\delta_p/4$.

\subsection{The event}
Let $\Omega^{\mathrm{loc}}_{n,\Gamma,\delta}$ be the intersection of the following three events:
\begin{enumerate}[label=(E\arabic*)]
\item\label{it:E1} $U$ satisfies the global $(\varepsilon_0,\delta_p,2\alpha_0)$-deep-hole scenario;
\item\label{it:E2} the lower bound of \ref{it:M4} holds with $t=\delta_p$;
\item\label{it:E3} for all distinct $x,x'\in\mathcal R_{n,\hat\delta}$, one has
\begin{equation}
\mathrm{dist}\Bigl(\mathrm{spec}\,H_{B_m(x)}\cap(-\infty,\Lambda_{\hat\delta}),\ \mathrm{spec}\,H_{B_m(x')}\cap(-\infty,\Lambda_{\hat\delta})\Bigr)>r_n.
\end{equation}
\end{enumerate}
The events \ref{it:E1} and \ref{it:E2} constitute the event $\Omega_N(\Gamma,\delta,\alpha)$ of the proof of \cite[Theorem~1.5]{Manai_2023} (with $2\alpha_0$ in place of $\alpha$), and \ref{it:E3} is the analogue of the non-resonance event \cite[(4.26)]{Manai_2023}, imposed here on the set $\mathcal R_{n,\hat\delta}$, which contains all configurations whose ball energies can be close to those of configurations in $\mathcal R_{n,\delta}$ (see \stepref{4}). By \ref{it:M1} (with $t=\delta_p$ and $\alpha=2\alpha_0$), \ref{it:M4} (with $t=\delta_p$, using $\Gamma<\beta_c-\delta_p$) and \ref{it:M3} (with $t=\hat\delta$, $\alpha=\alpha_0$ and $r=r_n$), there is $c>0$ such that, for all sufficiently large $n$,
\begin{equation}
\mathbb P\bigl((\Omega^{\mathrm{loc}}_{n,\Gamma,\delta})^c\bigr)\leq 3e^{-cn}+4e^{-(c_{\mathrm b}/2-2\beta_c\hat\delta)n}.
\end{equation}
Since $2\beta_c\hat\delta=4\beta_c\delta<c_{\mathrm b}/2$, this proves the probability bound in \autoref{thm:Manai}. From now on we work on $\Omega^{\mathrm{loc}}_{n,\Gamma,\delta}$, and ``$n$ large'' means that $n$ exceeds a deterministic threshold depending only on $\Gamma$ and $\delta$ (and also on $\alpha$ in \stepref{6}); this threshold may increase from line to line.

\subsection{Step 1: geometry and ball ground states}\label{sec:step1}
By \ref{it:E1}, for every $x\in\mathcal R$ the local $(\varepsilon_0,\delta_p,\alpha_0)$-deep-hole scenario holds at $x$, since $B_{\alpha_0n}(x)\subseteq B_{2\alpha_0n}(x)$. Moreover, the balls $B_{\lfloor2\alpha_0n\rfloor}(x)$, $x\in\mathcal R$, are pairwise disjoint, hence distinct $x,x'\in\mathcal R$ satisfy
\begin{equation}
d(x,x')\geq2\lfloor2\alpha_0n\rfloor+1\geq4\alpha_0n-1\geq2m+3
\end{equation}
for $n$ large. Consequently,
\begin{equation}\label{eq:Bm1-disjoint}
\text{the sets } B_{m+1}(x),\ x\in\mathcal R,\ \text{are pairwise disjoint.}
\end{equation}
For $x\in\mathcal R$ let $e_x$ and $|\varphi_x\rangle$ be the ground-state energy and the normalized ground state of $H_{B_m(x)}$. As in \cite[p.~1288, item~1]{Manai_2023}, \ref{it:M2} with $t=\delta_p$ and $\alpha=\alpha_0$ gives, for $n$ large and all $x\in\mathcal R$:
\begin{enumerate}[label=(\roman*)]
\item\label{it:s1-i} all eigenvalues of $H_{B_m(x)}$ other than $e_x$ are at least $\Lambda_{\delta_p}$;
\item\label{it:s1-ii} $e_x\leq\langle x|H|x\rangle=U(x)$ (variational principle; $T$ has zero diagonal), and, for a constant $C_0=C_0(\Gamma)$,
\begin{equation}\label{eq:ex-asymptotics}
\left|e_x-U(x)-\frac{\Gamma^2n}{U(x)}\right|\leq C_0n^{-1/4};
\end{equation}
\item\label{it:s1-iii} by \eqref{eq:MW-ball-decay} with $\alpha'=\alpha_0/2$, since $S_m(x)\cap B_{\alpha_0n/2}(x)=\varnothing$ for $n$ large,
\begin{equation}\label{eq:sphere-weight}
\|\Pi_{S_m(x)}\varphi_x\|^2\leq C_{\mathrm b}n\,e^{-c_{\mathrm b}n}.
\end{equation}
\end{enumerate}
Since
\begin{equation}
|U(x)|\geq(\beta_c-\delta_p)n\geq\frac{15}{16}\beta_cn\qquad(x\in\mathcal R),
\end{equation}
item~\ref{it:s1-ii} implies, for $n$ large,
\begin{equation}\label{eq:ex-lower}
e_x\geq U(x)-\frac{16\,\Gamma^2}{15\,\beta_c}-C_0n^{-1/4}\geq U(x)-K_\Gamma+\frac12\qquad(x\in\mathcal R).
\end{equation}
The vectors $|\varphi_x\rangle$, $x\in\mathcal R$, are orthonormal, because their supports $B_m(x)$ are pairwise disjoint.

\subsection{Step 2: block decomposition}\label{sec:step2}
Let
\begin{equation}
\mathcal O:=Q_n\setminus\bigcup_{x\in\mathcal R}B_m(x),
\end{equation}
let $H'$ be the block-diagonal operator
\begin{equation}
H':=\bigoplus_{x\in\mathcal R}H_{B_m(x)}\oplus H_{\mathcal O}
\end{equation}
with respect to
\begin{equation}
\ell^2(Q_n)=\bigoplus_{x\in\mathcal R}\mathrm{Ran}\,\Pi_{B_m(x)}\oplus\mathrm{Ran}\,\Pi_{\mathcal O},
\end{equation}
and let
\begin{equation}
V:=H-H'.
\end{equation} This is the auxiliary Hamiltonian \cite[(4.20)]{Manai_2023} of Manai and Warzel, and $V$ is the operator $-\Gamma A$ of \cite[p.~1287]{Manai_2023}. Since $U$ is diagonal, $V$ consists of the matrix elements of $\Gamma T$ between different blocks:
\begin{equation}
\langle y|V|y'\rangle=\begin{cases}-\Gamma&\text{if $d(y,y')=1$ and $y,y'$ belong to different blocks},\\
0&\text{otherwise}.\end{cases}
\end{equation} By \eqref{eq:Bm1-disjoint}, no edge of the hypercube joins $B_m(x)$ and $B_m(x')$ for $x\neq x'$, and every edge leaving $B_m(x)$ joins $S_m(x)$ to $S_{m+1}(x)\subseteq\mathcal O$. Hence
\begin{equation}
V=-\Gamma\sum_{x\in\mathcal R}J_x,
\end{equation}
where $J_x$ is the adjacency operator of the bipartite graph formed by the edges between $S_m(x)$ and $S_{m+1}(x)$, i.e., the operator $A_\sigma$ of \cite[p.~1287]{Manai_2023}. In place of the bound \cite[(4.21)]{Manai_2023} on $\|A\|$ we use the following elementary estimate, based on the Schur test: a matrix $K=(K_{yy'})$ with nonnegative entries satisfies
\begin{equation}\label{eq:Schur-test}
\|K\|\leq\Bigl(\max_{y}\sum_{y'}K_{yy'}\Bigr)^{1/2}\Bigl(\max_{y'}\sum_{y}K_{yy'}\Bigr)^{1/2},
\end{equation}
i.e., its norm is at most the geometric mean of its largest row sum and its largest column sum. Let
\begin{equation}
K_x:=\Pi_{S_m(x)}J_x\Pi_{S_{m+1}(x)},\qquad\text{so that}\qquad J_x=K_x+K_x^\dag.
\end{equation}
Since $K_x^2=0$, we have
\begin{equation}
J_x^2=K_xK_x^\dag+K_x^\dag K_x,
\end{equation}
where $K_xK_x^\dag$ and $K_x^\dag K_x$ act on the orthogonal subspaces $\mathrm{Ran}\,\Pi_{S_m(x)}$ and $\mathrm{Ran}\,\Pi_{S_{m+1}(x)}$, respectively; hence
\begin{equation}
\|J_x\|^2=\|J_x^2\|=\|K_x\|^2.
\end{equation}
As a matrix with rows indexed by $S_m(x)$ and columns indexed by $S_{m+1}(x)$, $K_x$ has entries
\begin{equation}
\langle y|K_x|y'\rangle=\begin{cases}1&\text{if }d(y,y')=1,\\
0&\text{otherwise}.\end{cases}
\end{equation} Every vertex of $S_m(x)$ has exactly $n-m$ neighbours in $S_{m+1}(x)$ and every vertex of $S_{m+1}(x)$ has exactly $m+1$ neighbours in $S_m(x)$, i.e., every row of $K_x$ sums to $n-m$ and every column of $K_x$ sums to $m+1$. Therefore, \eqref{eq:Schur-test} applied to $K_x$ gives
\begin{equation}\label{eq:Jx-norm}
\|J_x\|\mathrel{=}\|K_x\|\leq\sqrt{(n-m)(m+1)}\leq n\,,
\end{equation}
where the last inequality holds because $n-m\leq n$ and $m+1\leq n$. (Applied directly to the symmetric matrix $J_x$, whose row sums are $n-m$ on $S_m(x)$ and $m+1$ on $S_{m+1}(x)$, the Schur test would only give the weaker bound $\|J_x\|\leq n-m$.)
By \eqref{eq:Bm1-disjoint}, the sets $S_m(x)\cup S_{m+1}(x)\subseteq B_{m+1}(x)$, $x\in\mathcal R$, are pairwise disjoint, and each $J_x$ maps $\mathrm{Ran}\,\Pi_{S_m(x)\cup S_{m+1}(x)}$ into itself and vanishes on its orthogonal complement. Hence, for every $\xi\in\ell^2(Q_n)$, the vectors $J_x\xi=J_x\Pi_{S_m(x)\cup S_{m+1}(x)}\xi$, $x\in\mathcal R$, are pairwise orthogonal, and \eqref{eq:Jx-norm} gives
\begin{equation}\label{eq:V-xi}
\|V\xi\|^2=\Gamma^2\sum_{x\in\mathcal R}\bigl\|J_x\Pi_{S_m(x)\cup S_{m+1}(x)}\xi\bigr\|^2\leq\Gamma^2(n-m)(m+1)\|\xi\|^2.
\end{equation}
Since $u\mapsto u(1-u)$ is increasing on $[0,1/2]$ and $m/n\leq\alpha_0<1/2$,
\begin{equation}
(n-m)m\leq n^2\alpha_0(1-\alpha_0),
\end{equation}
and hence
\begin{equation}\label{eq:nm-bound}
(n-m)(m+1)=(n-m)m+(n-m)\leq n^2\alpha_0(1-\alpha_0)+n.
\end{equation}
By \eqref{eq:V-xi}, \eqref{eq:nm-bound}, the inequality $\sqrt{a+b}\leq\sqrt a+\sqrt b$ for $a,b\geq0$, and the second condition in \eqref{eq:alpha0-choice}, we obtain
\begin{equation}\label{eq:V-norm}
\|V\|\leq\Gamma\sqrt{(n-m)(m+1)}\leq\Gamma n\sqrt{\alpha_0(1-\alpha_0)}+\Gamma\sqrt n\leq\frac{\delta_p n}{16}+\Gamma\sqrt n.
\end{equation}
Moreover, the vector
\begin{equation}
V|\varphi_x\rangle=-\Gamma J_x\Pi_{S_m(x)}|\varphi_x\rangle
\end{equation}
is supported in $S_{m+1}(x)$, and by \eqref{eq:Jx-norm} and \stepref{1}\ref{it:s1-iii}, as in \cite[(4.23)]{Manai_2023},
\begin{equation}\label{eq:V-phi}
\|V\varphi_x\|\leq\Gamma n\,\|\Pi_{S_m(x)}\varphi_x\|\leq\tau_n:=\Gamma n\,(C_{\mathrm b}n)^{1/2}e^{-c_{\mathrm b}n/2}.
\end{equation}
Let
\begin{equation}
P:=\sum_{x\in\mathcal R}|\varphi_x\rangle\langle\varphi_x|,\qquad P^\perp:=I-P;
\end{equation}
this is the projection \cite[(4.22)]{Manai_2023}, except that we include all $x\in\mathcal R$. Then, following \cite[pp.~1288--1289]{Manai_2023}:
\begin{enumerate}[label=(\alph*)]
\item\label{it:s2-a} $PH'P^\perp=0$, since each $|\varphi_x\rangle$ is an eigenvector of $H'$;
\item\label{it:s2-b} $PVP=0$, since $V|\varphi_x\rangle$ is supported in $S_{m+1}(x)\subseteq\mathcal O$, which is disjoint from the supports of all $|\varphi_{x'}\rangle$;
\item\label{it:s2-c} $\|VP\|\leq\tau_n$, since the vectors $V|\varphi_x\rangle$ have pairwise disjoint supports, so that
\begin{equation}
\|VP\xi\|^2=\sum_x|\langle\varphi_x|\xi\rangle|^2\|V\varphi_x\|^2\leq\tau_n^2\|\xi\|^2;
\end{equation}
\item\label{it:s2-d} $P^\perp HP^\perp\geq\Lambda_{3\delta_p/4}$ on $\mathrm{Ran}\,P^\perp$, for $n$ large.
\end{enumerate}
To prove \ref{it:s2-d}, note that $\mathrm{Ran}\,P^\perp$ is the orthogonal sum of $\mathrm{Ran}\,\Pi_{\mathcal O}$ and, for $x\in\mathcal R$, of the orthogonal complement of $|\varphi_x\rangle$ in $\mathrm{Ran}\,\Pi_{B_m(x)}$; all these subspaces are invariant under $H'$. On the latter, $H'\geq\Lambda_{\delta_p}$ by \stepref{1}\ref{it:s1-i}. Every $y\in\mathcal O$ satisfies $y\notin\mathcal R$, i.e., $U(y)>-(\beta_c-\delta_p)n$; hence $H_{\mathcal O}$ is the compression of $\Gamma T+U\1_{\{U\geq-(\beta_c-\delta_p)n\}}$ to $\mathrm{Ran}\,\Pi_{\mathcal O}$, and \ref{it:E2} together with the variational principle gives
\begin{equation}
H_{\mathcal O}\geq-(\beta_c-\delta_p)n-o(n)\geq-(\beta_c-\frac78\delta_p)n
\end{equation}
for $n$ large. Since $\Lambda_{\delta_p}\geq-(\beta_c-\frac78\delta_p)n$, we get
\begin{equation}
P^\perp H'P^\perp\geq-(\beta_c-\frac78\delta_p)n\qquad\text{on }\mathrm{Ran}\,P^\perp,
\end{equation}
and \eqref{eq:V-norm} yields, for $n$ large,
\begin{equation}
P^\perp HP^\perp\geq-\Bigl(\beta_c-\frac{13}{16}\delta_p\Bigr)n-\Gamma\sqrt n\geq-\Bigl(\beta_c-\frac34\delta_p\Bigr)n=\Lambda_{3\delta_p/4}.
\end{equation}

\subsection{Step 3: Schur complement and eigenvalue counting}\label{sec:step3}
This step is the Schur complement method of \cite[pp.~1288--1289]{Manai_2023} (see \cite[(4.24) and (4.25)]{Manai_2023}), where (in our notation) the eigenvalues of $H$ below $\Lambda_{\delta_p/8}$ are compared with the ball energies $e_x$ by a standard perturbation result. To make the role of the thresholds explicit, we formulate this comparison as the counting inequality \eqref{eq:counting}, which we derive from Sylvester's law of inertia. 
Let $E\leq\Lambda_{\delta_p/2}$. By \ref{it:s2-d}, on $\mathrm{Ran}\,P^\perp$ we have
\begin{equation}
D(E):=P^\perp(H-E)P^\perp\geq\Lambda_{3\delta_p/4}-\Lambda_{\delta_p/2}=\delta_pn/4>0;
\end{equation}
hence it is invertible and, for $n$ large,
\begin{equation}
\|D(E)^{-1}\|\leq\frac{4}{\delta_pn}\leq1.
\end{equation} By \ref{it:s2-a} and \ref{it:s2-b}, with respect to $\mathrm{Ran}\,P\oplus\mathrm{Ran}\,P^\perp$,
\begin{equation}
H-E=\begin{pmatrix}P(H'-E)P&PVP^\perp\\ P^\perp VP&D(E)\end{pmatrix}
=X^{\dag}\begin{pmatrix}\mathcal K(E)&0\\0&D(E)\end{pmatrix}X,\qquad
X:=\begin{pmatrix}I&0\\ D(E)^{-1}P^\perp VP&I\end{pmatrix},
\end{equation}
where, on $\mathrm{Ran}\,P$,
\begin{equation}
\mathcal K(E):=P(H'-E)P-W(E),\qquad W(E):=PVP^\perp D(E)^{-1}P^\perp VP.
\end{equation}
Since $X$ is invertible and $D(E)>0$, Sylvester's law of inertia shows that $H-E$ and $\mathcal K(E)$ have the same number of negative eigenvalues. Moreover, by \ref{it:s2-c},
\begin{equation}
0\leq W(E)\leq\|VP\|^2\|D(E)^{-1}\|\leq\nu_n:=\tau_n^2.
\end{equation} On $\mathrm{Ran}\,P$, the operator $P(H'-E)P$ has the eigenvalues $e_x-E$, $x\in\mathcal R$, and
\begin{equation}
P(H'-E-\nu_n)P\leq\mathcal K(E)\leq P(H'-E)P.
\end{equation} Writing $\mathcal N_H(E)$ for the number of eigenvalues of $H$ in $(-\infty,E)$, counted with multiplicity, and
\begin{equation}
\mathcal N_{\mathcal R}(E):=|\{x\in\mathcal R:e_x<E\}|,
\end{equation}
the min-max principle therefore gives
\begin{equation}\label{eq:counting}
\mathcal N_{\mathcal R}(E)\leq\mathcal N_H(E)\leq\mathcal N_{\mathcal R}(E+\nu_n)\qquad\text{for all }E\leq\Lambda_{\delta_p/2}.
\end{equation}
Let $\lambda_1\leq\lambda_2\leq\dots\leq\lambda_{2^n}$ be the eigenvalues of $H$, repeated according to multiplicity, and enumerate $\mathcal R=\{x_1,\dots,x_M\}$ so that $e_k:=e_{x_k}$ is nondecreasing in $k$. Then \eqref{eq:counting} implies:
\begin{enumerate}[label=(3\alph*)]
\item\label{it:s3-a} If $e_k<\Lambda_{\delta_p/2}$, then $\lambda_k\leq e_k$. Indeed, for $0<\eta\leq\Lambda_{\delta_p/2}-e_k$ we have
\begin{equation}
\mathcal N_H(e_k+\eta)\geq\mathcal N_{\mathcal R}(e_k+\eta)\geq k,
\end{equation}
i.e., $\lambda_k<e_k+\eta$.
\item\label{it:s3-b} If $\lambda_j<\Lambda_{\delta_p/2}$, then $j\leq M$ and $\lambda_j\geq e_j-\nu_n$. Indeed, for $0<\eta\leq\Lambda_{\delta_p/2}-\lambda_j$ we have
\begin{equation}
\mathcal N_{\mathcal R}(\lambda_j+\eta+\nu_n)\geq\mathcal N_H(\lambda_j+\eta)\geq j,
\end{equation}
so that $M\geq j$ and $e_j<\lambda_j+\eta+\nu_n$.
\end{enumerate}

\subsection{Step 4: non-resonance}\label{sec:step4}
As in \cite[p.~1289]{Manai_2023}, resonances are excluded by means of the spectral-averaging estimate \ref{it:M3}, through \ref{it:E3}; the first case below, which relies on the energy asymptotics \eqref{eq:MW-ball-energy} of \cite[Lemma~4.3]{Manai_2023}, treats the configurations outside $\mathcal R_{n,\hat\delta}$. In this step our argument departs from the paragraph ``Eigenvectors'' on \cite[p.~1289]{Manai_2023}, see \autoref{it:mw-nonres} of \autoref{rem:MW-threshold}: the separation \eqref{eq:nonres} below is established with respect to all $x'\in\mathcal R$, i.e., with respect to the ground-state energies of all the balls whose ground states belong to the range of $P$, and not only with respect to the configurations in $\mathcal R_{n,\delta}$.
Let $x\in\mathcal R_{n,\delta}$. Since $\delta<\hat\delta<\delta_p$, we have $x\in\mathcal R_{n,\hat\delta}\subseteq\mathcal R$ and
\begin{equation}
e_x\leq U(x)\leq\Lambda_\delta=\Lambda_{\hat\delta}-\delta n.
\end{equation}
We claim that
\begin{equation}\label{eq:nonres}
|e_{x'}-e_x|>r_n\qquad\text{for all }x'\in\mathcal R\setminus\{x\}.
\end{equation}
If $x'\in\mathcal R\setminus\mathcal R_{n,\hat\delta}$, then $U(x')>\Lambda_{\hat\delta}$, and \eqref{eq:ex-lower} gives
\begin{equation}
e_{x'}-e_x>\Lambda_{\hat\delta}-K_\Gamma-\Lambda_\delta=\delta n-K_\Gamma>r_n.
\end{equation}
If $x'\in\mathcal R_{n,\hat\delta}$ and $e_{x'}\geq\Lambda_{\hat\delta}$, then $e_{x'}-e_x\geq\delta n>r_n$. If $x'\in\mathcal R_{n,\hat\delta}$ and $e_{x'}<\Lambda_{\hat\delta}$, then $e_x$ and $e_{x'}$ are eigenvalues in $(-\infty,\Lambda_{\hat\delta})$ of $H_{B_m(x)}$ and $H_{B_m(x')}$, respectively, and \ref{it:E3} gives $|e_x-e_{x'}|>r_n$. Note also that $r_n\geq2\nu_n$ for $n$ large, since
\begin{equation}
\nu_n=\Gamma^2C_{\mathrm b}n^3e^{-c_{\mathrm b}n}.
\end{equation}

\subsection{\texorpdfstring{Step 5: items \ref{it:loc-label}, \ref{it:loc-energy} and \ref{it:loc-threshold}}{Step 5: items 1, 2 and 4}}\label{sec:step5}
This step corresponds to the paragraph ``Spectrum'' of the proof of \cite[Theorem~1.5]{Manai_2023} (see \cite[p.~1289]{Manai_2023}); the two thresholds in items~\ref{it:loc-label} and~\ref{it:loc-threshold} replace the one-to-one correspondence of \cite[Theorem~1.5, item~1]{Manai_2023}, see \autoref{it:mw-statement} of \autoref{rem:MW-threshold}. 
Let $x\in\mathcal R_{n,\delta}$. By \eqref{eq:nonres}, $e_x$ differs from all other $e_{x'}$, so $x=x_k$ for an index $k=k(x)$ that does not depend on the choice of the enumeration, and
\begin{equation}
e_{k-1}\leq e_k-r_n\quad\text{if }k>1,\qquad e_{k+1}\geq e_k+r_n\quad\text{if }k<M.
\end{equation} Since $e_k\leq\Lambda_\delta<\Lambda_{\delta_p/2}$, \ref{it:s3-a} and \ref{it:s3-b} give
\begin{equation}\label{eq:lambda-k}
e_k-\nu_n\leq\lambda_k\leq e_k\leq U(x).
\end{equation}
The eigenvalue $\lambda_k$ is simple: if $k>1$, then \ref{it:s3-a} gives
\begin{equation}
\lambda_{k-1}\leq e_{k-1}\leq e_k-r_n<e_k-\nu_n\leq\lambda_k;
\end{equation}
and $\lambda_{k+1}>e_k\geq\lambda_k$, because either $\lambda_{k+1}\geq\Lambda_{\delta_p/2}>e_k$, or \ref{it:s3-b} applies and gives
\begin{equation}
\lambda_{k+1}\geq e_{k+1}-\nu_n\geq e_k+r_n-\nu_n>e_k.
\end{equation}
We set
\begin{equation}
E(x):=\lambda_{k(x)}
\end{equation}
and let $|\psi_x\rangle$ be a normalized eigenvector. Distinct configurations have distinct indices and hence distinct eigenvalues, and \eqref{eq:lambda-k} gives \eqref{eq:E-below-U}; this proves \autoref{it:loc-label}. Item~\ref{it:loc-energy} follows from $|E(x)-e_x|\leq\nu_n$ and \stepref{1}\ref{it:s1-ii}.

For \autoref{it:loc-threshold}, let $\lambda_j$ be an eigenvalue of $H$ that is not of the form $E(x)$ with $x\in\mathcal R_{n,\delta}$. If $\lambda_j\geq\Lambda_{\delta_p/2}$, then $\lambda_j>\Lambda_\delta$. Otherwise, \ref{it:s3-b} gives $j\leq M$ and $\lambda_j\geq e_j-\nu_n$. Moreover, $x_j\notin\mathcal R_{n,\delta}$, for otherwise $j=k(x_j)$ and $\lambda_j=E(x_j)$. Hence $U(x_j)>\Lambda_\delta$, and \eqref{eq:ex-lower} yields
\begin{equation}
\lambda_j\geq U(x_j)-K_\Gamma+\frac12-\nu_n>\Lambda_\delta-K_\Gamma.
\end{equation}

\subsection{\texorpdfstring{Step 6: item \ref{it:loc-decay}}{Step 6: item 3}}\label{sec:step6}
This step is the eigenvector estimate of \cite[pp.~1289--1290]{Manai_2023}, based on \cite[(4.24) and (4.27)]{Manai_2023}. 
Let $x=x_k\in\mathcal R_{n,\delta}$, $\lambda:=E(x)=\lambda_k$ and $|\psi\rangle:=|\psi_x\rangle$. Since $\lambda\leq\Lambda_\delta<\Lambda_{\delta_p/2}$, \stepref{3} applies with $E=\lambda$. Applying $P^\perp$ and $P$ to $(H-\lambda)\psi=0$ and using \ref{it:s2-a} and \ref{it:s2-b}, we get
\begin{equation}
P^\perp\psi=-D(\lambda)^{-1}P^\perp VP\psi,\qquad P(H'-\lambda)P\psi=-PVP^\perp\psi=W(\lambda)P\psi.
\end{equation}
Hence
\begin{equation}
\|P^\perp\psi\|\leq\tau_n,\qquad\|P(H'-\lambda)P\psi\|\leq\nu_n.
\end{equation}
Writing
\begin{equation}
P\psi=\sum_jc_j\varphi_{x_j},
\end{equation}
the latter bound reads
\begin{equation}
\sum_j(e_j-\lambda)^2|c_j|^2\leq\nu_n^2.
\end{equation}
For $j\neq k$, \eqref{eq:nonres} and $|\lambda-e_k|\leq\nu_n$ give
\begin{equation}
|e_j-\lambda|\geq r_n-\nu_n\geq\frac{r_n}{2},
\end{equation}
so that
\begin{equation}
\sum_{j\neq k}|c_j|^2\leq\frac{4\nu_n^2}{r_n^2}.
\end{equation} Choose the phase of $\psi$ such that $c_k\geq0$. Since
\begin{equation}
\psi-c_k\varphi_x=P^\perp\psi+\sum_{j\neq k}c_j\varphi_{x_j}
\end{equation}
is orthogonal to $\varphi_x$,
\begin{equation}
\|\psi-c_k\varphi_x\|^2\leq\vartheta_n^2:=\tau_n^2+\frac{4\nu_n^2}{r_n^2},\qquad c_k^2=1-\|\psi-c_k\varphi_x\|^2\geq1-\vartheta_n^2,
\end{equation}
hence
\begin{equation}
0\leq1-c_k\leq1-c_k^2\leq\vartheta_n^2,\qquad\|\psi-\varphi_x\|\leq\vartheta_n+\vartheta_n^2\leq2\vartheta_n.
\end{equation}
As
\begin{equation}
\vartheta_n^2=\Gamma^2C_{\mathrm b}n^3e^{-c_{\mathrm b}n}+4\Gamma^4C_{\mathrm b}^2n^6e^{-c_{\mathrm b}n},
\end{equation}
we have $\vartheta_n\leq e^{-c_{\mathrm b}n/3}$ for $n$ large.

Finally, as in \cite[p.~1290]{Manai_2023}, the localization of $\psi_x$ is inherited from the decay of the ball ground state established in \cite[Lemma~4.3]{Manai_2023}. Fix $\alpha\in(0,1)$ and set $\alpha':=\min\{\alpha,\alpha_0\}$. By \eqref{eq:MW-ball-decay} with $t=\delta_p$ and $\alpha_0$ in place of $\alpha$,
\begin{equation}
\sum_{y\notin B_{\alpha n}(x)}|\langle y|\varphi_x\rangle|^2\leq\sum_{y\notin B_{\alpha'n}(x)}|\langle y|\varphi_x\rangle|^2\leq C_{\mathrm b}n\,e^{-c(\alpha_0,\alpha')n}.
\end{equation}
By the triangle inequality in $\ell^2(Q_n\setminus B_{\alpha n}(x))$,
\begin{equation}
\sum_{y\notin B_{\alpha n}(x)}|\langle y|\psi_x\rangle|^2\leq\Bigl(2e^{-c_{\mathrm b}n/3}+\bigl(C_{\mathrm b}n\,e^{-c(\alpha_0,\alpha')n}\bigr)^{1/2}\Bigr)^2\leq e^{-c_\alpha n}
\end{equation}
for any fixed $c_\alpha\in(0,\min\{2c_{\mathrm b}/3,c(\alpha_0,\alpha')\})$ and all $n$ large. The left-hand side does not depend on the phase of $|\psi_x\rangle$. This proves \autoref{it:loc-decay} and completes the proof of \autoref{thm:Manai}.

\section*{Use of AI-assisted tools}
An AI-assisted language model was used during manuscript preparation for literature discovery and consistency checks.
Responsibility for the final text and for every mathematical claim remains with the authors.

\section*{Acknowledgements}
GDP has been supported by the UNA EUROPA SeedFunding project QUANTUMUnaE (CUP J37G25000380006).
GDP and EM are members of the ``Gruppo Nazionale per la Fisica Matematica (GNFM)'' of the ``Istituto Nazionale di Alta Matematica ``Francesco Severi'' (INdAM)''.

\bibliography{biblio}

@misc{gamarnik2024slowmixingquantumgibbs,
      title={Slow Mixing of Quantum Gibbs Samplers}, 
      author={David Gamarnik and Bobak T. Kiani and Alexander Zlokapa},
      year={2024},
      eprint={2411.04300},
      archivePrefix={arXiv},
      primaryClass={quant-ph},
      url={https://arxiv.org/abs/2411.04300},
      note={Version~2}
}

@article{Manai_2023,
   title={Spectral Analysis of the Quantum Random Energy Model},
   volume={402},
   ISSN={1432-0916},
   url={http://dx.doi.org/10.1007/s00220-023-04743-4},
   DOI={10.1007/s00220-023-04743-4},
   number={2},
   journal={Communications in Mathematical Physics},
   publisher={Springer Science and Business Media LLC},
   author={Manai, Chokri and Warzel, Simone},
   year={2023},
   month=jun, pages={1259--1306} }

@article{Manai_2020,
   title={Phase Diagram of the Quantum Random Energy Model},
   volume={180},
   ISSN={1572-9613},
   url={http://dx.doi.org/10.1007/s10955-020-02492-5},
   DOI={10.1007/s10955-020-02492-5},
   number={1--6},
   journal={Journal of Statistical Physics},
   publisher={Springer Science and Business Media LLC},
   author={Manai, Chokri and Warzel, Simone},
   year={2020},
   month=jan, pages={654--664} }

@article{Derrida81,
  title={Random-energy model: An exactly solvable model of disordered systems},
  author={Derrida, Bernard},
  journal={Physical Review B},
  volume={24},
  number={5},
  pages={2613--2626},
  year={1981},
  publisher={APS},
  doi={10.1103/PhysRevB.24.2613}
}

@article{Ruelle87,
author = {David Ruelle},
title = {{A mathematical reformulation of Derrida's REM and GREM}},
volume = {108},
journal = {Communications in Mathematical Physics},
number = {2},
publisher = {Springer},
pages = {225 -- 239},
year = {1987},
doi = {10.1007/BF01210613}
}

@article{Picco,
author = {Enzo Olivieri and Pierre Picco},
title = {{On the existence of thermodynamics for the random energy model}},
volume = {96},
journal = {Communications in Mathematical Physics},
number = {1},
publisher = {Springer},
pages = {125 -- 144},
year = {1984},
doi = {10.1007/BF01217351}
}

@book{BovierBook,
author = {Bovier, Anton},
address = {Cambridge},
booktitle = {Statistical mechanics of disordered systems a mathematical perspective},
isbn = {9780521849913},
language = {eng},
publisher = {Cambridge University Press},
series = {Cambridge series in statistical and probabilistic mathematics},
title = {Statistical mechanics of disordered systems : a mathematical perspective},
year = {2006},
}

@article{PY97,
author = {Jim Pitman and Marc Yor},
title = {{The two-parameter Poisson-Dirichlet distribution derived from a stable subordinator}},
volume = {25},
journal = {The Annals of Probability},
number = {2},
publisher = {Institute of Mathematical Statistics},
pages = {855 -- 900},
year = {1997},
doi = {10.1214/aop/1024404422},
URL = {https://doi.org/10.1214/aop/1024404422}
}

@book{davies1976quantum,
  title={Quantum theory of open systems},
  author={Davies, Edward Brian},
  year={1976},
  publisher={Academic Press London}
}

@article{Goldschmidt1990,
  author = {Goldschmidt, Yadin Y.},
  title = {Solvable model of the quantum spin glass in a transverse field},
  journal = {Physical Review B},
  volume = {41},
  number = {7},
  pages = {4858--4861},
  year = {1990},
  doi = {10.1103/PhysRevB.41.4858}
}

@article{JorgKrzakalaKurchanMaggs2008,
  author = {J{\"o}rg, Thomas and Krzakala, Florent and Kurchan, Jorge and Maggs, A. C.},
  title = {Simple Glass Models and Their Quantum Annealing},
  journal = {Physical Review Letters},
  volume = {101},
  pages = {147204},
  year = {2008},
  doi = {10.1103/PhysRevLett.101.147204}
}

@article{AdameWarzel2015,
  author = {Adame, J. and Warzel, S.},
  title = {Exponential vanishing of the ground-state gap of the quantum random energy model via adiabatic quantum computing},
  journal = {Journal of Mathematical Physics},
  volume = {56},
  number = {11},
  pages = {113301},
  year = {2015},
  doi = {10.1063/1.4934723}
}

@article{LaumannPalScardicchio2014,
  author = {Laumann, C. R. and Pal, A. and Scardicchio, A.},
  title = {Many-body mobility edge in a mean-field quantum spin glass},
  journal = {Physical Review Letters},
  volume = {113},
  pages = {200405},
  year = {2014},
  doi = {10.1103/PhysRevLett.113.200405}
}

@article{BaldwinLaumannPalScardicchio2016,
  author = {Baldwin, C. L. and Laumann, C. R. and Pal, A. and Scardicchio, A.},
  title = {The many-body localized phase of the quantum random energy model},
  journal = {Physical Review B},
  volume = {93},
  pages = {024202},
  year = {2016},
  doi = {10.1103/PhysRevB.93.024202}
}

@article{FaoroFeigelmanIoffe2019,
  author = {Faoro, Lara and Feigel'man, Mikhail V. and Ioffe, Lev B.},
  title = {Non-ergodic extended phase of the Quantum Random Energy model},
  journal = {Annals of Physics},
  volume = {409},
  pages = {167916},
  year = {2019},
  doi = {10.1016/j.aop.2019.167916}
}

@article{BiroliFacoettiSchiroTarziaVivo2021,
  author = {Biroli, Giulio and Facoetti, Davide and Schir{\`o}, Marco and Tarzia, Marco and Vivo, Pierpaolo},
  title = {Out-of-equilibrium phase diagram of the quantum random energy model},
  journal = {Physical Review B},
  volume = {103},
  pages = {014204},
  year = {2021},
  doi = {10.1103/PhysRevB.103.014204}
}

@article{BalducciEtAl2025,
  author = {Balducci, Federico and Bracci Testasecca, Giacomo and Niedda, Jacopo and Scardicchio, Antonello and Vanoni, Carlo},
  title = {Scaling analysis and renormalization group on the mobility edge in the quantum random energy model},
  journal = {Physical Review B},
  volume = {111},
  pages = {214206},
  year = {2025},
  doi = {10.1103/64m5-m9ty}
}

@misc{KouraichManaiWarzel2025,
  author = {Kouraich, Anouar and Manai, Chokri and Warzel, Simone},
  title = {The Quantum Random Energy Model is the Limit of Quantum $p$-Spin Glasses},
  year = {2025},
  eprint = {2505.02458},
  archivePrefix = {arXiv},
  primaryClass = {math-ph}
}

@article{FontesIsopiKohayakawaPicco1998,
  author = {Fontes, L. R. and Isopi, M. and Kohayakawa, Y. and Picco, P.},
  title = {The Spectral Gap of the REM under Metropolis Dynamics},
  journal = {The Annals of Applied Probability},
  volume = {8},
  number = {3},
  pages = {917--943},
  year = {1998},
  doi = {10.1214/aoap/1028903457}
}

@article{BenArousBovierGayrard2002,
  author = {Ben Arous, G{\'e}rard and Bovier, Anton and Gayrard, V{\'e}ronique},
  title = {Aging in the Random Energy Model},
  journal = {Physical Review Letters},
  volume = {88},
  pages = {087201},
  year = {2002},
  doi = {10.1103/PhysRevLett.88.087201}
}

@article{CernyWassmer2017,
  author = {{\v C}ern{\'y}, Ji{\v r}{\'i} and Wassmer, Tobias},
  title = {Aging of the Metropolis dynamics on the Random Energy Model},
  journal = {Probability Theory and Related Fields},
  volume = {167},
  number = {1--2},
  pages = {253--303},
  year = {2017},
  doi = {10.1007/s00440-015-0681-1}
}

@article{KastoryanoBrandao2016,
  author = {Kastoryano, Michael J. and Brand{\~a}o, Fernando G. S. L.},
  title = {Quantum Gibbs Samplers: The Commuting Case},
  journal = {Communications in Mathematical Physics},
  volume = {344},
  pages = {915--957},
  year = {2016},
  doi = {10.1007/s00220-016-2641-8}
}

@misc{ChenKastoryanoGilyen2023,
  author = {Chen, Chi-Fang and Kastoryano, Michael J. and Gily{\'e}n, Andr{\'a}s},
  title = {An efficient and exact noncommutative quantum Gibbs sampler},
  year = {2023},
  eprint = {2311.09207},
  archivePrefix = {arXiv},
  primaryClass = {quant-ph}
}

@article{DingLiLin2025,
  author = {Ding, Zhiyan and Li, Bowen and Lin, Lin},
  title = {Efficient Quantum Gibbs Samplers with Kubo--Martin--Schwinger Detailed Balance Condition},
  journal = {Communications in Mathematical Physics},
  volume = {406},
  pages = {67},
  year = {2025},
  doi = {10.1007/s00220-025-05235-3}
}

@misc{GilyenChenDoriguelloKastoryano2024,
  author = {Gily{\'e}n, Andr{\'a}s and Chen, Chi-Fang and Doriguello, Joao F. and Kastoryano, Michael J.},
  title = {Quantum generalizations of Glauber and Metropolis dynamics},
  year = {2024},
  eprint = {2405.20322},
  archivePrefix = {arXiv},
  primaryClass = {quant-ph}
}

@article{RakovszkyPlackeBreuckmannKhemani2026,
  author = {Rakovszky, Tibor and Placke, Benedikt and Breuckmann, Nikolas P. and Khemani, Vedika},
  title = {Bottlenecks in Quantum Channels and Finite Temperature Phases of Matter},
  journal = {Physical Review Letters},
  volume = {137},
  pages = {050402},
  year = {2026},
  doi = {10.1103/y9z2-52rl}
}

@article{ScandiAlhambra2026,
  author = {Scandi, Matteo and Alhambra, {\'A}lvaro M.},
  title = {Thermalization in Open Many-Body Systems and KMS Detailed Balance},
  journal = {Physical Review X},
  volume = {16},
  pages = {011040},
  year = {2026},
  doi = {10.1103/sfp3-3sqf}
}

@article{Derrida80,
  author = {Derrida, Bernard},
  title = {Random-Energy Model: Limit of a Family of Disordered Models},
  journal = {Physical Review Letters},
  volume = {45},
  number = {2},
  pages = {79--82},
  year = {1980},
  doi = {10.1103/PhysRevLett.45.79}
}

@article{ManaiWarzel2021RMP,
  author = {Manai, Chokri and Warzel, Simone},
  title = {The quantum random energy model as a limit of $p$-spin interactions},
  journal = {Reviews in Mathematical Physics},
  volume = {33},
  number = {1},
  pages = {2060013},
  year = {2021},
  doi = {10.1142/S0129055X20600132}
}

@article{ManaiWarzelGREM2022,
  author = {Manai, Chokri and Warzel, Simone},
  title = {Generalized random energy models in a transversal magnetic field: free energy and phase diagrams},
  journal = {Probability and Mathematical Physics},
  volume = {3},
  number = {2},
  pages = {215--245},
  year = {2022},
  doi = {10.2140/pmp.2022.3.215}
}

@article{ManaiWarzelAT2022,
  author = {Manai, Chokri and Warzel, Simone},
  title = {The de {A}lmeida--{T}houless Line in Hierarchical Quantum Spin Glasses},
  journal = {Journal of Statistical Physics},
  volume = {186},
  pages = {14},
  year = {2022},
  doi = {10.1007/s10955-021-02860-9}
}

@article{ManaiWarzelParisi2025,
  author = {Manai, Chokri and Warzel, Simone},
  title = {A {P}arisi formula for quantum spin glasses},
  journal = {Electronic Journal of Probability},
  volume = {30},
  pages = {1--39},
  year = {2025},
  doi = {10.1214/25-EJP1359}
}

@article{LeschkeManaiRuderWarzel2021,
  author = {Leschke, Hajo and Manai, Chokri and Ruder, Rainer and Warzel, Simone},
  title = {Existence of Replica-Symmetry Breaking in Quantum Glasses},
  journal = {Physical Review Letters},
  volume = {127},
  pages = {207204},
  year = {2021},
  doi = {10.1103/PhysRevLett.127.207204}
}

@article{Crawford2007,
  author = {Crawford, Nicholas},
  title = {Thermodynamics and Universality for Mean Field Quantum Spin Glasses},
  journal = {Communications in Mathematical Physics},
  volume = {274},
  number = {3},
  pages = {821--839},
  year = {2007},
  doi = {10.1007/s00220-007-0263-x}
}

@article{AdhikariBrennecke2020,
  author = {Adhikari, Arka and Brennecke, Christian},
  title = {Free energy of the quantum {S}herrington--{K}irkpatrick spin-glass model with transverse field},
  journal = {Journal of Mathematical Physics},
  volume = {61},
  number = {8},
  pages = {083302},
  year = {2020},
  doi = {10.1063/5.0009291}
}

@article{GarrahanManaiWarzel2023,
  author = {Garrahan, Juan P. and Manai, Chokri and Warzel, Simone},
  title = {Trajectory phase transitions in non-interacting systems: all-to-all dynamics and the random energy model},
  journal = {Philosophical Transactions of the Royal Society A},
  volume = {381},
  number = {2241},
  pages = {20210415},
  year = {2023},
  doi = {10.1098/rsta.2021.0415}
}

@misc{ManaiWarzel2025dyn,
  author = {Manai, Chokri and Warzel, Simone},
  title = {Dynamical Phase Diagram of the {REM} under independent spin-flips},
  year = {2025},
  eprint = {2506.15353},
  archivePrefix = {arXiv},
  primaryClass = {math-ph}
}

@misc{KouraichWarzel2026,
  author = {Kouraich, Anouar and Warzel, Simone},
  title = {Lower bound on the mixing time of $p$-spin glasses},
  year = {2026},
  eprint = {2605.14681},
  archivePrefix = {arXiv},
  primaryClass = {math.PR}
}

@article{JorgEtAl2010,
  author = {J{\"o}rg, Thomas and Krzakala, Florent and Kurchan, Jorge and Maggs, A. C. and Pujos, J.},
  title = {Energy gaps in quantum first-order mean-field-like transitions: The problems that quantum annealing cannot solve},
  journal = {EPL (Europhysics Letters)},
  volume = {89},
  number = {4},
  pages = {40004},
  year = {2010},
  doi = {10.1209/0295-5075/89/40004}
}

@article{BapstEtAl2013,
  author = {Bapst, Victor and Foini, Laura and Krzakala, Florent and Semerjian, Guilhem and Zamponi, Francesco},
  title = {The quantum adiabatic algorithm applied to random optimization problems: The quantum spin glass perspective},
  journal = {Physics Reports},
  volume = {523},
  number = {3},
  pages = {127--205},
  year = {2013},
  doi = {10.1016/j.physrep.2012.10.002}
}

@article{BaldwinLaumann2018,
  author = {Baldwin, C. L. and Laumann, C. R.},
  title = {Quantum algorithm for energy matching in hard optimization problems},
  journal = {Physical Review B},
  volume = {97},
  pages = {224201},
  year = {2018},
  doi = {10.1103/PhysRevB.97.224201}
}

@misc{ParoliniMossi2020,
  author = {Parolini, Tommaso and Mossi, Gianni},
  title = {Multifractal Dynamics of the {QREM}},
  year = {2020},
  eprint = {2007.00315},
  archivePrefix = {arXiv},
  primaryClass = {cond-mat.dis-nn}
}

@article{ScoquartGornyiMirlin2024,
  author = {Scoquart, Thibault and Gornyi, Igor V. and Mirlin, Alexander D.},
  title = {Role of {F}ock-space correlations in many-body localization},
  journal = {Physical Review B},
  volume = {109},
  pages = {214203},
  year = {2024},
  doi = {10.1103/PhysRevB.109.214203}
}

@article{ScoquartGornyiMirlin2025,
  author = {Scoquart, Thibault and Gornyi, Igor V. and Mirlin, Alexander D.},
  title = {Scaling of many-body localization transitions: Quantum dynamics in {F}ock space and real space},
  journal = {Physical Review B},
  volume = {112},
  pages = {064203},
  year = {2025},
  doi = {10.1103/qcvd-n2yk}
}

@article{DobrosavljevicThirumalai1990,
  author = {Dobrosavljevi{\'c}, V. and Thirumalai, D.},
  title = {$1/p$ expansion for a $p$-spin interaction spin-glass model in a transverse field},
  journal = {Journal of Physics A: Mathematical and General},
  volume = {23},
  number = {15},
  pages = {L767--L774},
  year = {1990},
  doi = {10.1088/0305-4470/23/15/013}
}

@article{NieuwenhuizenRitort1998,
  author = {Nieuwenhuizen, Theo M. and Ritort, Felix},
  title = {Quantum phase transition in spin glasses with multi-spin interactions},
  journal = {Physica A},
  volume = {250},
  pages = {8--45},
  year = {1998},
  doi = {10.1016/S0378-4371(97)00546-3}
}

@article{ObuchiNishimoriSherrington2007,
  author = {Obuchi, Tomoyuki and Nishimori, Hidetoshi and Sherrington, David},
  title = {Phase Diagram of the $p$-Spin-Interacting Spin Glass with Ferromagnetic Bias and a Transverse Field in the Infinite-$p$ Limit},
  journal = {Journal of the Physical Society of Japan},
  volume = {76},
  number = {5},
  pages = {054002},
  year = {2007},
  doi = {10.1143/JPSJ.76.054002}
}

@article{BenArousBovierGayrard2003a,
  author = {Ben Arous, G{\'e}rard and Bovier, Anton and Gayrard, V{\'e}ronique},
  title = {Glauber dynamics of the random energy model. {I}. {M}etastable motion on the extreme states},
  journal = {Communications in Mathematical Physics},
  volume = {235},
  number = {3},
  pages = {379--425},
  year = {2003},
  doi = {10.1007/s00220-003-0798-4}
}

@article{BenArousBovierGayrard2003b,
  author = {Ben Arous, G{\'e}rard and Bovier, Anton and Gayrard, V{\'e}ronique},
  title = {Glauber dynamics of the random energy model. {II}. {A}ging below the critical temperature},
  journal = {Communications in Mathematical Physics},
  volume = {236},
  number = {1},
  pages = {1--54},
  year = {2003},
  doi = {10.1007/s00220-003-0799-3}
}

@article{BenArousBovierCerny2008,
  author = {Ben Arous, G{\'e}rard and Bovier, Anton and {\v C}ern{\'y}, Ji{\v r}{\'i}},
  title = {Universality of the {REM} for Dynamics of Mean-Field Spin Glasses},
  journal = {Communications in Mathematical Physics},
  volume = {282},
  number = {3},
  pages = {663--695},
  year = {2008},
  doi = {10.1007/s00220-008-0565-7}
}

@article{Gayrard2019,
  author = {Gayrard, V{\'e}ronique},
  title = {Aging in {M}etropolis dynamics of the {REM}: a proof},
  journal = {Probability Theory and Related Fields},
  volume = {174},
  number = {1--2},
  pages = {501--551},
  year = {2019},
  doi = {10.1007/s00440-018-0873-6}
}

@incollection{BenArousCerny2006,
  author = {Ben Arous, G{\'e}rard and {\v C}ern{\'y}, Ji{\v r}{\'i}},
  title = {Dynamics of trap models},
  booktitle = {Mathematical Statistical Physics (Les Houches, Session LXXXIII, 2005)},
  publisher = {Elsevier},
  address = {Amsterdam},
  pages = {331--394},
  year = {2006},
  doi = {10.1016/S0924-8099(06)80045-4}
}

@article{BenArousJagannath2018,
  author = {Ben Arous, G{\'e}rard and Jagannath, Aukosh},
  title = {Spectral Gap Estimates in Mean Field Spin Glasses},
  journal = {Communications in Mathematical Physics},
  volume = {361},
  number = {1},
  pages = {1--52},
  year = {2018},
  doi = {10.1007/s00220-018-3152-6}
}

@article{GheissariJagannath2019,
  author = {Gheissari, Reza and Jagannath, Aukosh},
  title = {On the spectral gap of spherical spin glass dynamics},
  journal = {Annales de l'Institut Henri Poincar{\'e} Probabilit{\'e}s et Statistiques},
  volume = {55},
  number = {2},
  pages = {756--776},
  year = {2019},
  doi = {10.1214/18-AIHP897}
}

@book{LevinPeresWilmer2017,
  author = {Levin, David A. and Peres, Yuval},
  title = {Markov Chains and Mixing Times},
  edition = {Second},
  note = {With contributions by E. L. Wilmer},
  publisher = {American Mathematical Society},
  address = {Providence, RI},
  year = {2017},
  doi = {10.1090/mbk/107}
}

@article{SinclairJerrum1989,
  author = {Sinclair, Alistair and Jerrum, Mark},
  title = {Approximate counting, uniform generation and rapidly mixing {M}arkov chains},
  journal = {Information and Computation},
  volume = {82},
  number = {1},
  pages = {93--133},
  year = {1989},
  doi = {10.1016/0890-5401(89)90067-9}
}

@article{LawlerSokal1988,
  author = {Lawler, Gregory F. and Sokal, Alan D.},
  title = {Bounds on the {$L^2$} spectrum for {M}arkov chains and {M}arkov processes: a generalization of {C}heeger's inequality},
  journal = {Transactions of the American Mathematical Society},
  volume = {309},
  number = {2},
  pages = {557--580},
  year = {1988},
  doi = {10.1090/S0002-9947-1988-0930082-9}
}

@article{BovierKurkovaLowe2002,
  author = {Bovier, Anton and Kurkova, Irina and L{\"o}we, Matthias},
  title = {Fluctuations of the free energy in the {REM} and the $p$-spin {SK} models},
  journal = {The Annals of Probability},
  volume = {30},
  number = {2},
  pages = {605--651},
  year = {2002},
  doi = {10.1214/aop/1023481004}
}

@article{Davies1974,
  author = {Davies, E. B.},
  title = {Markovian master equations},
  journal = {Communications in Mathematical Physics},
  volume = {39},
  number = {2},
  pages = {91--110},
  year = {1974},
  doi = {10.1007/BF01608389}
}

@article{Lindblad1976,
  author = {Lindblad, G.},
  title = {On the generators of quantum dynamical semigroups},
  journal = {Communications in Mathematical Physics},
  volume = {48},
  number = {2},
  pages = {119--130},
  year = {1976},
  doi = {10.1007/BF01608499}
}

@article{GoriniKossakowskiSudarshan1976,
  author = {Gorini, Vittorio and Kossakowski, Andrzej and Sudarshan, E. C. G.},
  title = {Completely positive dynamical semigroups of {$N$}-level systems},
  journal = {Journal of Mathematical Physics},
  volume = {17},
  number = {5},
  pages = {821--825},
  year = {1976},
  doi = {10.1063/1.522979}
}

@article{KossakowskiFrigerioGoriniVerri1977,
  author = {Kossakowski, A. and Frigerio, A. and Gorini, V. and Verri, M.},
  title = {Quantum detailed balance and {KMS} condition},
  journal = {Communications in Mathematical Physics},
  volume = {57},
  number = {2},
  pages = {97--110},
  year = {1977},
  doi = {10.1007/BF01625769}
}

@article{Alicki1976,
  author = {Alicki, Robert},
  title = {On the detailed balance condition for non-{H}amiltonian systems},
  journal = {Reports on Mathematical Physics},
  volume = {10},
  number = {2},
  pages = {249--258},
  year = {1976},
  doi = {10.1016/0034-4877(76)90046-X}
}

@article{CarlenMaas2017,
  author = {Carlen, Eric A. and Maas, Jan},
  title = {Gradient flow and entropy inequalities for quantum {M}arkov semigroups with detailed balance},
  journal = {Journal of Functional Analysis},
  volume = {273},
  number = {5},
  pages = {1810--1869},
  year = {2017},
  doi = {10.1016/j.jfa.2017.05.003}
}

@article{FagnolaUmanita2007,
  author = {Fagnola, Franco and Umanit{\`a}, Veronica},
  title = {Generators of detailed balance quantum {M}arkov semigroups},
  journal = {Infinite Dimensional Analysis, Quantum Probability and Related Topics},
  volume = {10},
  number = {3},
  pages = {335--363},
  year = {2007},
  doi = {10.1142/S0219025707002762}
}

@article{TemmeEtAl2010,
  author = {Temme, K. and Kastoryano, M. J. and Ruskai, M. B. and Wolf, M. M. and Verstraete, F.},
  title = {The $\chi^2$-divergence and mixing times of quantum {M}arkov processes},
  journal = {Journal of Mathematical Physics},
  volume = {51},
  number = {12},
  pages = {122201},
  year = {2010},
  doi = {10.1063/1.3511335}
}

@article{KastoryanoTemme2013,
  author = {Kastoryano, Michael J. and Temme, Kristan},
  title = {Quantum logarithmic {S}obolev inequalities and rapid mixing},
  journal = {Journal of Mathematical Physics},
  volume = {54},
  number = {5},
  pages = {052202},
  year = {2013},
  doi = {10.1063/1.4804995}
}

@article{Spohn1980,
  author = {Spohn, Herbert},
  title = {Kinetic equations from {H}amiltonian dynamics: {M}arkovian limits},
  journal = {Reviews of Modern Physics},
  volume = {52},
  number = {3},
  pages = {569--615},
  year = {1980},
  doi = {10.1103/RevModPhys.52.569}
}

@book{AlickiLendi2007,
  author = {Alicki, Robert and Lendi, Karl},
  title = {Quantum Dynamical Semigroups and Applications},
  series = {Lecture Notes in Physics},
  volume = {717},
  publisher = {Springer},
  address = {Berlin},
  year = {2007},
  note = {First published as Lecture Notes in Physics 286 (1987)}
}

@book{Bhatia1997,
  author = {Bhatia, Rajendra},
  title = {Matrix Analysis},
  series = {Graduate Texts in Mathematics},
  volume = {169},
  publisher = {Springer},
  address = {New York},
  year = {1997},
  doi = {10.1007/978-1-4612-0653-8}
}

@article{AlickiHorodecki2010,
  author = {Alicki, R. and Horodecki, M. and Horodecki, P. and Horodecki, R.},
  title = {On Thermal Stability of Topological Qubit in {K}itaev's {4D} Model},
  journal = {Open Systems \& Information Dynamics},
  volume = {17},
  number = {1},
  pages = {1--20},
  year = {2010},
  doi = {10.1142/S1230161210000023}
}

@article{Temme2013,
  author = {Temme, Kristan},
  title = {Lower bounds to the spectral gap of {D}avies generators},
  journal = {Journal of Mathematical Physics},
  volume = {54},
  number = {12},
  pages = {122110},
  year = {2013},
  doi = {10.1063/1.4850896}
}

@article{Temme2017,
  author = {Temme, Kristan},
  title = {Thermalization Time Bounds for {P}auli Stabilizer {H}amiltonians},
  journal = {Communications in Mathematical Physics},
  volume = {350},
  number = {2},
  pages = {603--637},
  year = {2017},
  doi = {10.1007/s00220-016-2746-0}
}

@article{BardetEtAl2023,
  author = {Bardet, Ivan and Capel, {\'A}ngela and Gao, Li and Lucia, Angelo and P{\'e}rez-Garc{\'i}a, David and Rouz{\'e}, Cambyse},
  title = {Rapid Thermalization of Spin Chain Commuting {H}amiltonians},
  journal = {Physical Review Letters},
  volume = {130},
  number = {6},
  pages = {060401},
  year = {2023},
  doi = {10.1103/PhysRevLett.130.060401}
}

@article{KochanowskiEtAl2025,
  author = {Kochanowski, Jan and Alhambra, {\'A}lvaro M. and Capel, {\'A}ngela and Rouz{\'e}, Cambyse},
  title = {Rapid Thermalization of Dissipative Many-Body Dynamics of Commuting {H}amiltonians},
  journal = {Communications in Mathematical Physics},
  volume = {406},
  pages = {176},
  year = {2025},
  doi = {10.1007/s00220-025-05353-y}
}

@article{RouzeFrancaAlhambra2026,
  author = {Rouz{\'e}, Cambyse and Stilck Fran{\c c}a, Daniel and Alhambra, {\'A}lvaro M.},
  title = {Efficient thermalization and universal quantum computing with quantum {G}ibbs samplers},
  journal = {Nature Physics},
  volume = {22},
  pages = {1123--1127},
  year = {2026},
  doi = {10.1038/s41567-026-03246-y}
}

@inproceedings{BakshiEtAl2024,
  author = {Bakshi, Ainesh and Liu, Allen and Moitra, Ankur and Tang, Ewin},
  title = {High-Temperature {G}ibbs States are Unentangled and Efficiently Preparable},
  booktitle = {2024 IEEE 65th Annual Symposium on Foundations of Computer Science (FOCS)},
  pages = {1027--1036},
  year = {2024},
  doi = {10.1109/FOCS61266.2024.00068}
}

@inproceedings{BergamaschiChen2026,
  author = {Bergamaschi, Thiago and Chen, Chi-Fang},
  title = {Fast Mixing of Quantum Spin Chains at All Temperatures},
  booktitle = {Proceedings of the 58th Annual ACM Symposium on Theory of Computing (STOC)},
  pages = {835--844},
  year = {2026},
  doi = {10.1145/3798129.3800798}
}

@misc{BertoniDePalma2026,
  author = {Bertoni, Nicola and De Palma, Giacomo},
  title = {Rapid mixing for {G}ibbs samplers for quantum spin systems with all-to-all interactions},
  year = {2026},
  eprint = {2610.40361},
  archivePrefix = {arXiv},
  primaryClass = {quant-ph}
}

@article{TemmeEtAl2011,
  author = {Temme, K. and Osborne, T. J. and Vollbrecht, K. G. and Poulin, D. and Verstraete, F.},
  title = {Quantum {M}etropolis sampling},
  journal = {Nature},
  volume = {471},
  pages = {87--90},
  year = {2011},
  doi = {10.1038/nature09770}
}

@article{RallWangWocjan2023,
  author = {Rall, Patrick and Wang, Chunhao and Wocjan, Pawel},
  title = {Thermal State Preparation via Rounding Promises},
  journal = {Quantum},
  volume = {7},
  pages = {1132},
  year = {2023},
  doi = {10.22331/q-2023-10-10-1132}
}

@article{ChenKastoryanoBrandaoGilyen2025,
  author = {Chen, Chi-Fang and Kastoryano, Michael J. and Brand{\~a}o, Fernando G. S. L. and Gily{\'e}n, Andr{\'a}s},
  title = {Efficient quantum thermal simulation},
  journal = {Nature},
  volume = {646},
  pages = {561--566},
  year = {2025},
  doi = {10.1038/s41586-025-09583-x}
}

@article{HongGuoLucas2025,
  author = {Hong, Yifan and Guo, Jinkang and Lucas, Andrew},
  title = {Quantum memory at nonzero temperature in a thermodynamically trivial system},
  journal = {Nature Communications},
  volume = {16},
  pages = {316},
  year = {2025},
  doi = {10.1038/s41467-024-55570-7}
}

@misc{PlackeEtAl2024,
  author = {Placke, Benedikt and Rakovszky, Tibor and Breuckmann, Nikolas P. and Khemani, Vedika},
  title = {Topological Quantum Spin Glass Order and its realization in {qLDPC} codes},
  year = {2024},
  eprint = {2412.13248},
  archivePrefix = {arXiv},
  primaryClass = {quant-ph}
}
\bibliographystyle{unsrt}

\end{document}